\documentclass[12pt]{article}

\usepackage{amsmath}
\usepackage{amssymb}
\usepackage{newtxtext,newtxmath}

\usepackage[none]{hyphenat}
\usepackage[utf8]{inputenc}
\usepackage{graphicx}         
\usepackage{geometry}
\usepackage{authblk}

\usepackage{abstract}

\usepackage[numbers,sort&compress]{natbib}
\usepackage{hyperref}         
\usepackage{cleveref} %
\usepackage{lipsum}           
\usepackage{float}
\usepackage{subcaption}
\usepackage{xcolor}
\usepackage{soul}   %
\usepackage{booktabs}
\usepackage{subcaption}
\usepackage{threeparttable}
\usepackage{tabularx}
\usepackage{array}
\usepackage{multirow}
\usepackage[table]{xcolor} 
\usepackage{makecell} 
\usepackage[colorinlistoftodos]{todonotes} 
\usepackage{algorithm}
\usepackage{algorithmic}
\usepackage{float}
\usepackage{pdflscape}
\usepackage{tabularx}
\usepackage{adjustbox}
\usepackage{makecell}
\usepackage[font=small,labelfont=bf,labelsep=period]{caption}
\usepackage{enumitem}
\usepackage{calc}
\usepackage{url}

\begin{document}

\newcommand{\markTODO}[1]{%
    \phantomsection 
    \addcontentsline{tdo}{todo}{#1}
    \textcolor{red}{\textsuperscript{\textbf{?}}}
}

\hypersetup{
    colorlinks=true,
    linkcolor=blue,
    filecolor=magenta,      
    urlcolor=cyan,
    citecolor=blue,
}






\title{A Bayesian predictive framework for adaptive interim-analysis timing with robust borrowing in confirmatory trials}

\author[1]{Meihua Long}
\author[1]{Tianyu Zheng}
\author[1]{Jiali Song}
\author[1]{Leen Huang}
\author[2]{Cong Zhang}
\author[1]{Qimeng Che}
\author[1,3,*]{Yan Hou}

\affil[1]{Department of Biostatistics, School of Public Health, Peking University, Beijing, China}

\affil[2]{China Novartis Institutes for BioMedical Research Co., Shanghai, China}

\affil[3]{Key Laboratory of Carcinogenesis and Translational Research (Ministry of Education), Peking University Cancer Hospital \& Institute, Beijing, China}

\affil[*]{Correspondence: Yan Hou. Email: \href{mailto:houyan@bjmu.edu.cn}{houyan@bjmu.edu.cn}}

\date{}

\maketitle
\vspace{-2.2em}

\begin{abstract}

Confirmatory phase III trials require rigorous evidence, yet for first-in-class (FIC) therapies they must often be designed when same-mechanism evidence is scarce. This uncertainty motivates planned interim analyses and makes phase II data from the same therapy a relevant source of prior evidence. However, both borrowing and repeated interim analyses must be calibrated to control the overall type I error rate. Because borrowing changes the evidence available at interim analyses relative to a non-borrowing group sequential design (GSD), it also raises the question of whether interim analysis timing should be prospectively adapted to the borrowing-adjusted evidence base. We propose a prespecified adaptive interim-timing framework based on Bayesian information borrowing and Bayesian predictive probability (\(B^2\)-FIC). The borrowing model is calibrated against phase II--phase III discrepancy scenarios to control overall type I error rate. At the first interim analysis (IA1), the calibrated model combines phase II information with accumulating phase III data to update the posterior. Bayesian predictive probabilities from this posterior select the earliest information fraction for the second interim analysis (IA2) that meets the efficacy criterion. In simulations, \(B^2\)-FIC maintained empirical type I error and improved interim power across different scenarios. Predictive probabilities derived from phase II and phase III IA1 data selected earlier IA2 than GSD when evidence was favorable. Two oncology case studies illustrate the framework. Overall, \(B^2\)-FIC provides a calibrated framework for adapting interim timing to borrowing-adjusted evidence, an emerging design problem in confirmatory trials.

\end{abstract}



\newpage

\section{Introduction}

Recent International Council for Harmonisation of Technical Requirements for Pharmaceuticals for Human Use (ICH) E20 draft guidance\citep{ICH2025E20} highlights the importance of prospectively planned adaptations in confirmatory trials, which can be used to address uncertainty during trial conduct. 
For first-in-class (FIC) and other innovative therapies, this uncertainty is compounded by limited mature class-specific evidence. The most directly relevant external evidence may come
from the therapy's own phase II cohort. Such cohorts are often initially evaluated using short-term endpoints. With continued follow-up, they may yield longer-term time-to-event outcomes aligned with the subsequent phase~III primary endpoint.
Borrowing such information may improve efficiency, but it also changes the information available for interim monitoring: evidence at a phase~III analysis depends jointly on accumulating phase~III data and maturing phase~II follow-up. 
Standard group-sequential schedules, specified using phase III information times, do not
directly account for the additional and time-varying information introduced by
borrowing. The mismatch itself is dynamic, changing as phase II follow-up accrues and its commensurability with phase III data becomes clearer. 
This motivates a prespecified adaptive timing rule that uses borrowing-updated evidence to identify when a future interim analysis should be conducted, which we formulate as a Bayesian predictive decision problem.

The timing of interim analyses is usually specified as part of the trial
design in adaptive and group-sequential trials, either through fixed
information fractions or through rules determined before trial initiation
\citep{bauer2016twenty,boumendil2024two}. Related work has considered design-stage optimisation of
interim timing
\citep{togo2013optimal,feng2024robust,he2025optimal}.
The work closest to our setting is \citep{wu2020optimizing}, who combine
commensurate borrowing with interim-timing optimisation in a Bayesian
framework, but the timing decision is still selected prospectively at the design stage.
Dynamic-borrowing group-sequential designs address a complementary problem:
external borrowing may affect inference, and in some designs subsequent
allocation, at prespecified interim looks; however, the information-driven monitoring schedule remains fixed.
\citep{kotalik2022group,zhang2022bayesian,chiaruttini2025bayesian}. Thus, existing
approaches may incorporate external information into interim analyses, but do
not re-select the timing of a future interim look using the borrowing-informed
evidence available after an earlier analysis. We formulate this adaptive interim-analysis timing 
as a Bayesian predictive decision problem in which the predictive
distribution reflects both accruing phase III data and external
evidence.

This formulation builds on two established methodological components. First,
Bayesian predictive monitoring uses the posterior predictive distribution to
evaluate future trial success given current data, and has been used for efficacy
or futility decisions at prespecified analysis times
\citep{dmitrienko2006bayesian,saville2014utility,yin2012phase, chen2022bayesian}, for sample-size and
development decisions during accrual \citep{broglio2014not,rufibach2016sequentially},
and for event-time forecasting \citep{aubel2021bayesian, fu2025bayesian}. These methods provide a
natural measure of whether a future analysis is likely to satisfy a decision
criterion, but they have not generally been used as a scheduling rule for
selecting the next interim-analysis time after an earlier look. Second, the
predictive calculation must be coupled with a borrowing rule that remains
robust to phase II--phase III inconsistency. Power prior, meta-analytic-predictive (MAP) prior, unit information prior, commensurate prior, and
elastic priors \citep{chen2000power,neuenschwander2010summarizing,jin2021unit,hobbs2011hierarchical,jiang2023elastic} provide flexible
mechanisms for discounting external information when prior-data exchangeability is
limited. However, because external borrowing can affect frequentist operating
characteristics, its use in confirmatory adaptive monitoring requires
calibration against a prespecified type I error constraint. We therefore treat
type I error-calibrated borrowing as an admissibility requirement before
applying Bayesian predictive probability (BPP) to select the timing of a future
interim analysis.

Existing approaches do not provide a prespecified adaptive timing rule that
uses phase~III data and phase~II evidence to determine when a future interim
analysis should be conducted. To address this gap, we propose \(B^2\)-FIC, a
prespecified design framework for FIC confirmatory survival trials. The
notation \(B^2\) reflects its two Bayesian components: type~I error-controlled
Bayesian borrowing to construct the evidence base, and Bayesian predictive
probability to select the earliest acceptable future interim-analysis timing.
Our contributions are threefold. First, we define interim-analysis timing under external borrowing as a prespecified statistical design problem, rather than as a purely operational choice of calendar time or fixed information fraction. Second, we develop a Bayesian predictive-probability rule that selects the earliest candidate IA information fraction crossing a prespecified threshold and summarizes the resulting recommendations in a design-stage decision table.
Third, we calibrate candidate borrowing methods at the design stage to control
the trial-wise type~I error rate, reflecting the regulatory requirements of
confirmatory trials. In the simulation, we
compared performance of the borrowing methods across prespecified discrepancy and design scenarios in
terms of type~I error, power, and effective sample size.
The remainder of this article is organized as follows. Section~\ref{sec:methods} formulates the framework and calibration strategy.
Section~\ref{sec:simulation} reports the simulation study.
Section~\ref{sec:case-study} presents the case study.
Section~\ref{sec:discussion} provides concluding remarks and discusses the implications of the findings.

\section{Methods}
\label{sec:methods}

\subsection{Problem formulation and framework overview}

\begin{figure}[H]
    \centering
    \includegraphics[width=1\linewidth]{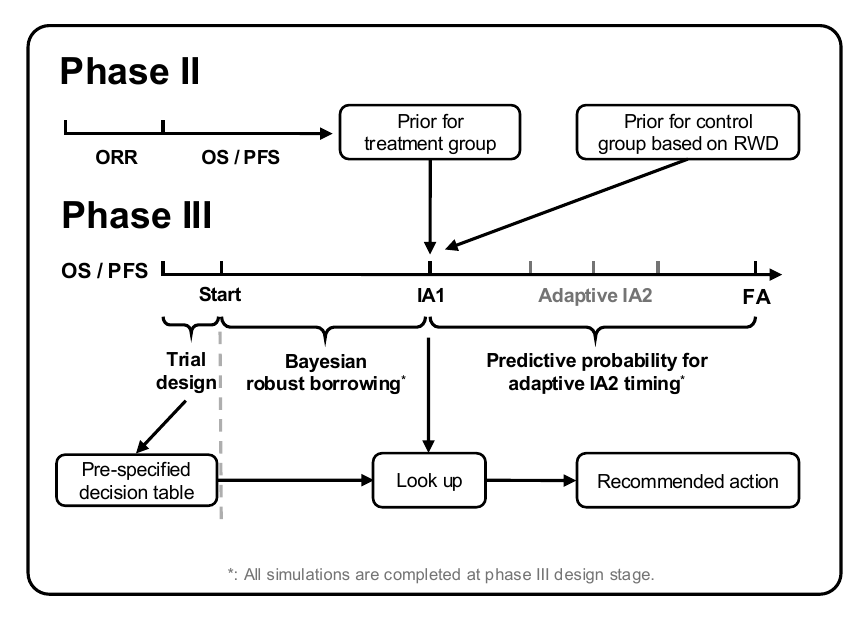}
    \caption{Overview of the proposed $B^2$-FIC framework. }
    \label{fig:workflow}
\end{figure}

We consider a confirmatory phase III trial whose primary endpoint is a long-term time-to-event outcome (e.g., progression-free or overall survival) and a prespecified sequence of planned analyses consisting of IA1, IA2, and a final analysis. The motivating development setting is one in which a preceding single-arm phase II study reads out on an earlier surrogate or short-term endpoint (e.g., objective response rate) to support the phase III go decision, while the phase II cohort continues to be followed for the same long-term time-to-event endpoint used in phase III. By the time phase~III interim analyses are conducted, additional
event-time information may therefore have accrued in the phase~II treatment
arm. This creates two linked methodological questions: how to incorporate the
continuing phase~II treatment-arm event-time information into phase~III
monitoring while controlling type~I error under possible phase~II--phase~III
prior--data conflict, and, conditional on a borrowing procedure that achieves
such calibration, how to use the resulting IA1 posterior evidence to determine
prospectively the earliest time at which IA2 is predicted to yield decisive
evidence.Figure~\ref{fig:workflow} summarizes the motivating trial setting and the design-stage and trial-conduct workflow of the proposed \(B^2\)-FIC framework.

The \(B^2\)-FIC framework rests on a central design principle: both the borrowing
mechanism and the timing decision rule are fully specified at the design stage,
whereas the realized influence of the phase~II evidence and the realized IA2
timing evolve as phase~III data accumulate. The framework therefore
prespecifies how continuing phase~II event-time information enters the IA1
posterior and how that posterior is translated into a recommended IA2 time,
without treating the phase~II data as a static historical input. 

At IA1, the design specifies an ordered candidate set of future IA2 times,
\[
  \mathcal{T}
  =
  \left\{
  \tau^{(1)} < \tau^{(2)} < \cdots < \tau^{(K)}
  \right\},
  \qquad
  \tau^{(1)}>\tau_1,
\]
where \(\tau_1\) is the calendar time of IA1. The candidate times may be
indexed by calendar time, target event count, or information fraction. The
corresponding IA1 action space is
\[
  \mathcal{A}
  =
  \left\{
  \text{stop for efficacy}
  \right\}
  \cup
  \mathcal{T}
  \cup
  \left\{
  \text{proceed directly to final analysis}
  \right\}.
\]
Choosing \(\tau^{(k)}\in\mathcal{T}\) corresponds to scheduling IA2 at time
\(\tau^{(k)}\). The action ``stop for efficacy'' terminates the trial at IA1
with a positive efficacy conclusion, whereas ``proceed directly to final
analysis'' corresponds to omitting IA2. All adaptive elements, including the
candidate set \(\mathcal{T}\), the borrowing specification, the posterior
efficacy boundaries, the predictive success threshold, and the mapping from
IA1 evidence to the recommended action, are fixed before phase~III initiation.

The remainder of this section develops the framework as follows. 
Section~\ref{sec:notation} establishes the outcome model and the common 
likelihood interface shared by all components. The framework then unfolds 
in two layers. The first layer 
(Section~\ref{sec:type1_calibrated_borrowing}) addresses inference under 
prespecified analysis times: it specifies the candidate borrowing 
procedures, the corresponding posterior 
decision rules, and the \(B^2\)-FIC calibration rule that ensures type~I error 
control under prior--data conflict. The second layer 
(Section~\ref{sec:predictive_timing}) builds adaptive IA2 timing on top 
of calibrated borrowing: the IA1 action is selected by evaluating 
Bayesian predictive probabilities over the candidate set \(\mathcal{T}\), 
with accelerated computation used to make the repeated predictive 
evaluation feasible.


\subsection{Notation and outcome model}
\label{sec:notation}
We adopt a Weibull proportional hazards parameterization for the phase III time-to-event endpoint. The notation used throughout the methodological development is as follows.

\begin{description}[
  leftmargin=!,
  labelwidth=0.7cm,
  labelsep=0.8em,
  align=right,
  font=\normalfont
]
  \item[\(g\):] Treatment group indicator, with \(g \in \{T,C\}\), where
  \(T\) denotes the treatment arm and \(C\) denotes the control arm;

  \item[\(s\):] Trial phase indicator, with \(s \in \{II,III\}\);

  \item[\(n_g^{(s)}\):] Sample size for group \(g\) in phase \(s\). When the
  group subscript is omitted,
  \(n^{(s)}=\sum_{g\in\{T,C\}} n_g^{(s)}\) denotes the total sample size in
  phase \(s\);

  \item[\(m_g^{(s)}\):] Median event time for group \(g\) in phase \(s\);

  \item[\(\lambda_g^{(s)}\):] Weibull scale parameter for group \(g\) in
  phase \(s\), with cumulative hazard
  \(\Lambda_g^{(s)}(t)=\lambda_g^{(s)}t^k\), hazard function
  \(h_g^{(s)}(t)=\lambda_g^{(s)}k t^{k-1}\), and survival function
  \(S_g^{(s)}(t)=\exp\{-\lambda_g^{(s)}t^k\}\);

  \item[\(k\):] Weibull shape parameter;

  \item[\(\mathrm{HR}^{(s)}\):] Treatment-to-control hazard ratio in
  phase \(s\), defined as
  \(\mathrm{HR}^{(s)}=\lambda_T^{(s)}/\lambda_C^{(s)}\);

  \item[\(\delta\):] Phase~II--phase~III treatment-arm discrepancy in the
treatment-arm Weibull scale parameter, defined on the log scale by
  \(\lambda_T^{(II)}=\lambda_T^{(III)}\exp(\delta)\). Under the shared
  Weibull shape parameter \(k\), this is equivalent to a discrepancy on the
  log-hazard scale up to a constant shift;

  \item[\(j\):] Planned analysis index, with \(j=1,\ldots,J\). In the main
  simulations, \(J=3\), corresponding to IA1, IA2, and the final analysis;

  \item[\(q_j\):] Information fraction for the \(j\)th planned analysis;

  \item[\(E_j\):] Target number of phase~III events at the \(j\)th planned
  analysis, \(E_j=\mathrm{round}(q_jE_{\mathrm{FA}})\), where
  \(E_{\mathrm{FA}}\) denotes the target number of events at the final
  analysis;

  \item[\(\mathcal{D}_{g,j}^{(s)}\):] Event-time data for group \(g\) in
  phase \(s\), administratively censored at the calendar time of the \(j\)th
  phase~III analysis. For phase~III data
  \(\mathcal{D}_{g,j}^{(III)}\), this is the usual interim or final analysis
  data set. For the phase~II treatment arm, \(\mathcal{D}_{T,j}^{(II)}\)
  accumulates additional events between phase~III analyses because the
  phase~II cohort remains under follow-up after the short-term phase~II
  readout;

  \item[\(\mathcal{D}_j\):] All event-time data available at the \(j\)th
  phase~III analysis,
  \[
    \mathcal{D}_j
    =
    \left\{
    \mathcal{D}_{T,j}^{(II)},
    \mathcal{D}_{T,j}^{(III)},
    \mathcal{D}_{C,j}^{(III)}
    \right\}.
  \]
\end{description}

External real world data (RWD) on the control-arm event-time distribution is incorporated through
an informative prior on \(\lambda_C^{(III)}\) rather than as a dynamically
accumulating data source, and is therefore not indexed by \(j\).

We treat the Weibull shape parameter \(k\) as known and common across phases
and arms, so proportional hazards follows from the group-specific
\(\lambda_g^{(s)}\) parameters. Censoring is administrative at each planned analysis; the continuing phase II treatment arm contributes accumulating events between phase III analyses, and loss-to-follow-up is not modelled.

\subsection{Bayesian robust borrowing}
\label{sec:type1_calibrated_borrowing}

Borrowing external evidence into a confirmatory trial entails a tension between efficiency under phase~II--phase~III congruence and type~I error control under prior--data conflict. We first include a conjugate prior as a full-borrowing benchmark. For calibrated treatment-arm borrowing, we focus on commensurate and elastic priors because they provide two complementary mechanisms for adapting the amount of borrowing to phase~II--phase~III discrepancy: the former models commensurability through a hierarchical parameter, whereas the latter discounts the phase~II contribution through an explicit discrepancy-dependent rule. Power-prior, MAP, and unit-information approaches provide important general frameworks for prior construction, but are less directly aligned with our setting, where borrowing is based on a single phase~II source and calibrated according to the phase~II--phase~III discrepancy.

To make the confirmatory type~I error constraint explicit, We use a modular \(B^2\)-FIC calibration rule that can be wrapped
around a candidate borrowing procedure. The rule anchors borrowing strength at
prespecified phase~II--phase~III discrepancy boundaries
\(\delta=\pm\delta^*\) on the treatment-arm log-rate scale. Under the shared
Weibull shape parameter \(k\), this is equivalent to calibration on the
log-hazard scale up to a constant shift. At these boundaries, the calibrated
procedure is required to maintain the overall one-sided type~I error rate at or
below a nominal level \(\alpha^*\). Within this constraint, the calibration
favors the most efficient borrowing rule, as measured by power under a
prespecified design alternative. Calibration is performed at the design stage,
so that the borrowing rule is fixed before phase~III initiation.

\subsubsection{Candidate borrowing methods}
\label{sec:candidate_methods}

We considered three candidate procedures for incorporating phase~II
treatment-arm information into the phase~III analysis of
\(\lambda_T^{(III)}\). Under the Weibull model with common shape
parameter \(k\), let
\[
  \mathcal{D}_{g,j}^{(s)}
  =
  \left\{
  (t_{ig,j}^{(s)}, e_{ig,j}^{(s)}):
  i=1,\ldots,n_g^{(s)}
  \right\}
\]
denote the event-time data for group \(g\) in phase \(s\), administratively
censored at analysis \(j\), where \(t_{ig,j}^{(s)}\) is the observed
follow-up time and \(e_{ig,j}^{(s)}\in\{0,1\}\) is the event indicator. The
likelihood contribution for \(\lambda_g^{(s)}\) has kernel
\[
  L\!\left(
  \lambda_g^{(s)}\mid \mathcal D_{g,j}^{(s)}
  \right)
  \propto
  \left(\lambda_g^{(s)}\right)^{\widehat E_{g,j}^{(s)}}
  \exp\left\{
    -\lambda_g^{(s)}\widehat T_{g,j}^{(s)}
  \right\},
\]
where
\[
  \widehat E_{g,j}^{(s)}
  =
  \sum_{i=1}^{n_g^{(s)}} e_{ig,j}^{(s)},
  \qquad
  \widehat T_{g,j}^{(s)}
  =
  \sum_{i=1}^{n_g^{(s)}}
  \left(t_{ig,j}^{(s)}\right)^k .
\]
Thus, conditional on \(k\), \(\widehat E_{g,j}^{(s)}\) and
\(\widehat T_{g,j}^{(s)}\) are sufficient statistics for
\(\lambda_g^{(s)}\), corresponding to the observed number of events and the
Weibull-adjusted total follow-up time, respectively. With a Gamma
distribution parameterized by shape and rate, conjugate updating follows
directly from this likelihood representation. The derivation of the Weibull
likelihood and the sufficient statistics is given in
Appendix~S1.

\paragraph{Conjugate prior.}
The conjugate prior corresponds to full borrowing of the continuing
phase~II treatment-arm data. At analysis \(j\), the phase~II
treatment-arm data induce the prior
\[
  \lambda_T^{(III)}
  \mid
  \mathcal D_{T,j}^{(II)}
  \sim
  \mathrm{Gamma}
  \left(
    \widehat E_{T,j}^{(II)},
    \widehat T_{T,j}^{(II)}
  \right).
\]
After incorporating the phase~III treatment-arm
data, the posterior is
\[
  \lambda_T^{(III)}
  \mid
  \mathcal D_{T,j}^{(II)},\mathcal D_{T,j}^{(III)}
  \sim
  \mathrm{Gamma}
  \left(
    A_{T,j}^{(II)}+\widehat E_{T,j}^{(III)},
    B_{T,j}^{(II)}+\widehat T_{T,j}^{(III)}
  \right).
\]
This procedure provides a full-borrowing benchmark, but it does not
discount phase~II information when the phase~II and phase~III
treatment-arm data are incongruent. The detailed derivation is given in Appendix~S2.

\paragraph{Commensurate prior.}
In contrast to the fixed full-borrowing conjugate prior, the commensurate
prior\citep{hobbs2011hierarchical} provides a dynamic borrowing approach by modelling the phase~II and
phase~III treatment-arm Weibull scale parameters as related but not
necessarily exchangeable.
Let \(\lambda_0=\lambda_T^{(II)}\),
\(\phi_0=\log(\lambda_0)\), and
\(\phi=\log\{\lambda_T^{(III)}\}\). A weak proper Gamma prior is assigned
to the phase~II treatment-arm scale parameter,
\[
  \lambda_0 \sim \mathrm{Gamma}(A_0,B_0),
\]
with \(A_0=B_0=0.01\) in the main implementation. Borrowing is then
introduced on the log scale through
\[
  \phi \mid \phi_0,\sigma_{\mathrm{bor}}
  \sim
  N\left(\phi_0,\sigma_{\mathrm{bor}}^2\right).
\]
Small values of \(\sigma_{\mathrm{bor}}\) imply strong commensurability
and therefore stronger borrowing, whereas larger values allow
\(\lambda_T^{(III)}\) to depart from the phase~II parameter. Under the
common Weibull shape parameter, this is equivalent to modelling
commensurability on the log-hazard scale up to an additive constant.
Posterior inference for \(\lambda_T^{(III)}\) is obtained by Markov chain Monte Carlo (MCMC).

\paragraph{Elastic prior.}
Building on the idea of dynamic borrowing, the elastic prior\citep{jiang2023elastic} makes the
borrowing rule explicit by discounting the phase~II likelihood-induced Gamma
contribution through a prespecified function of an analysis-specific
discrepancy statistic, thereby facilitating design-stage calibration of
operating characteristics such as type~I error and power. Let
\[
  A_{T,j}^{(II)}
  =
  \widehat E_{T,j}^{(II)},
  \qquad
  B_{T,j}^{(II)}
  =
  \widehat T_{T,j}^{(II)}
\]
denote the raw Gamma shape and rate parameters induced by the phase~II
treatment-arm likelihood. Let
\(C_j=C\{\mathcal D_{T,j}^{(II)},\mathcal D_{T,j}^{(III)}\}\) be a
nonnegative discrepancy statistic, with larger values indicating
greater phase~II--phase~III incongruence. The borrowing weight is
\[
  w_j(a_{\mathrm{el}},b_{\mathrm{el}})
  =
  \frac{1}{1+\exp\{a_{\mathrm{el}}+
  b_{\mathrm{el}}\log(C_j)\}},
  \qquad b_{\mathrm{el}}>0 .
\]
The treatment-arm prior is then
\[
  \lambda_T^{(III)}
  \mid
  \mathcal D_{T,j}^{(II)},C_j
  \sim
  \mathrm{Gamma}
  \left(
    w_j(a_{\mathrm{el}},b_{\mathrm{el}})A_{T,j}^{(II)},
    w_j(a_{\mathrm{el}},b_{\mathrm{el}})B_{T,j}^{(II)}
  \right).
\]
Thus, the phase~II prior mean is preserved, whereas its variance is
inflated when \(w_j(a_{\mathrm{el}},b_{\mathrm{el}})<1\), reducing the
effective contribution of phase~II information under incongruence.

\paragraph{Control arm prior.}
The same control-arm prior was used across all candidate treatment-arm
borrowing procedures:
\[
  \lambda_C^{(III)}
  \sim
  \mathrm{Gamma}
  \left(
    \alpha_C,\;
    \alpha_C/\lambda_C^{\mathrm{ext}}
  \right),
\]
where \(\lambda_C^{\mathrm{ext}}\) is the RWD-derived Weibull scale
parameter. This prior has mean \(\lambda_C^{\mathrm{ext}}\) under the
shape--rate parameterization and is updated conjugately using
\(\mathcal D_{C,j}^{(III)}\). Therefore, differences in operating
characteristics reflect the treatment-arm borrowing strategy.

\subsubsection{Posterior decision rule}
\label{sec:posterior_decision_rule}
Under the shared-shape Weibull model, we define the phase~III treatment
effect so that positive values indicate treatment benefit:
\[
  \theta
  =
  -\log\{\mathrm{HR}^{(III)}\}
  =
  \log\left\{
  \frac{\lambda_C^{(III)}}{\lambda_T^{(III)}}
  \right\},
\]
where \(\mathrm{HR}^{(III)}=\lambda_T^{(III)}/\lambda_C^{(III)}\).
Thus, \(\theta>0\) corresponds to a lower hazard in the treatment arm, and the
null hypothesis is \(H_0:\theta\leq 0\).

At analysis \(j\), each borrowing procedure yields the posterior probability
of phase~III treatment benefit,
\[
  p_j
  =
  \Pr\left(\theta>0\mid \mathcal D_j\right),
\]
where \(\mathcal D_j\) denotes all event-time data available at the \(j\)th
phase~III analysis. To align the Bayesian monitoring rule with the
frequentist group sequential design reference, the posterior probability
cutoff at analysis \(j\) was set to \(1-\gamma_j\), where \(\gamma_j\) is the
one-sided O'Brien--Fleming p-value boundary at information fraction \(q_j\).

The trial was declared positive at the first analysis \(j\) satisfying
\[
  p_j > 1-\gamma_j .
\]
If this criterion was not met, the trial continued to the next planned
analysis. All borrowing procedures used the same posterior decision rule and
the same boundary sequence \(\{\gamma_j\}_{j=1}^J\), so that differences in
operating characteristics reflected the borrowing specification rather than
the monitoring criterion.

\subsubsection{\(B^2\)-FIC calibration principle}

To better align Bayesian information borrowing with the regulatory
expectations for confirmatory phase~III trials, we embedded a design-stage
calibration step in the proposed framework \(B^2\)-FIC. Specifically, each borrowing
specification was calibrated before the phase~III trial using the trial-wise
one-sided type~I error rate as the key operating characteristic.
For each candidate borrowing
methods, the overall type~I error was evaluated under a null phase~III
treatment effect, with rejection accumulated across IA1, IA2, and the final
analysis. A specification was considered admissible only if the type~I error
at both boundary discrepancies was no larger than the nominal level
\(\alpha^*\). Efficiency was then assessed only within this admissible class.

This principle was applied to two dynamic borrowing priors. For the
commensurate prior, our calibrated \(B^2\)-CP replaces the conventional half-normal prior on
the commensurability scale with the robust mixture
\[
  \sigma_{\mathrm{bor}}
  \sim
  \frac{1}{2}\mathrm{HalfNormal}(s_{\mathrm{spike}})
  +
  \frac{1}{2}\mathrm{HalfNormal}(s_{\mathrm{slab}}),
  \qquad
  s_{\mathrm{spike}} < s_{\mathrm{slab}} .
\]
The spike component supports borrowing when the phase~II and phase~III
treatment-arm data are compatible, whereas the slab component permits
attenuation of borrowing under incongruence. The mixture scales are fixed
before the phase~III trial and verified by simulation to satisfy the boundary
type~I error requirement. Details of the
calibration procedure are provided in Appendix~S3.

For the elastic prior, \(B^2\)-EP selects the elastic-function
hyperparameters through a constrained search over the quantile anchors
\((q_0,q_1)\). The search was designed to retain only anchors satisfying the
prespecified type~I error requirement, while avoiding unnecessarily
conservative choices among the admissible candidates. Let
\(\delta=\pm\delta^*\) denote the prespecified boundary phase~II--phase~III
discrepancies used for calibration, and let
\(\alpha_{+}^{\mathrm{trial}}(q_0,q_1)\) and
\(\alpha_{-}^{\mathrm{trial}}(q_0,q_1)\) denote the corresponding overall
trial-wise one-sided type~I error rates. The calibrated feasible set is
\[
  \mathcal Q_{\mathrm{cal}}
  =
  \left\{
  (q_0,q_1):
  \max\left[
    \alpha_{+}^{\mathrm{trial}}(q_0,q_1),
    \alpha_{-}^{\mathrm{trial}}(q_0,q_1)
  \right]
  \leq
  \alpha^*
  \right\}.
\]
The final anchors are selected as
\[
  (q_0^{B^2},q_1^{B^2})
  =
  \arg\max_{(q_0,q_1)\in\mathcal Q_{\mathrm{cal}}}
  \mathrm{Power}(q_0,q_1),
\]
where power is evaluated under the prespecified design alternative. Thus,
\(B^2\)-EP uses type~I error control as a feasibility constraint and then
selects the most efficient anchors within the feasible region. In contrast,
\(B^2\)-CP imposes the calibration through the prior structure. In both cases,
the borrowing rule is fully prespecified and its operating characteristics are
evaluated before observing phase~III data.Details of the
calibration procedure are provided in Appendix~S4.

\subsection{Predictive probability for adaptive IA2 timing}
\label{sec:predictive_timing}

The first component of the framework calibrates the borrowing
specification to preserve type~I error control under prespecified
phase~II--phase~III boundary discrepancies. A consequence is that the
IA1 posterior carries information beyond what a phase~III-only group
sequential design would have at the same calendar time, while the
boundary type~I error guarantee remains in force. The next objective is to determine whether the information
contributed by borrowing can be used to improve monitoring efficiency.
We therefore use predictive probability to adapt the timing of
IA2, rather than to modify any efficacy rule. For each candidate
future information fraction, we evaluate the IA1 posterior-predictive
probability that the prespecified posterior efficacy criterion will be
met when the corresponding event count is reached. This probability is
used only to decide when to perform IA2; all efficacy decisions remain
based on the posterior boundary rule in
Section~\ref{sec:posterior_decision_rule} and the O'Brien--Fleming
boundary sequence.

\subsubsection{Bayesian predictive-probability timing rule}

At IA1, the calibrated Bayesian borrowing model yields the posterior
probability of phase~III treatment benefit,
\[
  p_1
  =
  \Pr\left(\theta>0\mid\mathcal D_1\right),
  \qquad
  \theta
  =
  \log\left\{
  \frac{\lambda_C^{(III)}}{\lambda_T^{(III)}}
  \right\}.
\]
If \(p_1>1-\gamma_1\), where \(\gamma_1\) is the prespecified IA1
one-sided efficacy boundary on the \(p\)-value scale, the trial is declared positive at IA1. Otherwise, IA2 timing is determined by Bayesian posterior prediction from \(\mathcal D_1\).

Let
\[
  \mathcal R=\{r_1,\ldots,r_L\},
  \qquad
  q_1<r_1<\cdots<r_L<1,
\]
denote the prespecified candidate IA2 information fractions, where \(q_1\) is
the fixed IA1 information fraction. For candidate \(r\in\mathcal R\), define
\[
  E(r)=\mathrm{round}\{rE_{\mathrm{FA}}\},
\]
the corresponding target number of phase~III events. Let \(\gamma_2(r)\) be
the one-sided O'Brien--Fleming p-value boundary at IA2 for the candidate
schedule \((q_1,r,1)\). On the posterior probability scale, the corresponding
efficacy threshold is \(1-\gamma_2(r)\).

For each candidate \(r\), we define the IA1 Bayesian predictive probability
of meeting the IA2 efficacy criterion as
\[
  \mathrm{BPP}(r)
  =
  \Pr\left[
  \Pr\left\{\theta>0\mid \mathcal D_r^{\mathrm{pred}}\right\}
  >
  1-\gamma_2(r)
  \,\middle|\,
  \mathcal D_1
  \right].
\]

Here \(\mathcal D_r^{\mathrm{pred}}\) denotes the predicted data available when
the phase~III trial reaches \(E(r)\) events,
\[
  \mathcal D_r^{\mathrm{pred}}
  =
  \left\{
  \mathcal D_{T,r}^{(II,\mathrm{pred})},
  \mathcal D_{T,r}^{(III,\mathrm{pred})},
  \mathcal D_{C,r}^{(III,\mathrm{pred})}
  \right\}.
\]
Thus, the inner probability is the posterior probability of treatment benefit
that would be computed at the candidate IA2 under the same calibrated Bayesian
borrowing model; the outer probability averages the corresponding efficacy
indicator over the posterior predictive distribution of future data conditional
on \(\mathcal D_1\). The phase~II treatment-arm component
is updated to the calendar time corresponding to \(E(r)\), consistent with the
data structure used for \(\mathcal D_j\).

Given a prespecified predictive-probability target
\(\eta_{\mathrm{PP}}\), the recommended IA2 information fraction is
\[
  r^*
  =
  \min\left\{
  r\in\mathcal R:
  \mathrm{PP}(r)\geq \eta_{\mathrm{PP}}
  \right\}.
\]
If no candidate satisfies this criterion, the trial proceeds to the final
analysis without recommending IA2.

\subsubsection{Accelerated computation}

A direct predictive-probability calculation requires nested simulation. For
each simulated IA1 data set, posterior samples are drawn from the calibrated
borrowing model. For each posterior draw, future event times and accrual are
simulated up to each candidate IA2 information fraction, and the posterior
efficacy criterion is recomputed at each candidate analysis. This
simulation-within-simulation structure is computationally demanding,
especially for commensurate-prior borrowing, where posterior inference requires
MCMC. In our implementation, a single design scenario with 1000 outer trial
simulations and 1000 inner posterior-predictive trajectories required
approximately 13 days when repeated MCMC refitting was used at each candidate
IA2 analysis.

We therefore used two complementary implementations. For \(B^2\)-EP, posterior
updating remains conjugate once the elastic borrowing weight has been
computed, so the imputed predictive probability estimator is computationally
feasible. For \(B^2\)-CP, individual posterior fits were accelerated using
PyMC with the nutpie sampler, but repeated MCMC refitting at all candidate
future analyses remained prohibitive. We therefore used an approximate
predictive probability based on the current posterior probability and a
borrowing-adjusted effective information ratio.

For \(B^2\)-EP, let
\[
  \left(
  \lambda_T^{(III,m)},\lambda_C^{(III,m)}
  \right)
  \sim
  \pi_{\mathrm{EP}}
  \left(
  \lambda_T^{(III)},\lambda_C^{(III)}
  \mid
  \mathcal D_1
  \right),
  \qquad m=1,\ldots,M .
\]
Conditional on this draw, future phase~III event times are generated under the
Weibull model. For subjects event-free at IA1 with elapsed follow-up \(u_i\),
the total event time is simulated from the conditional Weibull distribution,
\[
  Y_i^{(m)}
  =
  \left\{
  u_i^k
  -
  \frac{\log U_i}{\lambda_g^{(III,m)}}
  \right\}^{1/k},
  \qquad
  U_i\sim \mathrm{Uniform}(0,1),
  \qquad
  g\in\{T,C\}.
\]
For newly accrued subjects, entry times are generated according to the
prespecified accrual model and event times are sampled from the Weibull model.
At each candidate \(r\), the simulated data are administratively censored at
the calendar time when \(E(r)\) phase~III events have accrued. The elastic
discrepancy statistic, borrowing weight, and posterior probability of benefit
are then recomputed using the predicted data at that candidate analysis. The
imputed estimator is
\[
  \widehat{\mathrm{PP}}_{\mathrm{EP}}(r)
  =
  \frac{1}{M}
  \sum_{m=1}^M
  I\left[
  p_r^{(m)}>1-\gamma_2(r)
  \right],
\]
where
\[
  p_r^{(m)}
  =
  \Pr\left(
  \theta>0
  \mid
  \mathcal D_r^{(m)}
  \right).
\]

For \(B^2\)-CP, direct imputed predictive probability would require repeated
MCMC refitting over posterior-predictive trajectories and candidate IA2
analyses. We therefore used the approximate predictive-probability approach of
\citep{marion2025predictive}, which computes the predictive probability from
the current posterior probability and an information ratio. Because nominal
information fractions do not account for the information contributed by
borrowing, we used a borrowing-adjusted effective information fraction.

Let \(z_1=\Phi^{-1}(p_1)\). Let
\[
  V_{\mathrm{ref},1}
  =
  \mathrm{Var}_{\mathrm{ref}}
  \left(
  \theta\mid\mathcal D_1
  \right),
  \qquad
  V_{\mathrm{CP},1}
  =
  \mathrm{Var}_{\mathrm{CP}}
  \left(
  \theta\mid\mathcal D_1
  \right)
\]
denote the posterior variances of \(\theta\) under a no-borrowing reference
analysis and under \(B^2\)-CP, respectively. The effective IA1 information
fraction was estimated as
\[
  I_{1,\mathrm{eff}}
  =
  q_1
  \frac{V_{\mathrm{ref},1}}{V_{\mathrm{CP},1}},
  \qquad
  I_{\mathrm{bor}}
  =
  \max\left(I_{1,\mathrm{eff}}-q_1,0\right).
\]
For candidate \(r\), define
\[
  I_{r,\mathrm{eff}}=r+I_{\mathrm{bor}},
  \qquad
  \rho_r=\frac{I_{1,\mathrm{eff}}}{I_{r,\mathrm{eff}}},
  \qquad
  z_r^*=\Phi^{-1}\{1-\gamma_2(r)\}.
\]
The approximate predictive probability is then
\[
  \widetilde{\mathrm{PP}}_{\mathrm{CP}}(r)
  =
  \Phi
  \left[
  \frac{
    z_1-z_r^*\sqrt{\rho_r}
  }{
    \sqrt{1-\rho_r}
  }
  \right],
\]
where \(1-\gamma_2(r)\) is the posterior probability threshold corresponding
to the candidate IA2 O'Brien--Fleming boundary.

The same posterior-predictive data-generation mechanism as in \(B^2\)-EP was
used to estimate the expected calendar time associated with each candidate
information fraction and to validate the approximation. Full implementation details are provided in Appendix~S5 and
Appendix~S6.

In the benchmark scenario with 1000 outer simulations and 1000 inner
trajectories, the accelerated implementation reduced computation time from
approximately 13 days to about 3 hours. Paired comparisons with the direct
posterior-predictive estimator showed that the approximation reproduced the
average predictive probabilities reasonably well for design-stage screening,
although discrepancies were larger at the earliest candidate information
fraction. The mean difference between the direct and accelerated estimates
decreased from 0.070 at \(r=0.5\) to 0.005 at \(r=0.8\). We therefore used the
approximate calculation for design-stage search and scenario screening, with
direct posterior-predictive calculation retained for validation of selected
rules.

\subsubsection{Decision table construction}

To facilitate clinical implementation, the adaptive timing rule was implemented as a
design-stage decision table over a prespecified grid of IA1 evidence patterns,
indexed by phase~II and phase~III IA1 hazard-ratio intervals. Each cell maps
the observed IA1 evidence pattern to an action in the ordered set
\[
  \{\mathrm{IA1}, r_1,\ldots,r_L,\mathrm{FA}\},
\]
arranged by increasing information fraction.

For each Monte Carlo simulation, \(B^2\)-CP and \(B^2\)-EP each produced a
recommended action through the Bayesian predictive-probability rule. Because
the two procedures use the same calibration anchors but different borrowing
mechanisms, their paired recommendations were combined by taking the later
action on the ordered scale. This gives a conservative integrated action.

Within each hazard-ratio cell, the integrated actions define an empirical
distribution on the ordered action set. The cell-level recommendation was the
cumulative median, or rule-50 action, defined as the earliest action whose
cumulative empirical proportion reaches \(50\%\). This summary respects the
ordinal action scale, unlike the mode, which treats all disagreements between
actions as equally distant.

To enforce the design prior that less favourable hazard-ratio values should
not lead to earlier recommended actions, let \(\widehat F(k,i,j)\) denote the
empirical CDF of the integrated actions, indexed by action level \(k\) and
grid indices \((i,j)\) for
\((\mathrm{HR}_{II},\mathrm{HR}_{III})\). We projected \(\widehat F\) onto
\[
\mathcal C
=
\left\{
F:
F \text{ is non-decreasing in } k,\,
\text{non-increasing in } i,\,
\text{and non-increasing in } j
\right\}
\]
using Dykstra's cyclic projection algorithm
\citep{dykstra1983,robertson1988order}, with one-dimensional projections
computed by the pool-adjacent-violators algorithm. The rule-50 action was then
recomputed from the projected CDF \(\widehat F^*\), giving a decision table
that is monotone in both hazard-ratio axes by construction.

\section{Simulations}
\label{sec:simulation}
\subsection{Settings}
We conducted a simulation study to evaluate the performance of our proposed framework across realistic and challenging FIC phase III clinical trial scenarios. To clearly present the simulation settings, we structured the simulation settings according to the ADEMP framework.\citep{morris2019using}

\paragraph{Aims.}
To support prospective, evidence-adaptive selection of the
IA2 timing from phase~II and phase~III IA1 data, we
evaluated two linked components of the proposed framework.
The first concerned the calibrated borrowing module:
whether the proposed $B^2$-FIC calibration controls the
final-analysis false positive rate at a level comparable
to the group sequential design, while preserving efficiency
under clinically relevant alternatives. We assessed this
across variations in the degree of phase~II--phase~III
discrepancy, phase~II sample size, baseline event-time
scale, treatment-effect magnitude, and information-accrual
pattern. The second, conditional on the first, concerned
the adaptive IA2 timing module: whether the Bayesian
predictive probability rule, applied after calibrated
borrowing, yields an actionable interval-based decision
table for the earliest admissible IA2 timing.

\paragraph{Data-generating mechanisms.}
The phase~III primary endpoint was progression-free survival (PFS).
Individual PFS times were generated under the Weibull distribution, 
with a common shape parameter fixed at \(k=0.8\), representing a moderately decreasing hazard pattern. For each scenario, the phase~III
control-arm median PFS \(m_C^{(III)}\) was prespecified, and the
phase~III treatment-arm median PFS was determined by the target
hazard ratio \(\mathrm{HR}^{(III)}\). Phase~II was assumed to be a single-arm study of the investigational treatment, whereas the phase~III trial included both treatment and control arms. The external RWD control-arm median PFS was set equal to \(m_C^{(III)}\), and this RWD source was used to inform the phase~III control-arm prior. Phase~II--phase~III discrepancy was
introduced through the treatment arm by
\[
  \lambda_T^{(II)}=\lambda_T^{(III)}\exp(\delta),
\]
equivalently,
\[
  m_T^{(II)}=m_T^{(III)}\exp(-\delta/k).
\]

Phase~II enrollment began at calendar time zero and followed uniform
accrual over 12 months. Phase~III enrollment was assumed to start at
month~16 after an early phase~II response signal and also followed
uniform accrual over 12 months. The final-analysis event target
\(E_{\mathrm{FA}}\) was obtained from the corresponding
O'Brien--Fleming group sequential survival design. The total phase~III
sample size was set to \(E_{\mathrm{FA}}/0.70\), rounded to an even
number, and allocated equally between treatment and control arms.
Analyses were event driven: the \(j\)th analysis was conducted when
\(E_j=\mathrm{round}(q_jE_{\mathrm{FA}})\) phase~III events had accrued.
At each analysis time, both phase~III and continuing phase~II treatment group PFS data
were administratively censored on the same calendar-time scale.

The scenario-varying parameters are summarized in Table~\ref{tab:simulation_scenario_parameters}.

\paragraph{Estimands and other targets.}
The clinical estimand was the phase~III treatment effect on PFS,
expressed as the treatment-to-control hazard ratio
\(\mathrm{HR}^{(III)}=\lambda_T^{(III)}/\lambda_C^{(III)}\), or
equivalently the log-hazard contrast
\(\theta=\log\{\lambda_C^{(III)}/\lambda_T^{(III)}\}\). Treatment
benefit corresponds to \(\mathrm{HR}^{(III)}<1\), or \(\theta>0\).
At each planned analysis, Bayesian methods targeted the posterior
superiority probability
\[
  \Pr(\theta>0\mid \mathcal{D}_j)
  =
  \Pr\{\lambda_T^{(III)}<\lambda_C^{(III)}\mid \mathcal{D}_j\},
\]
where \(\mathcal{D}_j\) denotes the accumulated phase~III data and the
available phase~II treatment-arm data incorporated through the
borrowing model. The adaptive-timing estimand was the earliest
prespecified IA2 timing at which the predictive probability criterion
was satisfied.

\paragraph{Methods.}
The simulation study evaluated several analysis methods. The
frequentist benchmark was a group sequential design using a one-sided
Cox score/log-rank test with O'Brien--Fleming efficacy boundaries at
the planned analyses. All Bayesian implementations used the same
posterior decision rule defined in Section~X. The treatment-arm
external-borrowing models included a conjugate prior (Conj.), a
commensurate prior (CP), an elastic prior (EP), and their
\(B^2\)-FIC calibrated counterparts (\(B^2\)-CP and \(B^2\)-EP).

For CP, commensurability was modelled on the log-hazard scale with
a vague prior on the precision parameter. For \(B^2\)-CP,
commensurability was instead assigned a 50:50 spike--slab half-normal
prior, with scales 0.25 and 2.0, to induce robust borrowing. For both
EP and \(B^2\)-EP, the elastic function was calibrated using a clinically
meaningful discrepancy threshold \(\delta^*=0.2\). The two boundary scenarios
\(\delta=\pm\delta^*\), corresponding to hazard multipliers
\(\exp(\pm0.2)\), were used as incongruence anchors to constrain the
borrowing strength. For EP, the elastic-prior hyperparameters were then
selected by maximizing
\(U=\mathrm{Power}-w_1\alpha-w_2(\alpha-\eta)_+\), with
\(w_1=1\), \(w_2=2\), and \(\eta=0.10\). For \(B^2\)-EP, candidates
were first required to satisfy \(\alpha\leq0.025\), after which the one
with the largest power was selected.

The control arm was assigned a fixed informative Gamma conjugate prior,
centred at the scenario-specific control-arm median PFS, to reflect a
natural-history or approved comparator with abundant external evidence
(e.g., RWD). The prior shape parameter was set to \(\alpha_C=500\),
representing substantial event-scale prior information, so that
differences in operating characteristics primarily reflected the
treatment-arm borrowing strategy.

For each simulated trial, analyses were performed sequentially at the
prespecified information fractions. A trial was declared positive if
the corresponding efficacy criterion was met at any planned analysis;
otherwise it continued to the next analysis. Rejections were accumulated
over IA1, IA2, and the final analysis to estimate the overall
trial-wise false positive rate and time-specific power.

\paragraph{Performance measures.}
Performance measures were aligned with the aims. Reflecting regulatory emphasis on false-positive error control in
confirmatory trials, we evaluated the overall trial-wise false
positive rate, defined as the probability of declaring efficacy for
an ineffective treatment at any planned analysis\citep{CDE2026BayesianBorrowing}. In
the present setting, this quantity corresponds to the overall one-sided
family-wise error rate (FWER) across IA1, IA2, and the final analysis,
and was used as the empirical type I error for calibration of the
proposed framework \citep{FDA2026BayesianGuidance}. Because type I error control
is central in confirmatory phase~III settings, this quantity
was estimated from $10{,}000$ simulated trials per null
scenario and assessed against the nominal $0.025$ level. Power was estimated from
$1{,}000$ simulated trials per scenario.External borrowing was summarized by the event-scale using effective current sample size (ECSS) \citep{wiesenfarth2020quantification}, because it quantifies prior impact on the current-trial scale under phase\~II--phase\~III prior--data conflict. For the adaptive-timing
component, we reported the earliest IA2 timing at which the
predictive probability, applied after calibrated
borrowing, reached the prespecified threshold, together with
the corresponding calendar time to the IA2 decision.

\subsection{Operating characteristics of  borrowing}
\subsubsection{Type I error}

\begin{figure}[H]
  \centering
  \includegraphics[width=\textwidth]{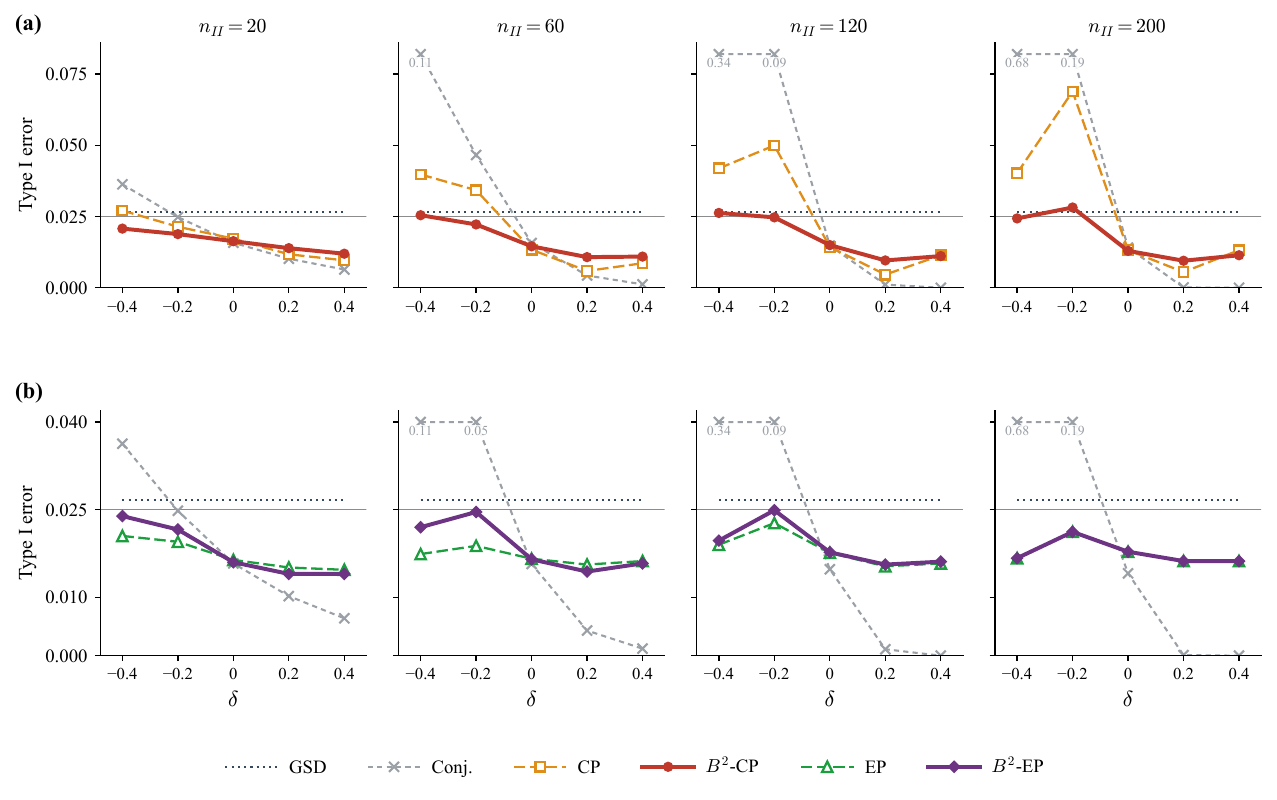}
\caption{Overall one-sided family-wise type I error under phase II--phase III discrepancy.}
  \label{fig:type1error_main}
\end{figure}

Figure~\ref{fig:type1error_main} illustrates the overall one-sided FWER in a representative null scenario with control-arm median PFS of
7 months and \(\mathrm{HR}^{(III)}=1\). The horizontal axis represents
the phase~II--phase~III treatment-arm discrepancy \(\delta\), and the
columns correspond to increasing phase~II treatment arm sample sizes.

The conjugate prior showed substantial FWER inflation when the phase~II
treatment effect was more favorable than the phase~III null effect
(\(\delta<0\)), and the inflation became more pronounced as the
phase~II sample size increased. When \(\delta>0\), the borrowed
phase~II treatment-arm information was unfavorable and the resulting
FWER was below the nominal level. 

The original commensurate prior
reduced this inflation relative to the conjugate prior, but still
exceeded the group sequential benchmark in several discrepant settings
with \(\delta<0\), especially for larger phase~II sample sizes.
Introducing the \(B^2\)-FIC robust mixture prior calibration markedly
attenuated this inflation, with \(B^2\)-CP yielding FWER close to
the group sequential benchmark across discrepancy levels.

The elastic prior was generally conservative, with FWER mostly below
the nominal level. The \(B^2\)-EP calibration was therefore intended
not to rescue type I error control, but to retain acceptable FWER
behavior while allowing less conservative borrowing.

This pattern was consistent in the sensitivity analyses. Across
different baseline event-time scales, phase~II sample sizes, and
phase~II--phase~III discrepancies, the conjugate prior remained highly
sensitive to prior--data conflict, whereas the calibrated borrowing
methods substantially reduced FWER inflation
(Table~\ref{tab:type1_median_pfs}). Because the proposed framework
allows IA2 timing to vary, we also evaluated FWER under alternative
IA2 information fractions. The calibrated methods remained close to
the non-borrowing GSD benchmark across IA2 information fractions of
0.6, 0.7, and 0.8, supporting their use in the subsequent adaptive IA2
scheduling analysis (Table~\ref{tab:type1_ia_schedule}). Type~I error rates at IA1, IA2, and the final analysis across the full set of design and borrowing scenarios are provided in Appendix~S7.

\subsubsection{Power}
Based on the type I error results above, the conjugate prior and the
uncalibrated commensurate prior were excluded from the subsequent power
comparison because they showed inflated false-positive rates under
phase II--phase III discrepancy. We therefore compared power among the
non-borrowing GSD, EP, \(B^2\)-CP, and \(B^2\)-EP, focusing on procedures that
either define the standard benchmark or maintained acceptable
false-positive control in the preceding analysis.

Table~\ref{tab:power_by_hr_timepoint} summarizes the time-specific empirical power of the three retained
borrowing methods relative to the non-borrowing GSD in the representative
setting with the control-arm median PFS fixed at 7 months and
\(n_{II}=20,60,\) and \(120\) per group.
Overall, EP, \(B^2\)-EP, and \(B^2\)-CP all improved power over GSD at IA1,
IA2, and FA under both \(\mathrm{HR}^{(III)}=0.6\) and
\(\mathrm{HR}^{(III)}=0.8\). The gains were most pronounced at the interim
analyses, when phase~III information was still limited and borrowing from
phase~II treatment-arm PFS contributed most to the decision. Under the stronger
treatment effect scenario \(\mathrm{HR}^{(III)}=0.6\), substantial gains were
already evident at IA1, whereas under \(\mathrm{HR}^{(III)}=0.8\) the
improvement at IA1 was more modest but remained clear at IA2. By the final
analysis, all borrowing procedures continued to outperform GSD, although the
incremental gains were smaller because the GSD itself already achieved high
final power.

Across methods, EP showed stable power gains with relatively limited
sensitivity to discrepancy, particularly at IA2 and FA. \(B^2\)-EP followed a
similar overall pattern and maintained high power across most discrepancy
settings. In contrast, \(B^2\)-CP showed clearer dependence on \(\delta\),
with power tending to decline as \(\delta\) became positive, reflecting less
favorable phase~II treatment-arm outcomes under the assumed discrepancy model.
The effect of phase~II sample size was most apparent at IA1, whereas
differences across \(n^{(II)}\) became less pronounced at IA2 and FA as more
phase~III events accrued and the borrowed phase~II information matured. The power at IA1, IA2, and the final analysis across the full set of design and borrowing scenarios is provided in Appendix~S8.

Figure~\ref{fig:power-trajectory-delta} examines whether
\(B^2\)-EP provided additional power over the original EP. In the representative
setting with control-arm median PFS fixed at 7 months and \(n_{II}=60\) per
group, the advantage of \(B^2\)-EP was concentrated at the interim analyses,
especially at IA1 under \(\mathrm{HR}^{(III)}=0.6\) when the phase~II
treatment-arm outcomes were concordant with or more favorable than those in
phase~III (\(\delta \leq 0\)). The difference was smaller at IA2 and was largely
attenuated by the final analysis. These results indicate that, after type~I
error control, the main benefit of \(B^2\)-EP is enhanced interim power and
earlier evidence accumulation, supporting its use in the subsequent adaptive
IA2 timing analysis.

These power gains should be interpreted relative to the prespecified non-borrowing GSD comparator, rather than relative to a theoretical UMP benchmark. This distinction is important because \citep{kopp2020power} showed that, when external information is conditioned upon and a UMP or UMP-unbiased test exists, borrowing cannot improve power while maintaining strict type I error control. In our simulations, the calibrated borrowing rules were retained only when the empirical overall one-sided type I error rate remained at or below \(\alpha=0.025\) within the simulated null scenarios.

\subsubsection{Effective sample size}
\begin{figure}[H]
    \centering
    \makebox[\linewidth][c]{%
    \includegraphics[
        width=1.2\linewidth,
        height=0.95\textheight,
        keepaspectratio,
        trim=2mm 2mm 2mm 2mm,
        clip
    ]{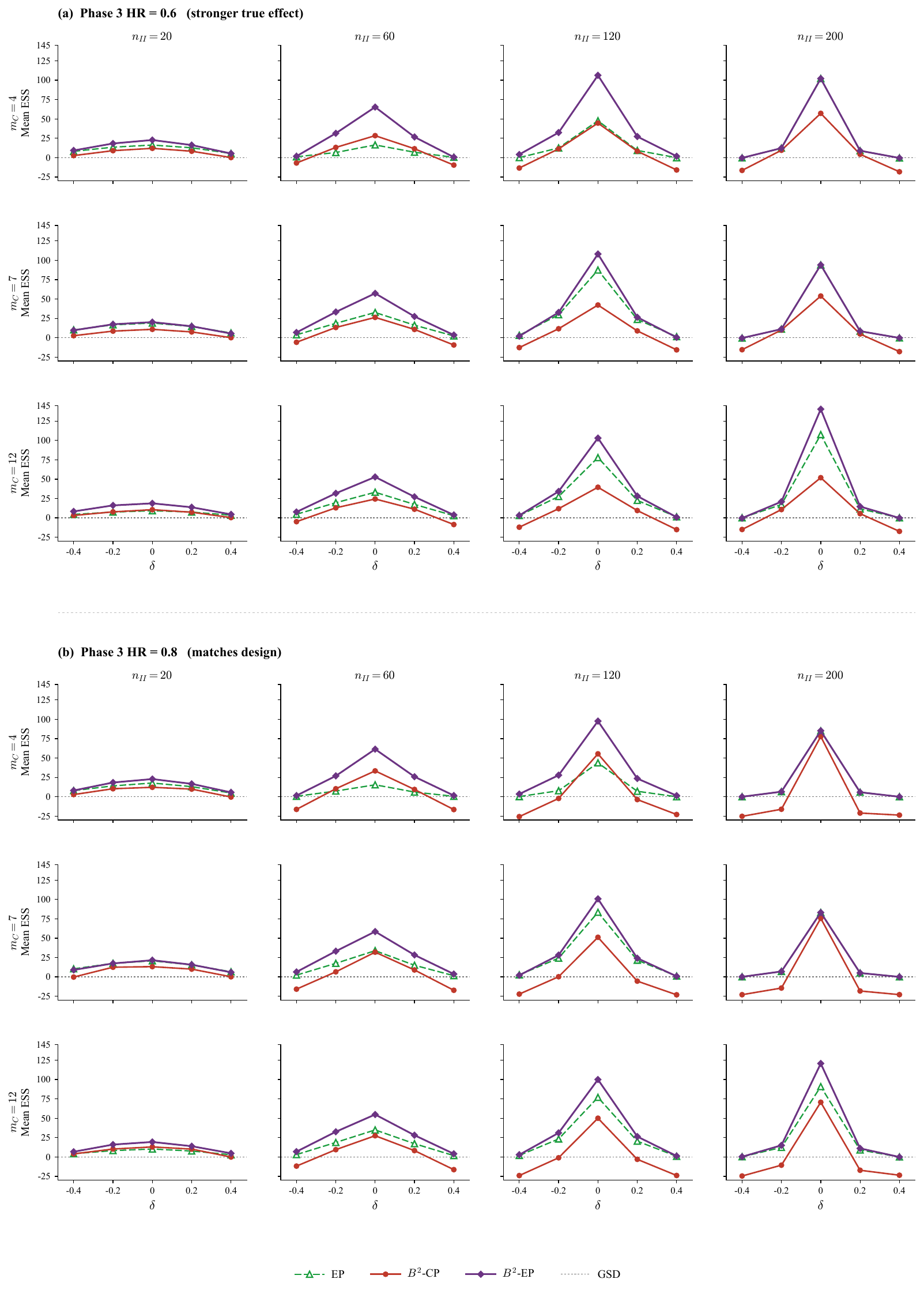}
    }
    \caption{Treatment-arm event-scale effective sample size across phase II--phase III discrepancy}
    \label{fig:ess-trajectory-delta}
\end{figure}

Figure~\ref{fig:ess-trajectory-delta} summarizes the treatment-arm
event-scale effective sample size (ESS) attributable to borrowing from the
phase~II treatment-arm PFS data. We quantified ESS following the effective
current sample size framework of \citep{wiesenfarth2020quantification}, which
interprets prior impact in terms of current-trial information and can reflect
loss of information under prior--data conflict. Across phase~III treatment
effects, control-arm median PFS settings, and phase~II sample sizes, ESS was
largest when the phase~II and phase~III treatment-arm hazards were
commensurate, with peaks generally occurring near \(\delta=0\). ESS decreased
as \(|\delta|\) increased, and was close to zero or slightly negative in several
discordant settings, suggesting that the borrowing procedures downweighted
inconsistent phase~II information.

Taken together, the ESS and power results indicate that \(B^2\)-EP was more
responsive than the original EP. It allowed greater effective borrowing when
the phase~II treatment-arm information was commensurate or more favorable,
while its ESS decreased to a level comparable to EP under unfavorable
discrepancy. This pattern is broadly consistent with the power results, where
the advantage of \(B^2\)-EP over EP was mainly observed at interim analyses
under concordant or favorable phase~II treatment-arm information. After
excluding procedures that failed to maintain acceptable type~I error control,
these findings support the use of \(B^2\)-EP as the primary elastic-prior
borrowing procedure in the subsequent adaptive IA2 timing analysis. The effective sample sizes at IA1, IA2, and the final analysis across the full set of design and borrowing scenarios are provided in Appendix~S9.

\subsection{Adaptive IA2 timing under calibrated borrowing}

The preceding operating-characteristic assessment determines the scope of the
adaptive IA2 scheduling analysis. We do not carry forward borrowing procedures
that either failed to control type~I error or were dominated by their calibrated
counterparts. In the commensurate-prior family, the uncalibrated CP was
sensitive to favorable phase~II--phase~III treatment-arm discrepancy and did
not maintain acceptable empirical type~I error control; \(B^2\)-CP is therefore
retained as the calibrated commensurate-prior implementation. In the
elastic-prior family, EP already controlled type~I error, but \(B^2\)-EP showed
a more responsive ESS profile and higher interim power under concordant or
favorable phase~II information, while retaining EP-like attenuation under
unfavorable discrepancy. Thus, \(B^2\)-EP is retained as the primary
elastic-prior implementation.

We next evaluate the end-to-end adaptive IA2 scheduling behavior of
\(B^2\)-CP and \(B^2\)-EP, using the standard GSD as the non-borrowing
reference. The focus shifts from fixed-analysis operating characteristics to
whether calibrated borrowing can support earlier IA2 recommendations under
prespecified phase~II--phase~III discrepancy scenarios.

\subsubsection{Scheduling decisions}

Tables~\ref{tab:decision_rules_compare} and
\ref{tab:decision_rules_compare_events} together specify the operational form
of the adaptive timing rule for the representative scenario with phase~III
control-arm median PFS \(m_C^{(III)}=7\) months. Corresponding decision tables
for \(m_C^{(III)}=4\) and \(m_C^{(III)}=12\) months are provided in Supplementary Appendix
Sections~S10 and S11, respectively. The decision table is computed
at the design stage and may be prespecified in the phase~III statistical
analysis plan. At IA1, the observed phase~II treatment-arm HR and phase~III
IA1 treatment-to-control HR determine the row and column, and the corresponding
cell gives the recommended action. Table~\ref{tab:decision_rules_compare}
expresses the recommendation on the information-fraction scale, where IA1
denotes efficacy stopping at IA1, \(0.60\)--\(0.80\) denote candidate IA2
information fractions, and \(1.0\) denotes proceeding to the final analysis.
Table~\ref{tab:decision_rules_compare_events} gives the corresponding
event-count targets, computed as
\(E(r)=\mathrm{round}\{rE_{\mathrm{FA}}\}\), where \(E_{\mathrm{FA}}\) is the
final-analysis event target for the scenario.
Intervals are left-open and right-closed unless otherwise stated.

The recommended timing became progressively later as either source of evidence
became less favourable. When both phase~II and phase~III IA1 evidence were
strong, the rule recommended stopping for efficacy at IA1 or scheduling an
early IA2. For example, when the phase~II HR was \(\leq 0.50\) and the
phase~III IA1 HR was \(\leq 0.60\), the recommended action was IA1. When the
phase~III IA1 evidence remained favourable but was weaker, the recommended IA2
shifted to later information fractions, moving from \(0.60\) to \(0.70\) and
then to \(0.80\). When the phase~III IA1 HR exceeded \(0.80\), or when the
phase~II HR exceeded \(0.80\), the rule recommended proceeding to the final
analysis. This pattern reflects the intended conservative behaviour of the
decision table: early analyses are recommended only when both historical and
current-trial evidence support a sufficiently high Bayesian predictive
probability of meeting the future efficacy criterion.

On the event-count scale, these recommendations translate into monitoring
targets for the event-driven trial. In the most favourable cells, the
recommended decision could be made after 36--98 phase~III events; intermediate
cells required 236--859 events, corresponding to information fractions of
\(0.70\)--\(0.80\); and cells assigned to the final analysis corresponded to
the full event target.

\subsubsection{Time to IA2 decision}

\begin{figure}[H]
\centering
\begin{adjustbox}{center}
\includegraphics[width=1\textwidth]{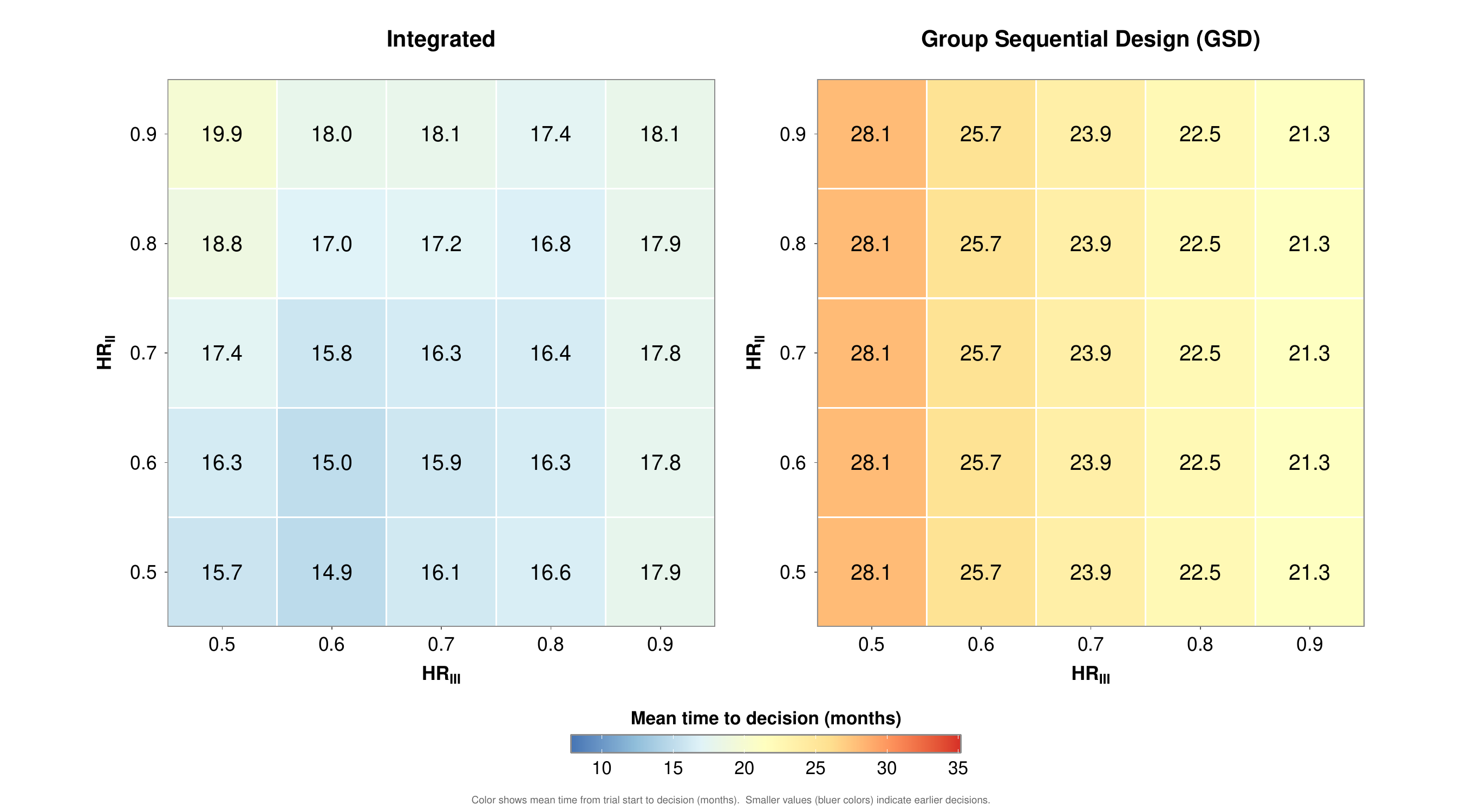}
\end{adjustbox}
\caption{Time to Phase III decision under the $B^2$-FIC framework versus Group Sequential Design}
\label{fig:heatmap_all_methods}
\end{figure}

Figure~\ref{fig:heatmap_all_methods} presents the mean calendar time from Phase~III trial initiation to the recommended decision for the same representative scenario with Phase~III control-arm median PFS $m_C^{(\text{III})} = 7$ months. This provides a complementary view of the decision tables in Tables~\ref{tab:decision_rules_compare} and~\ref{tab:decision_rules_compare_events}: whereas the tables present the operational lookup rule used at IA1, the figure translates each cell's recommended action into an expected time, offering a direct quantification of the time comparison and saving achieved by the proposed framework relative to GSD.

Both \(B^2\)-FIC implementations led to earlier decisions than the
non-borrowing group sequential design across the displayed grid. The mean time to decision ranged from approximately 14.9 to 19.9 months under the integrated rule, compared with 21.3 to 28.1 months under GSD. Averaged over the grid, the reduction relative to GSD was approximately 7--10 months depending on the scenario.

The acceleration was most pronounced when both Phase II and Phase III IA1 evidence were concordantly favourable. For example, when $\text{HR}^\text{(III)} \leq 0.60$ and $\text{HR}^\text{(II)} \leq 0.60$, the mean decision time under the integrated framework was approximately 14.9--16.3 months, compared with 25.7--28.1 months under GSD. As either evidence source became less favourable -- that is, as $\text{HR}^\text{(II)}$ or $\text{HR}^\text{(III)}$ increased toward 0.9 -- the integrated rule appropriately deferred the recommended decision, with mean times increasing toward approximately 17--20 months. This pattern is consistent with the monotone decision table: earlier decisions are concentrated in cells where both Phase II and Phase III evidence are strong, while weaker combined evidence leads to later IA2 or final analysis timing.

In contrast, the GSD panel is constant across the phase~II HR axis because the
standard group sequential comparator does not use phase~II treatment-arm
information. Its decision time is therefore driven only by the phase~III
scenario and the fixed monitoring schedule. The comparison indicates that the
proposed framework can shorten the calendar time to a phase~III decision while
retaining a prespecified, conservative rule for deferring decisions when the
combined evidence is weaker.

\section{Case Study}
\label{sec:case-study}
The first case study is motivated by the development of Palbociclib (IBRANCE), a first-in-class CDK4/6 inhibitor for the treatment of ER+/HER2$^{-}$ advanced breast cancer. The phase III trial (PALOMA-2) enrolled 666 patients in a randomized, double-blind, placebo-controlled design at a 2:1 ratio, with PFS as the primary endpoint \citep{Finn2016PALOMA2}. For the historical phase II data, we referenced the PALOMA-1 trial, which reported a hazard ratio of 0.488 for PFS \citep{Finn2015PALOMA1}. To better reflect a scenario where the framework's adaptive IA2 scheduling provides meaningful benefit, we adopted a control arm median PFS of 8.5 months and a phase II HR of 0.70, with candidate Phase III IA1 HR values set to $(0.5, 0.6, 0.7, 0.8, 0.9)$.The second case study concerns Apatinib, a novel vascular endothelial growth factor receptor 2 tyrosine kinase inhibitor, in patients with chemotherapy-refractory advanced gastric or gastroesophageal junction adenocarcinoma. The phase III trial of Apatinib is a randomized, double-blind, placebo-controlled, multi-center trial undertaken in 32 centers in China. Patients were randomly assigned at a 2:1 ratio to receive either oral apatinib or placebo between January 2011 and November 2012 \citep{Li2016ApatinibIII}. The primary endpoints were overall survival (OS) and progression-free survival. For the historical data, we used the randomized phase II trial of Apatinib, in which the hazard ratio for PFS was 0.18 and the control group median PFS was 1.8 months \citep{Li2013ApatinibII}. Candidate HR observed at phase III IA1 is set to $(0.3, 0.4, 0.5, 0.6, 0.7, 0.8, 0.9)$. Due to the unavailability of individual-level patient data, the time-to-event data were simulated using a Weibull distribution in accordance with the summary statistics observed in the two trials. Characteristics for the two trials are summarized in Table~\ref{tab:case_characteristics}.

The decision tables (Tables~\ref{tab:decision_rules_case_study}) present the pre-specified IA2 scheduling recommendations for the two case study trials across candidate phase III IA1 HR scenarios. For Trial 1 (HR$^\text{(II)}$ = 0.18), when the phase III IA1 HR is at its most favorable ($\leq 0.30$), the framework recommends early stop at IA1, consistent with the high posterior efficacy probability achieved through dynamic borrowing of phase II data. As the phase III IA1 HR increases to 0.4 or 0.5, the recommended IA2 is scheduled at information fraction 0.6, one step earlier than the GSD fixed IA2 at 0.7. For HR$^\text{(III)}$ values of 0.6 and 0.7, the framework recommends IA2 at information fractions 0.7 and 0.8 respectively. When the phase III IA1 HR is 0.8 or above, the framework defers to the final analysis. For Trial 2 (HR$^\text{(II)}$ = 0.70), representing a weaker phase II treatment effect, the framework does not recommend IA1 early stop under any of the candidate HR$^\text{(III)}$ scenarios. When phase III IA1 HR is 0.5 or 0.6, the recommended IA2 is at information fraction 0.7, consistent with the GSD fixed IA2. For phase III IA1 HR equals to 0.7, the framework recommends a more conservative IA2 at 0.8, and defers to the final analysis when phase III IA1 HR is 0.8 or above.

\section{Discussion}
\label{sec:discussion}

We proposed \(B^2\)-FIC, a Bayesian predictive framework for adaptive interim-analysis timing in confirmatory time-to-event trials, integrating type~I error-controlled robust borrowing with prospective selection of the second interim analysis (IA2) timing. The framework addresses two linked design problems
that are often considered separately: how to incorporate immature but accruing
phase~II evidence while preserving confirmatory operating characteristics, and
how to use the resulting IA1 posterior evidence to determine when the next
interim analysis is expected to become informative. All adaptive elements,
including the candidate IA2 times, borrowing model, posterior monitoring
boundaries, predictive-probability threshold, and mapping from IA1 evidence to
the recommended action, are fixed at the design stage. The resulting decision
table translates the rule into a prespecified IA2 scheduling procedure that can
be reviewed before the confirmatory trial begins.

A central contribution of this work is to formulate interim-analysis timing under Bayesian information borrowing as a statistical design problem. Existing borrowing methods focus on incorporating external information, whereas predictive monitoring methods evaluate future success at prespecified analysis times. In contrast, \(B^2\)-FIC uses borrowing-updated evidence to determine when a future interim analysis should be conducted. In the borrowing component, candidate dynamic borrowing
procedures are embedded within a design-stage calibration rule, so that
efficiency is considered only among methods with acceptable empirical
type I error performance under prespecified phase~II--phase~III discrepancy
scenarios. In the timing component, predictive probability is used not to change
the efficacy criterion, stop the trial, or re-estimate the sample size, but to
choose the earliest candidate IA2 time at which the prespecified posterior
efficacy rule is likely to be met. The observed power gains should therefore be
interpreted relative to this design framework. They do not contradict the result
of \citep{kopp2020power}, which concerns power gains from external borrowing
relative to a UMP or UMP-unbiased level-\(\alpha\) test when the external
information is conditioned upon. Our comparison is instead with a prespecified
non-borrowing group sequential design. The gains therefore represent
improvements relative to this clinically relevant comparator, not improvements
over a theoretical optimal test. Consistent with this objective, type I error
control was evaluated empirically over prespecified null scenarios and
discrepancy configurations, rather than established as a uniform analytical
guarantee over the full null parameter space.

The framework is intended for settings in which phase~II evidence is informative
but uncertain, such as first-in-class or other innovative oncology therapies. In
these settings, the phase~II cohort may continue to accrue PFS or OS information
while the phase~III trial is ongoing, making the external evidence dynamic
rather than fixed. The design-stage specification of the borrowing rule and
timing rule is aligned with the general regulatory principle that adaptive
features should be prospectively planned and supported by operating-characteristic
evaluation. The decision table also provides an operationally transparent
output: IA1 phase~III evidence primarily determines whether IA2 can be scheduled
early, whereas phase~II information influences borderline cases through the
calibrated borrowing model. The case-study examples illustrate this behaviour in
two different evidence settings. The palbociclib example, with a mild favourable
phase~II--phase~III discrepancy, shows how phase~II information can provide
moderate support for earlier IA2 timing. The apatinib example, with a much
stronger favourable discrepancy, illustrates how highly supportive phase~II
evidence is still filtered through the calibrated borrowing rule. Together,
these examples show that the same prespecified rule can adapt to different
strengths of external evidence without case-specific redesign.

Several extensions merit further investigation. First, although the current implementation focuses on time-to-event endpoints, the same design principle could be extended to binary or continuous endpoints by replacing the survival model and predictive calculation with endpoint-specific likelihoods and decision criteria. Second, the present framework borrows continuing phase~II information from the treatment arm and treats the control-arm external information as a fixed informative prior. This simplification was adopted to facilitate an initial investigation of the proposed design without introducing additional layers of complexity. The approach is most appropriate when the control-arm event process is well understood and the phase~II and phase~III treatment cohorts are sufficiently comparable. This similarity assumption should be justified at the design stage when external evidence is used prospectively. Third, the simulations used a shared-shape Weibull proportional hazards model with a decreasing hazard function. Extensions to exponential or increasing hazards, alternative survival distributions, and delayed or non-proportional treatment effects would further clarify the robustness of the proposed design.More broadly, external borrowing and interim-analysis timing are coupled design problems in confirmatory trials: the amount and commensurability of external information available at IA1 directly affect when the next analysis is expected to become informative.  \(B^2\)-FIC provides a prospective framework for linking these decisions while retaining explicit calibration of confirmatory operating characteristics.

\section{Acknowledgments}
This work was supported by the National Natural Science Foundation
of China (grant number 82373682) and the National Science and
Technology Major Project on the Prevention and Control of Emerging
and Major Infectious Diseases (grant number 2025ZD01906004).

The authors thank Lin Shen (Peking University Cancer Hospital \& Institute) for bringing this clinically relevant problem to our attention. Yanxun Xu (Johns Hopkins University), Fangrong Yan and Xin Chen (China Pharmaceutical University), Jie He (Nanjing University of Aeronautics and Astronautics), Wanqiu Xie (Heilongjiang University), Jun Zhang, Bo Fu, and Nan Song (Astellas), Xinyi Zhang (University of Chicago), and colleagues at LinkDoc Technology for helpful discussions during different stages of this work. The authors are solely responsible for the content of this article.

\section{Conflicts of Interest}
The authors declare no conflicts of interest.

\section{Data Availability Statement}
The data that support the findings of this study are available from the corresponding author upon reasonable request.

\bibliographystyle{ama} 
\bibliography{main}

\clearpage

\begin{table}[H]
\centering
\small
\caption{Scenario-varying parameters in the simulation study.}
\label{tab:simulation_scenario_parameters}
\begin{threeparttable}
\begin{tabularx}{\textwidth}{
  >{\raggedright\arraybackslash}p{0.18\textwidth}
  >{\raggedright\arraybackslash}p{0.42\textwidth}
  >{\raggedright\arraybackslash}X
}
\toprule
\textbf{Notation} & \textbf{Parameter} & \textbf{Values used} \\
\midrule

\(m_C^{(III)}\) &
Phase~III control-arm median PFS &
\(4,\ 7,\ 12\) months \\

\(\mathrm{HR}^{(III)}\) &
Phase~III treatment-to-control hazard ratio &
\(1.0\) under the global null; 
\(0.8,\ 0.6\) under alternatives; 
design HR \(=0.8\) \\

\(\delta\) &
Phase~II--phase~III treatment-arm discrepancy on the log-hazard scale &
\(-0.4,\ -0.2,\ 0,\ 0.2,\ 0.4\) \\

\(n^{(II)}\) &
Phase~II treatment group sample size  &
\(20,\ 60,\ 120,\ 200\) per arm \\

\(\boldsymbol{q}=(q_1,\ldots,q_J)\) &
Information fractions &
\((0.4,0.6,1.0)\),
\((0.4,0.7,1.0)\),
\((0.4,0.8,1.0)\) \\

\bottomrule
\end{tabularx}
\end{threeparttable}
\end{table}

\clearpage
\newgeometry{margin=0.1in, landscape}
\begin{landscape}

\begin{table}[H]
\centering
\tiny
\caption{Overall type I error across design and borrowing scenarios}
\label{tab:type1_median_pfs}
\setlength{\tabcolsep}{1.2pt}
\renewcommand{\arraystretch}{1.6}

\resizebox{1.05\textwidth}{!}{%
{\setlength{\tabcolsep}{2.8pt}%
\begin{tabular}{@{}c c
  *{4}{c}@{\hspace{8pt}}
  *{4}{c}@{\hspace{8pt}}
  *{4}{c}@{\hspace{8pt}}
  *{4}{c}@{\hspace{8pt}}
  *{4}{c}@{}}
\toprule
\multirow{3}{*}[-0.3em]{\textbf{\begin{tabular}{@{}c@{}}Median PFS\\(Control)\end{tabular}}} &
\multirow{3}{*}[-0.3em]{\textbf{Method}} &
\multicolumn{20}{c}{\textbf{Discrepancy $\delta$}} \\
\cmidrule(lr){3-22}
& & \multicolumn{4}{c}{\textbf{$\mathbf{-0.4}$}} & \multicolumn{4}{c}{\textbf{$\mathbf{-0.2}$}} & \multicolumn{4}{c}{\textbf{$\mathbf{0}$}} & \multicolumn{4}{c}{\textbf{$\mathbf{0.2}$}} & \multicolumn{4}{c}{\textbf{$\mathbf{0.4}$}} \\
\cmidrule(l{0pt}r{5.5pt}){3-6}
\cmidrule(l{0pt}r{5.5pt}){7-10}
\cmidrule(l{0pt}r{5.5pt}){11-14}
\cmidrule(l{0pt}r{5.5pt}){15-18}
\cmidrule(l{0pt}r{0pt}){19-22}
& & \textbf{20} & \textbf{60} & \textbf{120} & \textbf{200} &
    \textbf{20} & \textbf{60} & \textbf{120} & \textbf{200} &
    \textbf{20} & \textbf{60} & \textbf{120} & \textbf{200} &
    \textbf{20} & \textbf{60} & \textbf{120} & \textbf{200} &
    \textbf{20} & \textbf{60} & \textbf{120} & \textbf{200} \\
\midrule

\multirow{6}{*}{4 months}
& GSD          & 0.026 & 0.026 & 0.026 & 0.026 & 0.026 & 0.026 & 0.026 & 0.026 & 0.026 & 0.026 & 0.026 & 0.026 & 0.026 & 0.026 & 0.026 & 0.026 & 0.026 & 0.026 & 0.026 & 0.026 \\
\cmidrule(l{1pt}r{0pt}){2-22}
& Conj.        & 0.040 & 0.141 & 0.411 & 0.767 & 0.025 & 0.050 & 0.106 & 0.220 & 0.015 & 0.015 & 0.014 & 0.013 & 0.010 & 0.004 & 0.001 & 0.000 & 0.007 & 0.001 & 0.000 & 0.000 \\
\cmidrule(l{1pt}r{0pt}){2-22}
& CP        & 0.027 & 0.041 & 0.042 & 0.039 & 0.024 & 0.035 & 0.054 & 0.073 & 0.016 & 0.014 & 0.013 & 0.013 & 0.011 & 0.007 & 0.005 & 0.006 & 0.010 & 0.009 & 0.011 & 0.013 \\
\cmidrule(l{1pt}r{0pt}){2-22}
& EP           & 0.018 & 0.016 & 0.016 & 0.016 & 0.017 & 0.017 & 0.017 & 0.020 & 0.016 & 0.016 & 0.017 & 0.017 & 0.015 & 0.015 & 0.015 & 0.015 & 0.015 & 0.015 & 0.015 & 0.015 \\
\cmidrule(l{1pt}r{0pt}){2-22}
& $B^2$-CP  & 0.021 & 0.026 & 0.025 & 0.026 & 0.018 & 0.022 & 0.025 & 0.027 & 0.016 & 0.015 & 0.014 & 0.014 & 0.013 & 0.011 & 0.010 & 0.009 & 0.012 & 0.011 & 0.011 & 0.011 \\
\cmidrule(l{1pt}r{0pt}){2-22}
& $B^2$-EP     & 0.024 & 0.022 & 0.018 & 0.016 & 0.021 & 0.026 & 0.024 & 0.020 & 0.016 & 0.017 & 0.017 & 0.017 & 0.013 & 0.015 & 0.015 & 0.015 & 0.014 & 0.015 & 0.015 & 0.015 \\
\midrule

\multirow{6}{*}{7 months}
& GSD          & 0.027 & 0.027 & 0.027 & 0.027 & 0.027 & 0.027 & 0.027 & 0.027 & 0.027 & 0.027 & 0.027 & 0.027 & 0.027 & 0.027 & 0.027 & 0.027 & 0.027 & 0.027 & 0.027 & 0.027 \\
\cmidrule(l{1pt}r{0pt}){2-22}
& Conj.        & 0.036 & 0.113 & 0.341 & 0.685 & 0.025 & 0.047 & 0.092 & 0.193 & 0.016 & 0.016 & 0.015 & 0.014 & 0.010 & 0.004 & 0.001 & 0.000 & 0.006 & 0.001 & 0.000 & 0.000 \\
\cmidrule(l{1pt}r{0pt}){2-22}
& CP        & 0.027 & 0.040 & 0.042 & 0.040 & 0.021 & 0.034 & 0.050 & 0.069 & 0.017 & 0.013 & 0.014 & 0.013 & 0.012 & 0.006 & 0.005 & 0.005 & 0.010 & 0.009 & 0.011 & 0.013 \\
\cmidrule(l{1pt}r{0pt}){2-22}
& EP           & 0.021 & 0.017 & 0.019 & 0.017 & 0.020 & 0.019 & 0.023 & 0.021 & 0.016 & 0.017 & 0.018 & 0.018 & 0.015 & 0.016 & 0.015 & 0.016 & 0.015 & 0.016 & 0.016 & 0.016 \\
\cmidrule(l{1pt}r{0pt}){2-22}
& $B^2$-CP  & 0.021 & 0.025 & 0.026 & 0.024 & 0.019 & 0.022 & 0.025 & 0.028 & 0.016 & 0.015 & 0.015 & 0.013 & 0.014 & 0.011 & 0.010 & 0.009 & 0.012 & 0.011 & 0.011 & 0.011 \\
\cmidrule(l{1pt}r{0pt}){2-22}
& $B^2$-EP     & 0.024 & 0.022 & 0.020 & 0.017 & 0.022 & 0.025 & 0.025 & 0.021 & 0.016 & 0.017 & 0.018 & 0.018 & 0.014 & 0.014 & 0.016 & 0.016 & 0.014 & 0.016 & 0.016 & 0.016 \\
\midrule

\multirow{6}{*}{12 months}
& GSD          & 0.027 & 0.027 & 0.027 & 0.027 & 0.027 & 0.027 & 0.027 & 0.027 & 0.027 & 0.027 & 0.027 & 0.027 & 0.027 & 0.027 & 0.027 & 0.027 & 0.027 & 0.027 & 0.027 & 0.027 \\
\cmidrule(l{1pt}r{0pt}){2-22}
& Conj.        & 0.035 & 0.098 & 0.286 & 0.601 & 0.024 & 0.043 & 0.083 & 0.167 & 0.016 & 0.016 & 0.015 & 0.015 & 0.011 & 0.004 & 0.001 & 0.000 & 0.007 & 0.001 & 0.000 & 0.000 \\
\cmidrule(l{1pt}r{0pt}){2-22}
& CP        & 0.028 & 0.039 & 0.044 & 0.043 & 0.023 & 0.034 & 0.048 & 0.063 & 0.017 & 0.015 & 0.015 & 0.013 & 0.012 & 0.007 & 0.004 & 0.005 & 0.009 & 0.009 & 0.011 & 0.013 \\
\cmidrule(l{1pt}r{0pt}){2-22}
& EP           & 0.017 & 0.018 & 0.019 & 0.017 & 0.018 & 0.019 & 0.022 & 0.021 & 0.016 & 0.017 & 0.017 & 0.019 & 0.016 & 0.016 & 0.016 & 0.016 & 0.016 & 0.016 & 0.016 & 0.016 \\
\cmidrule(l{1pt}r{0pt}){2-22}
& $B^2$-CP  & 0.022 & 0.025 & 0.026 & 0.026 & 0.019 & 0.022 & 0.025 & 0.029 & 0.016 & 0.015 & 0.015 & 0.014 & 0.013 & 0.011 & 0.010 & 0.009 & 0.013 & 0.011 & 0.011 & 0.012 \\
\cmidrule(l{1pt}r{0pt}){2-22}
& $B^2$-EP     & 0.026 & 0.022 & 0.021 & 0.018 & 0.022 & 0.024 & 0.026 & 0.024 & 0.016 & 0.018 & 0.018 & 0.019 & 0.015 & 0.015 & 0.016 & 0.016 & 0.014 & 0.016 & 0.016 & 0.016 \\
\bottomrule
\end{tabular}%
}%
}

\begin{minipage}{\textwidth}
\footnotesize
GSD: group sequential design; Conj.: conjugate prior; CP: commensurate prior; EP: elastic prior; $B^2$-CP: $B^2$-FIC Commensurate prior; $B^2$-EP: $B^2$-FIC Elastic prior. Phase II sample sizes are 20, 60, 120, and 200 per group.
\end{minipage}
\end{table}

\clearpage

\begin{table}[H]
\addtocounter{table}{-1}
\centering
\tiny
\caption{Overall type I error across design and borrowing scenarios (Continued)}
\label{tab:type1_ia_schedule}
\setlength{\tabcolsep}{1.2pt}
\renewcommand{\arraystretch}{1.6}

\resizebox{1.05\textwidth}{!}{%
{\setlength{\tabcolsep}{2.8pt}%
\begin{tabular}{@{}c c
  *{4}{c}@{\hspace{8pt}}
  *{4}{c}@{\hspace{8pt}}
  *{4}{c}@{\hspace{8pt}}
  *{4}{c}@{\hspace{8pt}}
  *{4}{c}@{}}
\toprule
\multirow{3}{*}[-0.3em]{\textbf{IA schedule}} &
\multirow{3}{*}[-0.3em]{\textbf{Method}} &
\multicolumn{20}{c}{\textbf{Discrepancy $\delta$}} \\
\cmidrule(lr){3-22}
& & \multicolumn{4}{c}{\textbf{$\mathbf{-0.4}$}} & \multicolumn{4}{c}{\textbf{$\mathbf{-0.2}$}} & \multicolumn{4}{c}{\textbf{$\mathbf{0}$}} & \multicolumn{4}{c}{\textbf{$\mathbf{0.2}$}} & \multicolumn{4}{c}{\textbf{$\mathbf{0.4}$}} \\
\cmidrule(l{0pt}r{5.5pt}){3-6}
\cmidrule(l{0pt}r{5.5pt}){7-10}
\cmidrule(l{0pt}r{5.5pt}){11-14}
\cmidrule(l{0pt}r{5.5pt}){15-18}
\cmidrule(l{0pt}r{0pt}){19-22}
& & \textbf{20} & \textbf{60} & \textbf{120} & \textbf{200} &
    \textbf{20} & \textbf{60} & \textbf{120} & \textbf{200} &
    \textbf{20} & \textbf{60} & \textbf{120} & \textbf{200} &
    \textbf{20} & \textbf{60} & \textbf{120} & \textbf{200} &
    \textbf{20} & \textbf{60} & \textbf{120} & \textbf{200} \\
\midrule

\multirow{6}{*}{(0.4, 0.6, 1.0)}
& GSD          & 0.027 & 0.027 & 0.027 & 0.027 & 0.027 & 0.027 & 0.027 & 0.027 & 0.027 & 0.027 & 0.027 & 0.027 & 0.027 & 0.027 & 0.027 & 0.027 & 0.027 & 0.027 & 0.027 & 0.027 \\
\cmidrule(l{1pt}r{0pt}){2-22}
& Conj.        & 0.038 & 0.121 & 0.344 & 0.688 & 0.025 & 0.047 & 0.093 & 0.193 & 0.017 & 0.016 & 0.015 & 0.016 & 0.012 & 0.004 & 0.002 & 0.000 & 0.007 & 0.001 & 0.000 & 0.000 \\
\cmidrule(l{1pt}r{0pt}){2-22}
& CP        & 0.028 & 0.041 & 0.043 & 0.041 & 0.023 & 0.032 & 0.053 & 0.068 & 0.020 & 0.017 & 0.015 & 0.015 & 0.014 & 0.009 & 0.005 & 0.006 & 0.010 & 0.011 & 0.013 & 0.013 \\
\cmidrule(l{1pt}r{0pt}){2-22}
& EP           & 0.017 & 0.021 & 0.018 & 0.019 & 0.017 & 0.023 & 0.021 & 0.026 & 0.017 & 0.018 & 0.018 & 0.020 & 0.017 & 0.016 & 0.017 & 0.017 & 0.017 & 0.017 & 0.017 & 0.017 \\
\cmidrule(l{1pt}r{0pt}){2-22}
& $B^2$-CP  & 0.022 & 0.026 & 0.028 & 0.028 & 0.021 & 0.024 & 0.026 & 0.029 & 0.017 & 0.015 & 0.016 & 0.014 & 0.014 & 0.011 & 0.010 & 0.010 & 0.013 & 0.012 & 0.013 & 0.012 \\
\cmidrule(l{1pt}r{0pt}){2-22}
& $B^2$-EP     & 0.026 & 0.022 & 0.019 & 0.017 & 0.023 & 0.023 & 0.023 & 0.021 & 0.018 & 0.019 & 0.018 & 0.019 & 0.015 & 0.016 & 0.016 & 0.017 & 0.015 & 0.017 & 0.017 & 0.017 \\
\midrule

\multirow{6}{*}{(0.4, 0.7, 1.0)}
& GSD          & 0.027 & 0.027 & 0.027 & 0.027 & 0.027 & 0.027 & 0.027 & 0.027 & 0.027 & 0.027 & 0.027 & 0.027 & 0.027 & 0.027 & 0.027 & 0.027 & 0.027 & 0.027 & 0.027 & 0.027 \\
\cmidrule(l{1pt}r{0pt}){2-22}
& Conj.        & 0.036 & 0.113 & 0.341 & 0.685 & 0.025 & 0.047 & 0.092 & 0.193 & 0.016 & 0.016 & 0.015 & 0.014 & 0.010 & 0.004 & 0.001 & 0.000 & 0.006 & 0.001 & 0.000 & 0.000 \\
\cmidrule(l{1pt}r{0pt}){2-22}
& CP        & 0.027 & 0.040 & 0.042 & 0.040 & 0.021 & 0.034 & 0.050 & 0.069 & 0.017 & 0.013 & 0.014 & 0.013 & 0.012 & 0.006 & 0.005 & 0.005 & 0.010 & 0.009 & 0.011 & 0.013 \\
\cmidrule(l{1pt}r{0pt}){2-22}
& EP           & 0.021 & 0.017 & 0.019 & 0.017 & 0.020 & 0.019 & 0.023 & 0.021 & 0.016 & 0.017 & 0.018 & 0.018 & 0.015 & 0.016 & 0.015 & 0.016 & 0.015 & 0.016 & 0.016 & 0.016 \\
\cmidrule(l{1pt}r{0pt}){2-22}
& $B^2$-CP  & 0.021 & 0.025 & 0.026 & 0.024 & 0.019 & 0.022 & 0.025 & 0.028 & 0.016 & 0.015 & 0.015 & 0.013 & 0.014 & 0.011 & 0.010 & 0.009 & 0.012 & 0.011 & 0.011 & 0.011 \\
\cmidrule(l{1pt}r{0pt}){2-22}
& $B^2$-EP     & 0.024 & 0.022 & 0.020 & 0.017 & 0.022 & 0.025 & 0.025 & 0.021 & 0.016 & 0.017 & 0.018 & 0.018 & 0.014 & 0.014 & 0.016 & 0.016 & 0.014 & 0.016 & 0.016 & 0.016 \\
\midrule

\multirow{6}{*}{(0.4, 0.8, 1.0)}
& GSD          & 0.027 & 0.027 & 0.027 & 0.027 & 0.027 & 0.027 & 0.027 & 0.027 & 0.027 & 0.027 & 0.027 & 0.027 & 0.027 & 0.027 & 0.027 & 0.027 & 0.027 & 0.027 & 0.027 & 0.027 \\
\cmidrule(l{1pt}r{0pt}){2-22}
& Conj.        & 0.036 & 0.112 & 0.334 & 0.680 & 0.025 & 0.047 & 0.093 & 0.187 & 0.016 & 0.016 & 0.014 & 0.016 & 0.011 & 0.005 & 0.002 & 0.000 & 0.007 & 0.001 & 0.000 & 0.000 \\
\cmidrule(l{1pt}r{0pt}){2-22}
& CP        & 0.030 & 0.038 & 0.043 & 0.039 & 0.025 & 0.036 & 0.050 & 0.066 & 0.019 & 0.017 & 0.014 & 0.014 & 0.013 & 0.008 & 0.006 & 0.007 & 0.011 & 0.011 & 0.012 & 0.014 \\
\cmidrule(l{1pt}r{0pt}){2-22}
& EP           & 0.022 & 0.020 & 0.019 & 0.018 & 0.021 & 0.022 & 0.022 & 0.022 & 0.018 & 0.020 & 0.019 & 0.020 & 0.016 & 0.018 & 0.018 & 0.018 & 0.017 & 0.018 & 0.018 & 0.018 \\
\cmidrule(l{1pt}r{0pt}){2-22}
& $B^2$-CP  & 0.023 & 0.026 & 0.027 & 0.027 & 0.021 & 0.024 & 0.026 & 0.029 & 0.018 & 0.016 & 0.014 & 0.016 & 0.014 & 0.012 & 0.011 & 0.011 & 0.013 & 0.012 & 0.012 & 0.012 \\
\cmidrule(l{1pt}r{0pt}){2-22}
& $B^2$-EP     & 0.022 & 0.022 & 0.020 & 0.019 & 0.021 & 0.024 & 0.025 & 0.023 & 0.018 & 0.019 & 0.020 & 0.020 & 0.016 & 0.017 & 0.018 & 0.018 & 0.017 & 0.018 & 0.018 & 0.018 \\
\bottomrule
\end{tabular}%
}%
}

\begin{minipage}{\textwidth}
\footnotesize
GSD: group sequential design; Conj.: conjugate prior; CP: commensurate prior; EP: elastic prior; $B^2$-CP: $B^2$-FIC Commensurate prior; $B^2$-EP: $B^2$-FIC Elastic prior. Phase II sample sizes are 20, 60, 120, and 200 per group.
\end{minipage}
\end{table}

\restoregeometry
\end{landscape}

\restoregeometry

\clearpage

\begin{table}[H]
\centering
\caption{Time-specific empirical power by target hazard ratio and borrowing method}
\label{tab:power_by_hr_timepoint}
\tiny
\setlength{\tabcolsep}{2.4pt}
\renewcommand{\arraystretch}{1.12}

\begin{threeparttable}
\resizebox{0.98\textwidth}{!}{%
\begin{tabular}{@{}ccl*{15}{c}@{}}
\toprule
\multirow{3}{*}{\textbf{\(\mathrm{HR}^{(III)}\)}} &
\multirow{3}{*}{\textbf{Analysis}} &
\multirow{3}{*}{\textbf{Method}} &
\multicolumn{15}{c}{\textbf{Discrepancy $\delta$}} \\
\cmidrule(lr){4-18}
& & &
\multicolumn{3}{c}{$\mathbf{-0.4}$} &
\multicolumn{3}{c}{$\mathbf{-0.2}$} &
\multicolumn{3}{c}{$\mathbf{0}$} &
\multicolumn{3}{c}{$\mathbf{0.2}$} &
\multicolumn{3}{c}{$\mathbf{0.4}$} \\
\cmidrule(lr){4-6}
\cmidrule(lr){7-9}
\cmidrule(lr){10-12}
\cmidrule(lr){13-15}
\cmidrule(lr){16-18}
& & &
\textbf{20} & \textbf{60} & \textbf{120} &
\textbf{20} & \textbf{60} & \textbf{120} &
\textbf{20} & \textbf{60} & \textbf{120} &
\textbf{20} & \textbf{60} & \textbf{120} &
\textbf{20} & \textbf{60} & \textbf{120} \\
\midrule

\multirow{12}{*}{\textbf{0.6}} & \multirow{4}{*}{IA1} & GSD        & 0.082 & 0.082 & 0.082 & 0.082 & 0.082 & 0.082 & 0.082 & 0.082 & 0.082 & 0.082 & 0.082 & 0.082 & 0.082 & 0.082 & 0.082 \\
& & EP         & 0.490 & 0.385 & 0.440 & 0.457 & 0.462 & 0.569 & 0.395 & 0.461 & 0.629 & 0.316 & 0.357 & 0.422 & 0.263 & 0.270 & 0.263 \\
& & $B^2$-EP   & 0.564 & 0.536 & 0.459 & 0.508 & 0.643 & 0.610 & 0.416 & 0.596 & 0.696 & 0.321 & 0.409 & 0.459 & 0.246 & 0.268 & 0.265 \\
& & $B^2$-CP   & 0.392 & 0.427 & 0.436 & 0.364 & 0.412 & 0.442 & 0.331 & 0.374 & 0.398 & 0.272 & 0.275 & 0.306 & 0.230 & 0.187 & 0.181 \\
\cmidrule(l{0pt}r{0pt}){2-18}

& \multirow{4}{*}{IA2} & GSD        & 0.600 & 0.600 & 0.600 & 0.600 & 0.600 & 0.600 & 0.600 & 0.600 & 0.600 & 0.600 & 0.600 & 0.600 & 0.600 & 0.600 & 0.600 \\
& & EP         & 0.910 & 0.884 & 0.892 & 0.915 & 0.895 & 0.907 & 0.923 & 0.919 & 0.935 & 0.902 & 0.918 & 0.953 & 0.884 & 0.888 & 0.891 \\
& & $B^2$-EP   & 0.927 & 0.904 & 0.889 & 0.942 & 0.925 & 0.907 & 0.930 & 0.955 & 0.941 & 0.905 & 0.938 & 0.960 & 0.872 & 0.883 & 0.892 \\
& & $B^2$-CP   & 0.928 & 0.919 & 0.922 & 0.923 & 0.937 & 0.929 & 0.909 & 0.940 & 0.946 & 0.889 & 0.904 & 0.925 & 0.863 & 0.848 & 0.834 \\
\cmidrule(l{0pt}r{0pt}){2-18}

& \multirow{4}{*}{FA} & GSD        & 0.888 & 0.888 & 0.888 & 0.888 & 0.888 & 0.888 & 0.888 & 0.888 & 0.888 & 0.888 & 0.888 & 0.888 & 0.888 & 0.888 & 0.888 \\
& & EP         & 0.993 & 0.990 & 0.992 & 0.996 & 0.992 & 0.993 & 0.991 & 0.992 & 0.995 & 0.992 & 0.995 & 0.999 & 0.992 & 0.996 & 0.989 \\
& & $B^2$-EP   & 0.993 & 0.992 & 0.992 & 0.996 & 0.992 & 0.993 & 0.991 & 0.995 & 0.995 & 0.991 & 0.997 & 0.999 & 0.992 & 0.994 & 0.988 \\
& & $B^2$-CP   & 0.996 & 0.996 & 0.997 & 0.996 & 0.998 & 0.997 & 0.994 & 0.998 & 0.999 & 0.992 & 0.996 & 0.997 & 0.990 & 0.989 & 0.985 \\
\midrule

\multirow{12}{*}{\textbf{0.8}} & \multirow{4}{*}{IA1} & GSD        & 0.101 & 0.101 & 0.101 & 0.101 & 0.101 & 0.101 & 0.101 & 0.101 & 0.101 & 0.101 & 0.101 & 0.101 & 0.101 & 0.101 & 0.101 \\
& & EP         & 0.273 & 0.241 & 0.248 & 0.265 & 0.264 & 0.303 & 0.229 & 0.244 & 0.273 & 0.203 & 0.216 & 0.219 & 0.201 & 0.217 & 0.212 \\
& & $B^2$-EP   & 0.303 & 0.296 & 0.248 & 0.284 & 0.308 & 0.322 & 0.237 & 0.263 & 0.294 & 0.197 & 0.209 & 0.219 & 0.192 & 0.207 & 0.211 \\
& & $B^2$-CP   & 0.247 & 0.255 & 0.273 & 0.220 & 0.261 & 0.292 & 0.207 & 0.235 & 0.220 & 0.182 & 0.165 & 0.148 & 0.160 & 0.151 & 0.133 \\
\cmidrule(l{0pt}r{0pt}){2-18}

& \multirow{4}{*}{IA2} & GSD        & 0.633 & 0.633 & 0.633 & 0.633 & 0.633 & 0.633 & 0.633 & 0.633 & 0.633 & 0.633 & 0.633 & 0.633 & 0.633 & 0.633 & 0.633 \\
& & EP         & 0.845 & 0.828 & 0.828 & 0.848 & 0.838 & 0.838 & 0.841 & 0.835 & 0.866 & 0.819 & 0.822 & 0.819 & 0.812 & 0.822 & 0.818 \\
& & $B^2$-EP   & 0.861 & 0.838 & 0.826 & 0.855 & 0.860 & 0.840 & 0.843 & 0.847 & 0.874 & 0.815 & 0.813 & 0.817 & 0.804 & 0.813 & 0.820 \\
& & $B^2$-CP   & 0.856 & 0.864 & 0.862 & 0.850 & 0.865 & 0.873 & 0.836 & 0.845 & 0.860 & 0.825 & 0.806 & 0.811 & 0.807 & 0.787 & 0.770 \\
\cmidrule(l{0pt}r{0pt}){2-18}

& \multirow{4}{*}{FA} & GSD        & 0.902 & 0.902 & 0.902 & 0.902 & 0.902 & 0.902 & 0.902 & 0.902 & 0.902 & 0.902 & 0.902 & 0.902 & 0.902 & 0.902 & 0.902 \\
& & EP         & 0.974 & 0.969 & 0.972 & 0.975 & 0.969 & 0.973 & 0.973 & 0.972 & 0.975 & 0.967 & 0.969 & 0.966 & 0.968 & 0.969 & 0.968 \\
& & $B^2$-EP   & 0.975 & 0.971 & 0.971 & 0.977 & 0.972 & 0.974 & 0.975 & 0.978 & 0.977 & 0.963 & 0.968 & 0.966 & 0.966 & 0.966 & 0.969 \\
& & $B^2$-CP   & 0.977 & 0.977 & 0.979 & 0.977 & 0.977 & 0.980 & 0.971 & 0.980 & 0.979 & 0.965 & 0.969 & 0.964 & 0.961 & 0.958 & 0.957 \\
\bottomrule
\end{tabular}%
}

\vspace{2pt}
\begin{minipage}{0.98\textwidth}
\footnotesize
GSD: group sequential design; EP: elastic prior; \(B^2\)-EP: \(B^2\)-FIC calibrated elastic prior; \(B^2\)-CP: \(B^2\)-FIC calibrated commensurate prior. 
\end{minipage}

\end{threeparttable}
\end{table}

\clearpage

\begin{table}[H]
\centering
\caption{IA2 scheduling decision table on the information-fraction scale}
\label{tab:decision_rules_compare}
\footnotesize
\setlength{\tabcolsep}{3pt}
\renewcommand{\arraystretch}{1.15}

\begin{tabularx}{\textwidth}{
>{\centering\arraybackslash}p{2.6cm}
*{5}{>{\centering\arraybackslash}X}
}
\toprule
\multirow{2}{*}{\makecell{\textbf{Phase II}\\ \textbf{HR interval}}} &
\multicolumn{5}{c}{\textbf{Phase III IA1 HR interval}} \\
\cmidrule(lr){2-6}
&
\makecell{$\leq 0.50$} &
\makecell{$>0.50$\\ to $\leq 0.60$} &
\makecell{$>0.60$\\ to $\leq 0.70$} &
\makecell{$>0.70$\\ to $\leq 0.80$} &
\makecell{$>0.80$\\ to $\leq 0.90$} \\
\midrule

\rule{0pt}{2.3ex}$\leq 0.50$
& IA1 & IA1 & $0.70$ & $0.80$ & $1.0$ \\

\rule{0pt}{2.3ex}$>0.50$ to $\leq 0.60$
& $0.60$ & $0.60$ & $0.70$ & $0.80$ & $1.0$ \\

\rule{0pt}{2.3ex}$>0.60$ to $\leq 0.70$
& $0.60$ & $0.60$ & $0.70$ & $0.80$ & $1.0$ \\

\rule{0pt}{2.3ex}$>0.70$ to $\leq 0.80$
& $0.70$ & $0.70$ & $0.80$ & $0.80$ & $1.0$ \\

\rule{0pt}{2.3ex}$>0.80$ to $\leq 0.90$
& $1.0$ & $1.0$ & $1.0$ & $1.0$ & $1.0$ \\

\bottomrule
\end{tabularx}

\vspace{0.5ex}
\parbox{\textwidth}{%
\footnotesize
Results are shown for \(m_C^{(III)}=7\) months. 0.60--0.80
denote IA2 information fractions, and 1.0 denotes final analysis.
}
\end{table}

\clearpage

\begin{table}[H]
\centering
\caption{IA2 scheduling decision table on the event-count scale}
\label{tab:decision_rules_compare_events}
\footnotesize
\setlength{\tabcolsep}{3pt}
\renewcommand{\arraystretch}{1.15}

\begin{tabularx}{\textwidth}{
>{\centering\arraybackslash}p{2.6cm}
*{5}{>{\centering\arraybackslash}X}
}
\toprule
\multirow{2}{*}{\makecell{\textbf{Phase II}\\ \textbf{HR interval}}} &
\multicolumn{5}{c}{\textbf{Phase III IA1 HR interval}} \\
\cmidrule(lr){2-6}
&
\makecell{$\leq 0.50$} &
\makecell{$>0.50$\\ to $\leq 0.60$} &
\makecell{$>0.60$\\ to $\leq 0.70$} &
\makecell{$>0.70$\\ to $\leq 0.80$} &
\makecell{$>0.80$\\ to $\leq 0.90$} \\
\midrule

\rule{0pt}{2.3ex}$\leq 0.50$
& 36 & 66 & 236 & 688 & 3853 \\

\rule{0pt}{2.3ex}$>0.50$ to $\leq 0.60$
& 54 & 98 & 236 & 688 & 3853 \\

\rule{0pt}{2.3ex}$>0.60$ to $\leq 0.70$
& 54 & 98 & 236 & 688 & 3853 \\

\rule{0pt}{2.3ex}$>0.70$ to $\leq 0.80$
& 63 & 115 & 269 & 688 & 3853 \\

\rule{0pt}{2.3ex}$>0.80$ to $\leq 0.90$
& 89 & 163 & 336 & 859 & 3853 \\

\bottomrule
\end{tabularx}

\vspace{0.5ex}
\parbox{\textwidth}{%
\footnotesize
Results are shown for \(m_C^{(III)}=7\) months. Entries are the IA2 event
targets corresponding to Table~\ref{tab:decision_rules_compare}.
}
\end{table}

\clearpage
\begin{table}[ht]
\centering
\caption{Characteristics of two example clinical trials used in case study}
\label{tab:case_characteristics}
\renewcommand{\arraystretch}{1.4}
\small
\begin{tabular}{p{0.40\textwidth} >{\centering\arraybackslash}p{0.25\textwidth} >{\centering\arraybackslash}p{0.25\textwidth}}
\toprule
\textbf{Parameter} & \textbf{Trial 1} & \textbf{Trial 2} \\
\midrule
Randomization ratio & 2:1 & 2:1 \\
Pre-specified information fractions & (0.3, 0.7, 1.0) & (0.3, 0.7, 1.0) \\
Candidate IA2 info fractions & (0.4, 0.5, 0.6, 0.7, 0.8) & (0.4, 0.5, 0.6, 0.7, 0.8) \\
Candidate HR$^\text{(III)}$ at IA1& (0.5, 0.6, 0.7, 0.8, 0.9) & (0.3, 0.4, 0.5, 0.6, 0.7, 0.8, 0.9) \\
Posterior predictive threshold $\gamma$ & 0.9 & 0.9 \\
Control arm median PFS & 8.5 months & 1.8 months \\
HR$^\text{(II)}$& 0.70 & 0.18 \\
Phase II treatment arm $n$ & 80 & 47 \\
Phase III total events (FA) & 162 & 72 \\
\bottomrule
\end{tabular}
\end{table}

\clearpage

\begin{table}[H]
\centering
\caption{Case study: IA2 scheduling decision tables}
\label{tab:decision_rules_case_study}
\scriptsize
\renewcommand{\arraystretch}{1.15}

\begin{tabularx}{\textwidth}{
>{\centering\arraybackslash}m{2.8cm}
*{5}{>{\centering\arraybackslash}X}
}
\toprule
\multirow{2}{*}{\textbf{Trial 1}} &
\multicolumn{5}{c}{\textbf{Phase III IA1 HR interval}} \\
\cmidrule(lr){2-6}
& \makecell{\textbf{$>$0.4}\\ \textbf{to $\leq$ 0.5}} & \makecell{\textbf{$>$0.5}\\ \textbf{to $\leq$ 0.6}} & \makecell{\textbf{$>$0.6}\\ \textbf{to $\leq$ 0.7}} & \makecell{\textbf{$>$0.7}\\ \textbf{to $\leq$ 0.8}} & \makecell{\textbf{$>$0.8}\\ \textbf{to $\leq$ 0.9}} \\
\midrule
HR$^\text{(II)}$ = 0.70 & IA2 at $0.7$ & IA2 at $0.7$ & IA2 at $0.8$ & FA & FA \\
\bottomrule
\end{tabularx}

\begin{tabularx}{\textwidth}{
>{\centering\arraybackslash}m{2.8cm}
*{7}{>{\centering\arraybackslash}X}
}
\toprule
\multirow{2}{*}{\textbf{Trial 2}} &
\multicolumn{7}{c}{\textbf{Phase III IA1 HR interval}} \\
\cmidrule(lr){2-8}
& \makecell{\textbf{$\leq$ 0.3}} & \makecell{\textbf{$>$0.3}\\ \textbf{to $\leq$ 0.4}} & \makecell{\textbf{$>$0.4}\\ \textbf{to $\leq$ 0.5}} & \makecell{\textbf{$>$0.5}\\ \textbf{to $\leq$ 0.6}} & \makecell{\textbf{$>$0.6}\\ \textbf{to $\leq$ 0.7}} & \makecell{\textbf{$>$0.7}\\ \textbf{to $\leq$ 0.8}} & \makecell{\textbf{$>$0.8}\\ \textbf{to $\leq$ 0.9}} \\
\midrule
HR$^\text{(II)}$ = 0.18 & Early stop & IA2 at $0.6$ & IA2 at $0.6$ & IA2 at $0.7$ & IA2 at $0.8$ & FA & FA \\
\bottomrule
\end{tabularx}

\vspace{1em}

\vspace{0.5ex}
\end{table}

\clearpage

\clearpage

\clearpage
\begin{figure}[H]
    \centering
    \includegraphics[
        width=\textwidth,
        height=0.52\textheight,
        keepaspectratio,
        trim=2mm 2mm 2mm 2mm,
        clip
    ]{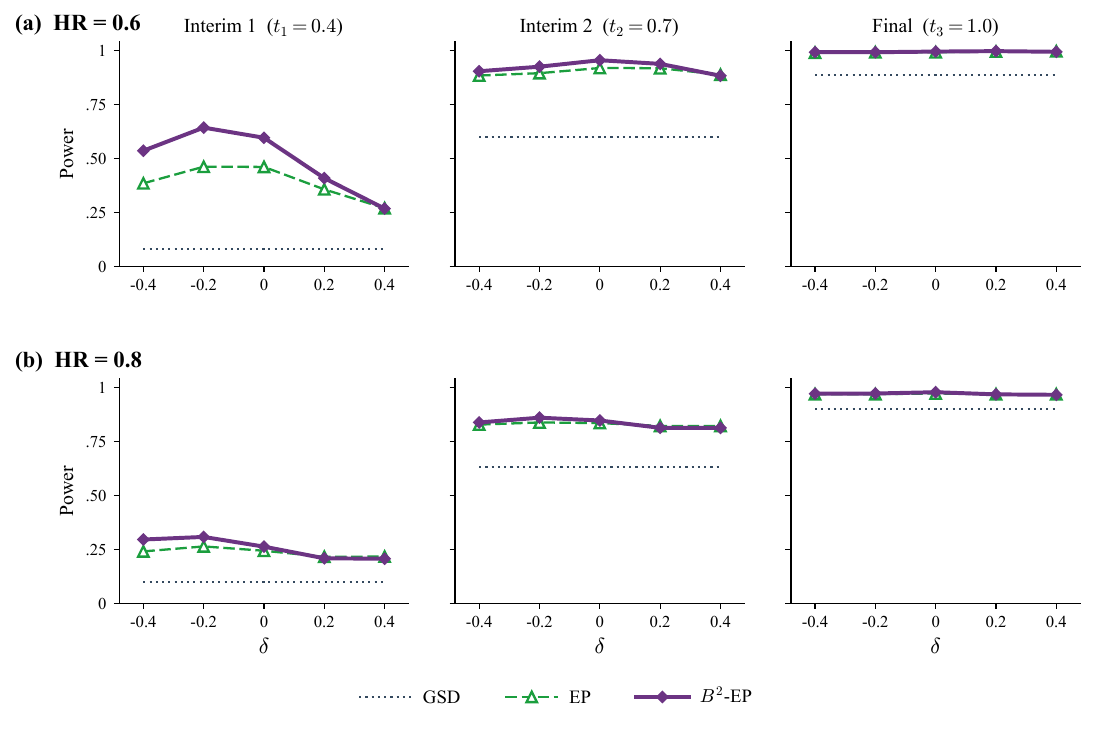}
    \caption{Empirical power across discrepancy scenarios}
    \label{fig:power-trajectory-delta}
\end{figure}

\clearpage

\clearpage

\clearpage
\appendix






\renewcommand{\thesection}{S\arabic{section}}
\renewcommand{\thesubsection}{S\arabic{section}.\arabic{subsection}}
\renewcommand{\thetable}{S\arabic{table}}
\renewcommand{\thefigure}{S\arabic{figure}}
\setcounter{section}{0}
\setcounter{table}{0}
\setcounter{figure}{0}


\section{Weibull likelihood and sufficient statistics}
\label{app:weibull_likelihood}

This appendix provides the derivation of the likelihood representation used in
Section~\ref{sec:notation_outcome_model}. For a given group \(g\), phase \(s\),
and analysis \(j\), we assume a Weibull model with shared shape parameter
\(k\) and group-specific Weibull scale parameter \(\lambda_g^{(s)}\) under the
parameterization used in this study. The distribution function is
\[
  F_g^{(s)}(t)
  =
  \Pr(T\leq t)
  =
  1-\exp\{-\lambda_g^{(s)}t^k\},
  \qquad t\geq 0.
\]
Therefore, the survival function is
\[
  S_g^{(s)}(t)
  =
  \Pr(T>t)
  =
  1-F_g^{(s)}(t)
  =
  \exp\{-\lambda_g^{(s)}t^k\}.
\]

Let
\[
  \mathcal{D}_{g,j}^{(s)}
  =
  \left\{
  (t_{ig,j}^{(s)}, e_{ig,j}^{(s)}):
  i=1,\ldots,n_g^{(s)}
  \right\},
\]
where \(t_{ig,j}^{(s)}\) is the observed follow-up time and
\(e_{ig,j}^{(s)}\in\{0,1\}\) is the event indicator. If
\(e_{ig,j}^{(s)}=1\), the event is observed at \(t_{ig,j}^{(s)}\), and the
likelihood contribution is the event-time density. Since
\[
  f_g^{(s)}(t)
  =
  \frac{d}{dt}F_g^{(s)}(t)
  =
  \lambda_g^{(s)}k t^{k-1}
  \exp\{-\lambda_g^{(s)}t^k\},
\]
the event contribution is
\[
  f_g^{(s)}\!\left(t_{ig,j}^{(s)}\right)
  =
  \lambda_g^{(s)}k
  \left\{t_{ig,j}^{(s)}\right\}^{k-1}
  \exp\left\{
    -\lambda_g^{(s)}
    \left\{t_{ig,j}^{(s)}\right\}^{k}
  \right\}.
\]
If \(e_{ig,j}^{(s)}=0\), the event has not been observed by
\(t_{ig,j}^{(s)}\), and the likelihood contribution is the survival
probability
\[
  S_g^{(s)}\!\left(t_{ig,j}^{(s)}\right)
  =
  \exp\left\{
    -\lambda_g^{(s)}
    \left\{t_{ig,j}^{(s)}\right\}^{k}
  \right\}.
\]

Therefore, the likelihood contribution from observation \(i\) can be written
as
\[
  \left[
    f_g^{(s)}\!\left(t_{ig,j}^{(s)}\right)
  \right]^{e_{ig,j}^{(s)}}
  \left[
    S_g^{(s)}\!\left(t_{ig,j}^{(s)}\right)
  \right]^{1-e_{ig,j}^{(s)}} .
\]
Substituting the event and censoring contributions gives
\[
  \left[
    \lambda_g^{(s)} k
    \left\{t_{ig,j}^{(s)}\right\}^{k-1}
    \exp\left\{
      -\lambda_g^{(s)}
      \left\{t_{ig,j}^{(s)}\right\}^{k}
    \right\}
  \right]^{e_{ig,j}^{(s)}}
  \left[
    \exp\left\{
      -\lambda_g^{(s)}
      \left\{t_{ig,j}^{(s)}\right\}^{k}
    \right\}
  \right]^{1-e_{ig,j}^{(s)}} .
\]
Since the exponential term appears in both the event and censoring
contributions, this simplifies to
\[
  \left[
    \lambda_g^{(s)} k
    \left\{t_{ig,j}^{(s)}\right\}^{k-1}
  \right]^{e_{ig,j}^{(s)}}
  \exp\left\{
    -\lambda_g^{(s)}
    \left\{t_{ig,j}^{(s)}\right\}^{k}
  \right\}.
\]

Thus, conditional on \(k\), the likelihood for \(\lambda_g^{(s)}\) is
\[
  L\left(\lambda_g^{(s)}\mid \mathcal{D}_{g,j}^{(s)}\right)
  =
  \prod_{i=1}^{n_g^{(s)}}
  \left[
    \lambda_g^{(s)} k
    \left\{t_{ig,j}^{(s)}\right\}^{k-1}
  \right]^{e_{ig,j}^{(s)}}
  \exp\left\{
    -\lambda_g^{(s)}
    \left\{t_{ig,j}^{(s)}\right\}^{k}
  \right\}.
\]
Collecting terms that depend on \(\lambda_g^{(s)}\), we obtain
\[
  L\left(\lambda_g^{(s)}\mid \mathcal{D}_{g,j}^{(s)}\right)
  =
  \left(\lambda_g^{(s)}\right)^{\sum_i e_{ig,j}^{(s)}}
  \exp\left\{
    -\lambda_g^{(s)}
    \sum_i
    \left(t_{ig,j}^{(s)}\right)^k
  \right\}
  \prod_{i=1}^{n_g^{(s)}}
  \left[
    k
    \left\{t_{ig,j}^{(s)}\right\}^{k-1}
  \right]^{e_{ig,j}^{(s)}} .
\]
The final product depends only on \(k\) and the observed follow-up times, and
does not depend on \(\lambda_g^{(s)}\). It can therefore be absorbed into the
proportionality constant. Defining
\[
  \widehat E_{g,j}^{(s)}
  =
  \sum_{i=1}^{n_g^{(s)}} e_{ig,j}^{(s)},
  \qquad
  \widehat T_{g,j}^{(s)}
  =
  \sum_{i=1}^{n_g^{(s)}}
  \left(t_{ig,j}^{(s)}\right)^k,
\]
the likelihood becomes
\[
  L\left(\lambda_g^{(s)}\mid \mathcal{D}_{g,j}^{(s)}\right)
  \propto
  \left(\lambda_g^{(s)}\right)^{\widehat E_{g,j}^{(s)}}
  \exp\left\{
    -\lambda_g^{(s)}\widehat T_{g,j}^{(s)}
  \right\}.
\]
Thus, conditional on the shape parameter \(k\), the likelihood for
\(\lambda_g^{(s)}\) factorizes into a term depending on the data only through
\(\widehat E_{g,j}^{(s)}\) and \(\widehat T_{g,j}^{(s)}\), multiplied by a term
free of \(\lambda_g^{(s)}\). By the Neyman--Fisher factorization theorem,
\(\{\widehat E_{g,j}^{(s)},\widehat T_{g,j}^{(s)}\}\) are sufficient statistics
for \(\lambda_g^{(s)}\).

The relationship between the median event time and the Weibull scale parameter
follows from the definition of the median. If \(m_g^{(s)}\) is the median, then
\[
  S_g^{(s)}\!\left(m_g^{(s)}\right)
  =
  \frac{1}{2}.
\]
Using
\[
  S_g^{(s)}(t)
  =
  \exp\{-\lambda_g^{(s)}t^k\},
\]
we have
\[
  \exp\left\{
    -\lambda_g^{(s)}
    \left(m_g^{(s)}\right)^k
  \right\}
  =
  \frac{1}{2}.
\]
Taking logarithms gives
\[
  \lambda_g^{(s)}
  \left(m_g^{(s)}\right)^k
  =
  \log(2),
\]
and therefore
\[
  \lambda_g^{(s)}
  =
  \frac{\log(2)}{\left(m_g^{(s)}\right)^k}.
\]

Finally, under the phase~II--phase~III discrepancy model used for the
treatment arm,
\[
  \lambda_T^{(II)}
  =
  \lambda_T^{(III)}\exp(\delta).
\]
Using the median--scale relationship above,
\[
  \frac{\log(2)}{\left(m_T^{(II)}\right)^k}
  =
  \frac{\log(2)}{\left(m_T^{(III)}\right)^k}
  \exp(\delta).
\]
Canceling \(\log(2)\) and rearranging gives
\[
  \left(m_T^{(II)}\right)^k
  =
  \left(m_T^{(III)}\right)^k
  \exp(-\delta),
\]
and hence
\[
  m_T^{(II)}
  =
  m_T^{(III)}\exp(-\delta/k).
\]
Thus, \(\delta<0\) implies \(m_T^{(II)}>m_T^{(III)}\), corresponding to a more
optimistic phase~II treatment arm, whereas \(\delta>0\) implies
\(m_T^{(II)}<m_T^{(III)}\).

\section{Conjugate prior borrowing}
\label{app:conjugate_borrowing}

This section summarizes the conjugate Gamma updating rule used for the
phase~III Weibull scale parameter. Let \(g\in\{T,C\}\) denote the phase~III
treatment or control arm, and let \(\lambda_g^{(III)}\) be the corresponding
Weibull scale parameter. At analysis \(j\), the current phase~III event-time
data are \(\mathcal D_{g,j}^{(III)}\). From Section~A, the phase~III likelihood
has kernel
\[
  L\!\left(
  \lambda_g^{(III)}
  \mid
  \mathcal D_{g,j}^{(III)}
  \right)
  \propto
  \left(\lambda_g^{(III)}\right)^{\widehat E_{g,j}^{(III)}}
  \exp\left\{
    -\lambda_g^{(III)}
    \widehat T_{g,j}^{(III)}
  \right\}.
\]

External information is incorporated through a Gamma prior for
\(\lambda_g^{(III)}\). We use the shape--rate parameterization of the Gamma
distribution: if
\[
  \lambda_g^{(III)}
  \sim
  \mathrm{Gamma}(A_g,B_g),
\]
then its density is
\[
  \pi\!\left(\lambda_g^{(III)}\right)
  =
  \frac{B_g^{A_g}}{\Gamma(A_g)}
  \left(\lambda_g^{(III)}\right)^{A_g-1}
  \exp\left\{
    -B_g\lambda_g^{(III)}
  \right\},
  \qquad
  \lambda_g^{(III)}>0,
\]
where \(A_g\) is the shape parameter and \(B_g\) is the rate parameter. Up to
a normalizing constant that does not depend on \(\lambda_g^{(III)}\), this has
kernel
\[
  \pi\!\left(\lambda_g^{(III)}\right)
  \propto
  \left(\lambda_g^{(III)}\right)^{A_g-1}
  \exp\left\{
    -B_g\lambda_g^{(III)}
  \right\}.
\]

Combining this Gamma prior with the phase~III likelihood gives
\[
\begin{aligned}
  \pi\!\left(
  \lambda_g^{(III)}
  \mid
  \mathcal D_{g,j}^{(III)}
  \right)
  &\propto
  \pi\!\left(\lambda_g^{(III)}\right)
  L\!\left(
  \lambda_g^{(III)}
  \mid
  \mathcal D_{g,j}^{(III)}
  \right)  \\
  &\propto
  \left(\lambda_g^{(III)}\right)^{A_g-1}
  \exp\left\{
    -B_g\lambda_g^{(III)}
  \right\}
  \left(\lambda_g^{(III)}\right)^{\widehat E_{g,j}^{(III)}}
  \exp\left\{
    -\lambda_g^{(III)}\widehat T_{g,j}^{(III)}
  \right\}  \\
  &=
  \left(\lambda_g^{(III)}\right)^{
    A_g+\widehat E_{g,j}^{(III)}-1
  }
  \exp\left\{
    -\left(
      B_g+\widehat T_{g,j}^{(III)}
    \right)
    \lambda_g^{(III)}
  \right\}.
\end{aligned}
\]
This is the kernel of a Gamma distribution. Therefore,
\[
  \lambda_g^{(III)}
  \mid
  \mathcal D_{g,j}^{(III)}
  \sim
  \mathrm{Gamma}
  \left(
    A_g+\widehat E_{g,j}^{(III)},
    B_g+\widehat T_{g,j}^{(III)}
  \right).
\]
Thus, for either phase~III arm, the posterior shape parameter is obtained by
adding the observed number of phase~III events, and the posterior rate
parameter is obtained by adding the Weibull-adjusted phase~III follow-up time.

For the phase~III treatment arm, the conjugate borrowing prior is
constructed directly from the likelihood contribution of the continuing
phase~II treatment-arm event-time data. At analysis \(j\), the prior
parameters are
\[
  A_T
  =
  \widehat E_{T,j}^{(II)},
  \qquad
  B_T
  =
  \widehat T_{T,j}^{(II)}.
\]
Equivalently,
\[
  \lambda_T^{(III)}
  \mid
  \mathcal D_{T,j}^{(II)}
  \sim
  \mathrm{Gamma}
  \left(
    \widehat E_{T,j}^{(II)},
    \widehat T_{T,j}^{(II)}
  \right).
\]
After incorporating the current phase~III treatment-arm data,
\[
  \lambda_T^{(III)}
  \mid
  \mathcal D_{T,j}^{(II)},\mathcal D_{T,j}^{(III)}
  \sim
  \mathrm{Gamma}
  \left(
    \widehat E_{T,j}^{(II)}+\widehat E_{T,j}^{(III)},
    \widehat T_{T,j}^{(II)}+\widehat T_{T,j}^{(III)}
  \right).
\]
This corresponds to full borrowing of the continuing phase~II treatment-arm
event-time information, with no additional baseline pseudo-events added to the
phase~II-induced Gamma contribution.

For the phase~III control arm, external RWD is incorporated as a fixed
informative Gamma prior rather than as a dynamically accumulating event-time
data source. The prior parameters are
\[
  A_C
  =
  \alpha_C,
  \qquad
  B_C
  =
  \frac{\alpha_C}{\lambda_C^{\mathrm{ext}}},
\]
where \(\lambda_C^{\mathrm{ext}}\) is the RWD-derived Weibull scale parameter.
Updating this prior with the current phase~III control-arm data gives
\[
  \lambda_C^{(III)}
  \mid
  \mathcal D_{C,j}^{(III)}
  \sim
  \mathrm{Gamma}
  \left(
    \alpha_C+\widehat E_{C,j}^{(III)},
    \frac{\alpha_C}{\lambda_C^{\mathrm{ext}}}
    +
    \widehat T_{C,j}^{(III)}
  \right).
\]

\section{\(B^2\)-FIC calibrated commensurate prior borrowing}
\label{app:b2_cp_borrowing}

The \(B^2\)-FIC calibrated commensurate prior models the phase~II and phase~III
treatment-arm Weibull scale parameters as related but not necessarily
exchangeable. Let
\[
  \lambda_T^{(II)} = \lambda_0
\]
denote the treatment-arm Weibull scale parameter underlying the continuing
phase~II cohort. A weak Gamma prior is assigned to \(\lambda_0\),
\[
  \lambda_0
  \sim
  \mathrm{Gamma}(A_0,B_0),
\]
where the Gamma distribution is parameterized by shape and rate. In the main
implementation, \(A_0=B_0=0.01\). The phase~II treatment-arm likelihood has
kernel
\[
  L\!\left(\lambda_0\mid \mathcal D_{T,j}^{(II)}\right)
  \propto
  \lambda_0^{\widehat E_{T,j}^{(II)}}
  \exp\left\{
    -\lambda_0\widehat T_{T,j}^{(II)}
  \right\}.
\]
Therefore, the phase~II-updated marginal contribution for \(\lambda_0\) is
proportional to
\[
  \lambda_0^{A_0-1+\widehat E_{T,j}^{(II)}}
  \exp\left\{
    -\lambda_0
    \left(
      B_0+\widehat T_{T,j}^{(II)}
    \right)
  \right\}.
\]

Borrowing is introduced on the log scale of the Weibull scale parameter. Let
\[
  \phi_0=\log(\lambda_0),
  \qquad
  \phi=\log\left(\lambda_T^{(III)}\right).
\]
The commensurate prior links the phase~III treatment-arm parameter to the
phase~II treatment-arm parameter through
\[
  \phi \mid \phi_0,\sigma_{\mathrm{bor}}
  \sim
  N\left(\phi_0,\sigma_{\mathrm{bor}}^2\right).
\]
Equivalently,
\[
  \lambda_T^{(III)}
  =
  \exp(\phi),
  \qquad
  \phi=\phi_0+\epsilon,
  \qquad
  \epsilon\mid\sigma_{\mathrm{bor}}
  \sim
  N(0,\sigma_{\mathrm{bor}}^2).
\]
Small values of \(\sigma_{\mathrm{bor}}\) imply strong commensurability and
therefore strong borrowing, whereas large values of \(\sigma_{\mathrm{bor}}\)
allow the phase~III treatment-arm parameter to depart from the phase~II
parameter.

The \(B^2\)-FIC calibrated version uses a robust spike--slab prior on
\(\sigma_{\mathrm{bor}}\),
\[
  \sigma_{\mathrm{bor}}
  \sim
  \frac{1}{2}
  \mathrm{HalfNormal}(s_{\mathrm{spike}})
  +
  \frac{1}{2}
  \mathrm{HalfNormal}(s_{\mathrm{slab}}),
\]
with
\[
  s_{\mathrm{spike}}=0.25,
  \qquad
  s_{\mathrm{slab}}=2.0
\]
in the main implementation. The spike component favors tight borrowing when
the phase~II and phase~III treatment-arm data are compatible, whereas the slab
component permits large phase~II--phase~III discrepancies and attenuates
borrowing.

Using the non-centered parameterization implemented in the MCMC sampler, the
commensurate prior can be written as
\[
  z\sim N(0,1),
  \qquad
  \phi=\phi_0+z\sigma_{\mathrm{bor}},
  \qquad
  \lambda_T^{(III)}=\exp(\phi).
\]
Here \(z\) is an auxiliary standard-normal variable used for posterior
sampling; marginally, this representation is equivalent to
\(\phi\mid\phi_0,\sigma_{\mathrm{bor}}\sim
N(\phi_0,\sigma_{\mathrm{bor}}^2)\).

The phase~III treatment-arm likelihood is
\[
  L\!\left(\lambda_T^{(III)}
  \mid
  \mathcal D_{T,j}^{(III)}
  \right)
  \propto
  \left(\lambda_T^{(III)}\right)^{\widehat E_{T,j}^{(III)}}
  \exp\left\{
    -\lambda_T^{(III)}\widehat T_{T,j}^{(III)}
  \right\}.
\]
Thus, with respect to the parameterization
\((\lambda_0,\phi,\sigma_{\mathrm{bor}})\), the joint posterior kernel is
\[
\begin{aligned}
  &\pi\left(
    \lambda_0,
    \phi,
    \sigma_{\mathrm{bor}}
    \mid
    \mathcal D_{T,j}^{(II)},\mathcal D_{T,j}^{(III)}
  \right)  \\
  &\quad \propto
  L\!\left(\lambda_0\mid \mathcal D_{T,j}^{(II)}\right)
  L\!\left(\exp(\phi)\mid \mathcal D_{T,j}^{(III)}\right)
  \pi\left(
    \phi \mid \log\lambda_0,\sigma_{\mathrm{bor}}
  \right)
  \pi(\lambda_0)
  \pi(\sigma_{\mathrm{bor}}).
\end{aligned}
\]
Posterior inference for \(\lambda_T^{(III)}=\exp(\phi)\) under this model is
obtained by MCMC.

\section{\(B^2\)-FIC calibrated elastic prior borrowing}
\label{app:b2_ep_borrowing}

The \(B^2\)-FIC calibrated elastic prior discounts the phase~II treatment-arm
Gamma contribution according to an analysis-specific discrepancy statistic.
At analysis \(j\), let
\[
  A_{T,j}^{(II)}
  =
  \widehat E_{T,j}^{(II)},
  \qquad
  B_{T,j}^{(II)}
  =
  \widehat T_{T,j}^{(II)}
\]
denote the raw Gamma shape and rate parameters induced directly by the
phase~II treatment-arm likelihood. Without discounting, these parameters would
define the phase~II-derived prior
\[
  \lambda_T^{(III)}
  \mid
  \mathcal D_{T,j}^{(II)}
  \sim
  \mathrm{Gamma}
  \left(
    A_{T,j}^{(II)},\,
    B_{T,j}^{(II)}
  \right).
\]

To quantify phase~II--phase~III treatment-arm congruence, let
\[
  C_j
  =
  C\left(
  \mathcal D_{T,j}^{(II)},\mathcal D_{T,j}^{(III)}
  \right)
\]
denote a nonnegative discrepancy statistic, with larger values indicating
greater inconsistency between the phase~II and phase~III treatment-arm
event-time data. In the main implementation, \(C_j\) is the chi-square
statistic from a log-rank comparison of the phase~II and phase~III
treatment-arm event-time data.

The elastic borrowing weight is
\[
  w_j(a_{\mathrm{el}},b_{\mathrm{el}})
  =
  \frac{1}{1+\exp\{a_{\mathrm{el}}+b_{\mathrm{el}}\log(C_j)\}},
  \qquad
  0\leq w_j(a_{\mathrm{el}},b_{\mathrm{el}})\leq 1,
\]
with \(b_{\mathrm{el}}>0\), so that the borrowing weight decreases as the
discrepancy statistic increases. The \(B^2\)-EP treatment-arm prior is obtained
by discounting both Gamma parameters by \(w_j(a_{\mathrm{el}},b_{\mathrm{el}})\):
\[
  \lambda_T^{(III)}
  \mid
  \mathcal D_{T,j}^{(II)}, C_j
  \sim
  \mathrm{Gamma}
  \left(
    w_j(a_{\mathrm{el}},b_{\mathrm{el}})A_{T,j}^{(II)},\,
    w_j(a_{\mathrm{el}},b_{\mathrm{el}})B_{T,j}^{(II)}
  \right).
\]
This construction preserves the phase~II prior mean but inflates its variance
when \(w_j(a_{\mathrm{el}},b_{\mathrm{el}})<1\), thereby reducing the effective
contribution of the phase~II treatment-arm data under incongruence.

After observing the phase~III treatment-arm data at analysis \(j\), the
posterior remains conjugate:
\[
  \lambda_T^{(III)}
  \mid
  \mathcal D_{T,j}^{(II)},\mathcal D_{T,j}^{(III)}, C_j
  \sim
  \mathrm{Gamma}
  \left(
    w_j(a_{\mathrm{el}},b_{\mathrm{el}})A_{T,j}^{(II)}
    +
    \widehat E_{T,j}^{(III)},\,
    w_j(a_{\mathrm{el}},b_{\mathrm{el}})B_{T,j}^{(II)}
    +
    \widehat T_{T,j}^{(III)}
  \right).
\]

The elastic-function hyperparameters \((a_{\mathrm{el}},b_{\mathrm{el}})\) are
fixed at the design stage. In the \(B^2\)-FIC calibrated implementation, they
are determined through two reference values of the discrepancy statistic. Let
\(C_0\) denote a congruent reference point and \(C_1\) denote an incongruent
reference point, with \(C_1>C_0\). These reference points are obtained from
prespecified quantile anchors \((q_0,q_1)\) of the simulated distributions of
\(C_j\). Specifically, \(C_0\) is taken as the \(q_0\) quantile under a
congruent phase~II--phase~III scenario, whereas \(C_1\) is taken as the smaller
of the \(q_1\) quantiles from the two boundary-incongruence scenarios
\(\delta=+\delta^*\) and \(\delta=-\delta^*\). The latter choice provides a
conservative incongruence reference point for calibrating the elastic function.

The elastic function is anchored by
\[
  w(C_0)=0.99,
  \qquad
  w(C_1)=0.01.
\]
Solving
\[
  \frac{1}{1+\exp\{a_{\mathrm{el}}+b_{\mathrm{el}}\log(C_0)\}}=0.99,
  \qquad
  \frac{1}{1+\exp\{a_{\mathrm{el}}+b_{\mathrm{el}}\log(C_1)\}}=0.01
\]
gives
\[
  b_{\mathrm{el}}
  =
  \frac{2\log(99)}
       {\log(C_1)-\log(C_0)},
  \qquad
  a_{\mathrm{el}}
  =
  -\log(99)-b_{\mathrm{el}}\log(C_0).
\]
Thus, the calibrated elastic prior borrows nearly fully when the phase~II and
phase~III treatment-arm data are close to the congruent reference point, and
borrows minimally when their discrepancy approaches the boundary-incongruence
reference point.

To make type~I error protection explicit in the confirmatory phase~III setting,
the quantile anchors \((q_0,q_1)\), and hence the corresponding
\((a_{\mathrm{el}},b_{\mathrm{el}})\), are selected using the \(B^2\)-FIC
hard-constrained calibration rule rather than the original elastic-prior
utility criterion. For each candidate pair \((q_0,q_1)\) on a prespecified
grid, the corresponding \((a_{\mathrm{el}},b_{\mathrm{el}})\) is obtained from
the anchor equations above. Let
\(\alpha_{+}^{\mathrm{trial}}(q_0,q_1)\) and
\(\alpha_{-}^{\mathrm{trial}}(q_0,q_1)\) denote the overall trial-wise
one-sided type~I error rates under the two boundary discrepancy scenarios
\(\delta=+\delta^*\) and \(\delta=-\delta^*\), respectively, and let
\(\mathrm{Power}(q_0,q_1)\) denote power under the prespecified design
alternative. The calibrated feasible set is
\[
  \mathcal Q_{\mathrm{cal}}
  =
  \left\{
  (q_0,q_1):
  \max\left[
    \alpha_{+}^{\mathrm{trial}}(q_0,q_1),
    \alpha_{-}^{\mathrm{trial}}(q_0,q_1)
  \right]
  \leq
  \alpha^*
  \right\}.
\]
The \(B^2\)-EP anchors are selected as
\[
  (q_0^{B^2},q_1^{B^2})
  =
  \arg\max_{(q_0,q_1)\in\mathcal Q_{\mathrm{cal}}}
  \mathrm{Power}(q_0,q_1).
\]
The final \(B^2\)-EP model uses the elastic-function parameters
\((a_{\mathrm{el}}^{B^2},b_{\mathrm{el}}^{B^2})\) determined by these selected
anchors. Thus, unlike the original EP, which uses a soft utility penalty for
type~I error inflation, the \(B^2\)-EP implementation imposes type~I error
control as a hard feasibility constraint and optimizes power only within the
feasible region.

\section{Predictive probability for adaptive IA2 timing under \(B^2\)-CP}
\label{app:b2_cp_predictive_probability}

For the \(B^2\)-CP implementation, the IA1 posterior for
\((\lambda_T^{(III)},\lambda_C^{(III)})\) is obtained from the calibrated
commensurate prior model described in
Appendix~\ref{app:b2_cp_borrowing}. At IA1, let
\[
  p_1
  =
  \Pr\left(\theta>0\mid \mathcal D_1\right),
  \qquad
  \theta
  =
  \log\left\{
  \frac{\lambda_C^{(III)}}{\lambda_T^{(III)}}
  \right\}.
\]
If \(p_1>1-\gamma_1\), the trial is declared positive at IA1. Otherwise, the
predictive probability for candidate IA2 timings is evaluated using a fast
normal approximation on the posterior-probability scale.

Let
\[
  z_1=\Phi^{-1}(p_1),
\]
where \(\Phi(\cdot)\) is the standard normal distribution function. To account
for the additional information induced by borrowing, we define an effective
IA1 information fraction. Let
\[
  V_{\mathrm{ref},1}
  =
  \mathrm{Var}_{\mathrm{ref}}
  \left(
  \theta\mid \mathcal D_1
  \right)
\]
denote the posterior variance of \(\theta\) under a no-borrowing reference
analysis with weak Gamma priors, and let
\[
  V_{\mathrm{CP},1}
  =
  \mathrm{Var}_{\mathrm{CP}}
  \left(
  \theta\mid \mathcal D_1
  \right)
\]
denote the posterior variance under \(B^2\)-CP. The effective IA1 information
fraction is
\[
  I_{1,\mathrm{eff}}
  =
  q_1
  \frac{V_{\mathrm{ref},1}}{V_{\mathrm{CP},1}},
\]
and the borrowing-induced information increment is
\[
  I_{\mathrm{bor}}
  =
  \max\left(I_{1,\mathrm{eff}}-q_1,\,0\right).
\]
For each candidate \(r\in\mathcal R\), the effective information fraction is
defined as
\[
  I_{r,\mathrm{eff}}
  =
  r+I_{\mathrm{bor}},
\]
with information ratio
\[
  \rho_r
  =
  \frac{I_{1,\mathrm{eff}}}{I_{r,\mathrm{eff}}}.
\]
In implementation, \(\rho_r\) is truncated to lie within \((0,1)\) for
numerical stability.

Let
\[
  z_r^*
  =
  \Phi^{-1}\{1-\gamma_2(r)\},
\]
where \(\gamma_2(r)\) is the one-sided O'Brien--Fleming p-value boundary for
the candidate IA2 analysis under schedule \((q_1,r,1)\). The approximate
predictive probability of success at candidate \(r\) is
\[
  \widetilde{\mathrm{BPP}}_{\mathrm{CP}}(r)
  =
  \Phi
  \left[
  \frac{
    z_1-z_r^*\sqrt{\rho_r}
  }{
    \sqrt{1-\rho_r}
  }
  \right].
\]
This approximation treats the posterior evidence statistic on the normal
scale as evolving according to the planned information fraction, with the
current information augmented by the effective information contributed by
borrowing.

The recommended IA2 timing is the earliest candidate satisfying the
prespecified predictive-probability target:
\[
  r_{\mathrm{CP}}^*
  =
  \min\left\{
  r\in\mathcal R:
  \widetilde{\mathrm{BPP}}_{\mathrm{CP}}(r)
  \geq
  \eta_{\mathrm{BPP}}
  \right\}.
\]
If no candidate satisfies this criterion, the recommendation is to proceed to
the final analysis.

For reporting the expected calendar timing, posterior-predictive future
trajectories are generated from the IA1 \(B^2\)-CP posterior using the same
Weibull simulation mechanism as in Appendix~\ref{app:b2_ep_predictive_probability}.
These simulations are used to estimate the mean calendar time
\[
  \bar \tau_r
  =
  \frac{1}{M}
  \sum_{m=1}^M
  \tau_r^{(m)}
\]
for each candidate \(r\), but the \(B^2\)-CP predictive probability itself is
computed from the fast approximation above rather than from repeated MCMC
posterior refitting at future candidate analyses.

\section{Predictive probability for adaptive IA2 timing under \(B^2\)-EP}
\label{app:b2_ep_predictive_probability}

Let \(q_1\) denote the fixed IA1 information fraction and let
\[
  \mathcal R=\{r_1,\ldots,r_L\},
  \qquad
  q_1<r_1<\cdots<r_L<1,
\]
denote the prespecified candidate IA2 information fractions. For each
candidate \(r\in\mathcal R\), the corresponding target number of phase~III
events is
\[
  E(r)=\mathrm{round}\{rE_{\mathrm{FA}}\},
\]
where \(E_{\mathrm{FA}}\) is the target number of events at the final analysis.
Let \(\gamma_2(r)\) denote the one-sided O'Brien--Fleming p-value boundary for
the candidate schedule \((q_1,r,1)\), evaluated at the IA2 analysis.

At IA1, the \(B^2\)-EP posterior for
\((\lambda_T^{(III)},\lambda_C^{(III)})\) is obtained as described in
Appendix~\ref{app:b2_ep_borrowing}. The IA1 posterior probability of treatment
benefit is
\[
  p_1
  =
  \Pr\left(\theta>0\mid\mathcal D_1\right),
  \qquad
  \theta
  =
  \log\left\{
  \frac{\lambda_C^{(III)}}{\lambda_T^{(III)}}
  \right\}.
\]
If \(p_1>1-\gamma_1\), where \(\gamma_1\) is the IA1 O'Brien--Fleming
boundary, the trial is declared positive at IA1. Otherwise, predictive
probabilities are computed for the candidate IA2 timings.

For \(m=1,\ldots,M\), draw
\[
  \left(
  \lambda_T^{(III,m)},
  \lambda_C^{(III,m)}
  \right)
  \sim
  \pi_{\mathrm{EP}}
  \left(
  \lambda_T^{(III)},\lambda_C^{(III)}
  \mid
  \mathcal D_1
  \right).
\]
Conditional on this draw, a complete future trajectory is generated from IA1
onward. For subjects who are event-free at IA1 with elapsed follow-up
\(u_i\), the residual event time is generated under the conditional Weibull
distribution. Equivalently, the total event time is sampled as
\[
  Y_i^{(m)}
  =
  \left\{
  u_i^k
  -
  \frac{\log U_i}{\lambda_g^{(III,m)}}
  \right\}^{1/k},
  \qquad
  U_i\sim \mathrm{Uniform}(0,1),
\]
for \(g\in\{T,C\}\). For newly accrued phase~III subjects, entry times are
generated according to the prespecified accrual model and event times are
sampled from the Weibull distribution with parameter
\(\lambda_g^{(III,m)}\). The continuing phase~II treatment-arm cohort is also
followed forward under the sampled treatment-arm parameter.

For each candidate \(r\in\mathcal R\), let \(\tau_r^{(m)}\) be the calendar
time at which the simulated phase~III data reach \(E(r)\) events. The
corresponding future data set,
\[
  \mathcal D_r^{(m)}
  =
  \left\{
  \mathcal D_{T,r}^{(II,m)},
  \mathcal D_{T,r}^{(III,m)},
  \mathcal D_{C,r}^{(III,m)}
  \right\},
\]
is obtained by administratively censoring all available records at
\(\tau_r^{(m)}\). The elastic discrepancy statistic
\(C_r^{(m)}\), borrowing weight
\(w_r^{(m)}(a_{\mathrm{el}}^{B^2},b_{\mathrm{el}}^{B^2})\), and posterior for
\((\lambda_T^{(III)},\lambda_C^{(III)})\) are then recomputed using
\(\mathcal D_r^{(m)}\). This gives the future posterior probability
\[
  p_r^{(m)}
  =
  \Pr\left(
  \theta>0
  \mid
  \mathcal D_r^{(m)}
  \right).
\]

The \(B^2\)-EP predictive probability of success at candidate IA2 timing \(r\)
is estimated by
\[
  \widehat{\mathrm{BPP}}_{\mathrm{EP}}(r)
  =
  \frac{1}{M_r}
  \sum_{m\in\mathcal M_r}
  I\left\{
  p_r^{(m)} > 1-\gamma_2(r)
  \right\},
\]
where \(\mathcal M_r\) is the set of valid posterior-predictive trajectories
for candidate \(r\), and \(M_r=|\mathcal M_r|\). In the conjugate EP
implementation, \(M_r=M\) unless a candidate data set is not evaluable.

Given a prespecified predictive-probability target \(\eta_{\mathrm{BPP}}\),
the recommended IA2 timing is the earliest candidate satisfying
\[
  \widehat{\mathrm{BPP}}_{\mathrm{EP}}(r)\geq \eta_{\mathrm{BPP}}.
\]
That is,
\[
  r_{\mathrm{EP}}^*
  =
  \min\left\{
  r\in\mathcal R:
  \widehat{\mathrm{BPP}}_{\mathrm{EP}}(r)
  \geq
  \eta_{\mathrm{BPP}}
  \right\}.
\]
If no candidate reaches the target, the recommendation is to proceed to the
final analysis. The expected calendar time corresponding to candidate \(r\) is
estimated by
\[
  \bar \tau_r
  =
  \frac{1}{M}
  \sum_{m=1}^M
  \tau_r^{(m)}.
\]

\section{Type I error at IA1, IA2 and FA across design and borrowing scenarios}
\label{sec:type1-error-ia1-ia2-fa}
\clearpage
\newgeometry{margin=0.1in, landscape}
\begin{landscape}

\begin{table}[H]
\centering
\tiny
\caption{\textbf{Cumulative type I error through interim 1 across design and borrowing scenarios by Median PFS}}
\label{tab:type1_by_timepoint_ia1_median_pfs}
\setlength{\tabcolsep}{1.2pt}
\renewcommand{\arraystretch}{1.6}

\resizebox{1.05\textwidth}{!}{%
{\setlength{\tabcolsep}{2.8pt}%
\begin{tabular}{@{}c c
  *{4}{c}@{\hspace{8pt}}
  *{4}{c}@{\hspace{8pt}}
  *{4}{c}@{\hspace{8pt}}
  *{4}{c}@{\hspace{8pt}}
  *{4}{c}@{}}
\toprule
\multirow{3}{*}[-0.3em]{\textbf{\begin{tabular}{@{}c@{}}Median PFS\\(Control)\end{tabular}}} &
\multirow{3}{*}[-0.3em]{\textbf{Method}} &
\multicolumn{20}{c}{\textbf{Discrepancy $\delta$}} \\
\cmidrule(lr){3-22}
& & \multicolumn{4}{c}{\textbf{$\mathbf{-0.4}$}} & \multicolumn{4}{c}{\textbf{$\mathbf{-0.2}$}} & \multicolumn{4}{c}{\textbf{$\mathbf{0}$}} & \multicolumn{4}{c}{\textbf{$\mathbf{0.2}$}} & \multicolumn{4}{c}{\textbf{$\mathbf{0.4}$}} \\
\cmidrule(l{0pt}r{5.5pt}){3-6}
\cmidrule(l{0pt}r{5.5pt}){7-10}
\cmidrule(l{0pt}r{5.5pt}){11-14}
\cmidrule(l{0pt}r{5.5pt}){15-18}
\cmidrule(l{0pt}r{0pt}){19-22}
& & \textbf{20} & \textbf{60} & \textbf{120} & \textbf{200} &
    \textbf{20} & \textbf{60} & \textbf{120} & \textbf{200} &
    \textbf{20} & \textbf{60} & \textbf{120} & \textbf{200} &
    \textbf{20} & \textbf{60} & \textbf{120} & \textbf{200} &
    \textbf{20} & \textbf{60} & \textbf{120} & \textbf{200} \\
\midrule

\multirow{6}{*}{4 months}
& GSD          & 0.000 & 0.000 & 0.000 & 0.000 & 0.000 & 0.000 & 0.000 & 0.000 & 0.000 & 0.000 & 0.000 & 0.000 & 0.000 & 0.000 & 0.000 & 0.000 & 0.000 & 0.000 & 0.000 & 0.000 \\
\cmidrule(l{1pt}r{0pt}){2-22}
& Conj.        & 0.001 & 0.012 & 0.096 & 0.378 & 0.000 & 0.002 & 0.007 & 0.022 & 0.000 & 0.000 & 0.000 & 0.000 & 0.000 & 0.000 & 0.000 & 0.000 & 0.000 & 0.000 & 0.000 & 0.000 \\
\cmidrule(l{1pt}r{0pt}){2-22}
& CP        & 0.000 & 0.002 & 0.002 & 0.003 & 0.000 & 0.001 & 0.002 & 0.003 & 0.000 & 0.000 & 0.000 & 0.000 & 0.000 & 0.000 & 0.000 & 0.000 & 0.000 & 0.000 & 0.000 & 0.000 \\
\cmidrule(l{1pt}r{0pt}){2-22}
& EP           & 0.000 & 0.000 & 0.000 & 0.001 & 0.000 & 0.000 & 0.000 & 0.001 & 0.000 & 0.000 & 0.000 & 0.000 & 0.000 & 0.000 & 0.000 & 0.000 & 0.000 & 0.000 & 0.000 & 0.000 \\
\cmidrule(l{1pt}r{0pt}){2-22}
& $B^2$ CP  & 0.001 & 0.001 & 0.000 & 0.001 & 0.000 & 0.001 & 0.000 & 0.001 & 0.000 & 0.000 & 0.000 & 0.000 & 0.000 & 0.000 & 0.000 & 0.000 & 0.000 & 0.000 & 0.000 & 0.000 \\
\cmidrule(l{1pt}r{0pt}){2-22}
& $B^2$ EP     & 0.001 & 0.002 & 0.001 & 0.001 & 0.000 & 0.001 & 0.001 & 0.001 & 0.000 & 0.000 & 0.000 & 0.000 & 0.000 & 0.000 & 0.000 & 0.000 & 0.000 & 0.000 & 0.000 & 0.000 \\
\midrule

\multirow{6}{*}{7 months}
& GSD          & 0.000 & 0.000 & 0.000 & 0.000 & 0.000 & 0.000 & 0.000 & 0.000 & 0.000 & 0.000 & 0.000 & 0.000 & 0.000 & 0.000 & 0.000 & 0.000 & 0.000 & 0.000 & 0.000 & 0.000 \\
\cmidrule(l{1pt}r{0pt}){2-22}
& Conj.        & 0.001 & 0.007 & 0.057 & 0.237 & 0.000 & 0.002 & 0.004 & 0.014 & 0.000 & 0.000 & 0.000 & 0.000 & 0.000 & 0.000 & 0.000 & 0.000 & 0.000 & 0.000 & 0.000 & 0.000 \\
\cmidrule(l{1pt}r{0pt}){2-22}
& CP        & 0.000 & 0.000 & 0.002 & 0.003 & 0.000 & 0.000 & 0.001 & 0.001 & 0.000 & 0.000 & 0.000 & 0.000 & 0.000 & 0.000 & 0.000 & 0.000 & 0.000 & 0.000 & 0.000 & 0.000 \\
\cmidrule(l{1pt}r{0pt}){2-22}
& EP           & 0.000 & 0.000 & 0.001 & 0.000 & 0.000 & 0.000 & 0.001 & 0.001 & 0.000 & 0.000 & 0.000 & 0.000 & 0.000 & 0.000 & 0.000 & 0.000 & 0.000 & 0.000 & 0.000 & 0.000 \\
\cmidrule(l{1pt}r{0pt}){2-22}
& $B^2$ CP  & 0.000 & 0.000 & 0.000 & 0.000 & 0.000 & 0.000 & 0.000 & 0.000 & 0.000 & 0.000 & 0.000 & 0.000 & 0.000 & 0.000 & 0.000 & 0.000 & 0.000 & 0.000 & 0.000 & 0.000 \\
\cmidrule(l{1pt}r{0pt}){2-22}
& $B^2$ EP     & 0.000 & 0.001 & 0.001 & 0.000 & 0.000 & 0.000 & 0.001 & 0.001 & 0.000 & 0.000 & 0.000 & 0.000 & 0.000 & 0.000 & 0.000 & 0.000 & 0.000 & 0.000 & 0.000 & 0.000 \\
\midrule

\multirow{6}{*}{12 months}
& GSD          & 0.001 & 0.001 & 0.001 & 0.001 & 0.001 & 0.001 & 0.001 & 0.001 & 0.001 & 0.001 & 0.001 & 0.001 & 0.001 & 0.001 & 0.001 & 0.001 & 0.001 & 0.001 & 0.001 & 0.001 \\
\cmidrule(l{1pt}r{0pt}){2-22}
& Conj.        & 0.001 & 0.005 & 0.032 & 0.133 & 0.000 & 0.001 & 0.003 & 0.009 & 0.000 & 0.000 & 0.000 & 0.000 & 0.000 & 0.000 & 0.000 & 0.000 & 0.000 & 0.000 & 0.000 & 0.000 \\
\cmidrule(l{1pt}r{0pt}){2-22}
& CP        & 0.001 & 0.001 & 0.003 & 0.002 & 0.000 & 0.000 & 0.001 & 0.003 & 0.000 & 0.000 & 0.000 & 0.000 & 0.000 & 0.000 & 0.000 & 0.000 & 0.000 & 0.000 & 0.000 & 0.000 \\
\cmidrule(l{1pt}r{0pt}){2-22}
& EP           & 0.000 & 0.000 & 0.001 & 0.001 & 0.000 & 0.000 & 0.001 & 0.001 & 0.000 & 0.000 & 0.000 & 0.000 & 0.000 & 0.000 & 0.000 & 0.000 & 0.000 & 0.000 & 0.000 & 0.000 \\
\cmidrule(l{1pt}r{0pt}){2-22}
& $B^2$ CP  & 0.000 & 0.001 & 0.001 & 0.001 & 0.000 & 0.000 & 0.000 & 0.000 & 0.000 & 0.000 & 0.000 & 0.000 & 0.000 & 0.000 & 0.000 & 0.000 & 0.000 & 0.000 & 0.000 & 0.000 \\
\cmidrule(l{1pt}r{0pt}){2-22}
& $B^2$ EP     & 0.001 & 0.001 & 0.001 & 0.002 & 0.000 & 0.001 & 0.001 & 0.001 & 0.000 & 0.000 & 0.000 & 0.000 & 0.000 & 0.000 & 0.000 & 0.000 & 0.000 & 0.000 & 0.000 & 0.000 \\
\bottomrule
\end{tabular}%
}%
}

\begin{minipage}{\textwidth}
\footnotesize
GSD: group sequential design; Conj.: conjugate prior; CP: commensurate prior; EP: elastic prior; $B^2$ CP: $B^2$-FIC Commensurate prior; $B^2$ EP: $B^2$-FIC Elastic prior. Values are cumulative type I error estimated from 10000 null simulation replicates. Phase II sample sizes are 20, 60, 120, and 200 per group.
\end{minipage}
\end{table}

\restoregeometry
\end{landscape}

\clearpage
\newgeometry{margin=0.1in, landscape}
\begin{landscape}

\begin{table}[H]
\centering
\tiny
\caption{\textbf{Cumulative type I error through interim 2 across design and borrowing scenarios by Median PFS}}
\label{tab:type1_by_timepoint_ia2_median_pfs}
\setlength{\tabcolsep}{1.2pt}
\renewcommand{\arraystretch}{1.6}

\resizebox{1.05\textwidth}{!}{%
{\setlength{\tabcolsep}{2.8pt}%
\begin{tabular}{@{}c c
  *{4}{c}@{\hspace{8pt}}
  *{4}{c}@{\hspace{8pt}}
  *{4}{c}@{\hspace{8pt}}
  *{4}{c}@{\hspace{8pt}}
  *{4}{c}@{}}
\toprule
\multirow{3}{*}[-0.3em]{\textbf{\begin{tabular}{@{}c@{}}Median PFS\\(Control)\end{tabular}}} &
\multirow{3}{*}[-0.3em]{\textbf{Method}} &
\multicolumn{20}{c}{\textbf{Discrepancy $\delta$}} \\
\cmidrule(lr){3-22}
& & \multicolumn{4}{c}{\textbf{$\mathbf{-0.4}$}} & \multicolumn{4}{c}{\textbf{$\mathbf{-0.2}$}} & \multicolumn{4}{c}{\textbf{$\mathbf{0}$}} & \multicolumn{4}{c}{\textbf{$\mathbf{0.2}$}} & \multicolumn{4}{c}{\textbf{$\mathbf{0.4}$}} \\
\cmidrule(l{0pt}r{5.5pt}){3-6}
\cmidrule(l{0pt}r{5.5pt}){7-10}
\cmidrule(l{0pt}r{5.5pt}){11-14}
\cmidrule(l{0pt}r{5.5pt}){15-18}
\cmidrule(l{0pt}r{0pt}){19-22}
& & \textbf{20} & \textbf{60} & \textbf{120} & \textbf{200} &
    \textbf{20} & \textbf{60} & \textbf{120} & \textbf{200} &
    \textbf{20} & \textbf{60} & \textbf{120} & \textbf{200} &
    \textbf{20} & \textbf{60} & \textbf{120} & \textbf{200} &
    \textbf{20} & \textbf{60} & \textbf{120} & \textbf{200} \\
\midrule

\multirow{6}{*}{4 months}
& GSD          & 0.008 & 0.008 & 0.008 & 0.008 & 0.008 & 0.008 & 0.008 & 0.008 & 0.008 & 0.008 & 0.008 & 0.008 & 0.008 & 0.008 & 0.008 & 0.008 & 0.008 & 0.008 & 0.008 & 0.008 \\
\cmidrule(l{1pt}r{0pt}){2-22}
& Conj.        & 0.013 & 0.069 & 0.285 & 0.657 & 0.007 & 0.018 & 0.048 & 0.121 & 0.004 & 0.004 & 0.004 & 0.003 & 0.002 & 0.001 & 0.000 & 0.000 & 0.001 & 0.000 & 0.000 & 0.000 \\
\cmidrule(l{1pt}r{0pt}){2-22}
& CP        & 0.007 & 0.015 & 0.017 & 0.018 & 0.006 & 0.010 & 0.017 & 0.027 & 0.002 & 0.003 & 0.003 & 0.003 & 0.002 & 0.001 & 0.001 & 0.001 & 0.002 & 0.002 & 0.003 & 0.002 \\
\cmidrule(l{1pt}r{0pt}){2-22}
& EP           & 0.005 & 0.004 & 0.004 & 0.004 & 0.004 & 0.005 & 0.006 & 0.007 & 0.004 & 0.004 & 0.004 & 0.005 & 0.004 & 0.004 & 0.004 & 0.004 & 0.004 & 0.004 & 0.004 & 0.004 \\
\cmidrule(l{1pt}r{0pt}){2-22}
& $B^2$ CP  & 0.005 & 0.007 & 0.007 & 0.007 & 0.004 & 0.005 & 0.007 & 0.007 & 0.004 & 0.003 & 0.003 & 0.003 & 0.003 & 0.002 & 0.002 & 0.002 & 0.003 & 0.002 & 0.003 & 0.003 \\
\cmidrule(l{1pt}r{0pt}){2-22}
& $B^2$ EP     & 0.008 & 0.009 & 0.005 & 0.004 & 0.005 & 0.009 & 0.009 & 0.007 & 0.004 & 0.004 & 0.005 & 0.005 & 0.003 & 0.004 & 0.004 & 0.004 & 0.004 & 0.004 & 0.004 & 0.004 \\
\midrule

\multirow{6}{*}{7 months}
& GSD          & 0.007 & 0.007 & 0.007 & 0.007 & 0.007 & 0.007 & 0.007 & 0.007 & 0.007 & 0.007 & 0.007 & 0.007 & 0.007 & 0.007 & 0.007 & 0.007 & 0.007 & 0.007 & 0.007 & 0.007 \\
\cmidrule(l{1pt}r{0pt}){2-22}
& Conj.        & 0.011 & 0.051 & 0.205 & 0.528 & 0.007 & 0.013 & 0.039 & 0.095 & 0.004 & 0.004 & 0.004 & 0.003 & 0.002 & 0.001 & 0.000 & 0.000 & 0.001 & 0.000 & 0.000 & 0.000 \\
\cmidrule(l{1pt}r{0pt}){2-22}
& CP        & 0.007 & 0.013 & 0.016 & 0.017 & 0.006 & 0.010 & 0.015 & 0.025 & 0.005 & 0.003 & 0.004 & 0.003 & 0.003 & 0.001 & 0.001 & 0.001 & 0.002 & 0.002 & 0.003 & 0.003 \\
\cmidrule(l{1pt}r{0pt}){2-22}
& EP           & 0.007 & 0.006 & 0.007 & 0.006 & 0.006 & 0.006 & 0.008 & 0.009 & 0.005 & 0.005 & 0.005 & 0.005 & 0.004 & 0.005 & 0.005 & 0.005 & 0.004 & 0.005 & 0.005 & 0.005 \\
\cmidrule(l{1pt}r{0pt}){2-22}
& $B^2$ CP  & 0.005 & 0.008 & 0.008 & 0.008 & 0.005 & 0.006 & 0.007 & 0.008 & 0.004 & 0.004 & 0.003 & 0.003 & 0.003 & 0.003 & 0.002 & 0.002 & 0.003 & 0.002 & 0.003 & 0.003 \\
\cmidrule(l{1pt}r{0pt}){2-22}
& $B^2$ EP     & 0.008 & 0.009 & 0.008 & 0.006 & 0.006 & 0.007 & 0.009 & 0.009 & 0.005 & 0.005 & 0.005 & 0.005 & 0.004 & 0.005 & 0.005 & 0.005 & 0.004 & 0.005 & 0.005 & 0.005 \\
\midrule

\multirow{6}{*}{12 months}
& GSD          & 0.007 & 0.007 & 0.007 & 0.007 & 0.007 & 0.007 & 0.007 & 0.007 & 0.007 & 0.007 & 0.007 & 0.007 & 0.007 & 0.007 & 0.007 & 0.007 & 0.007 & 0.007 & 0.007 & 0.007 \\
\cmidrule(l{1pt}r{0pt}){2-22}
& Conj.        & 0.010 & 0.038 & 0.145 & 0.401 & 0.006 & 0.014 & 0.031 & 0.073 & 0.004 & 0.004 & 0.004 & 0.004 & 0.003 & 0.001 & 0.000 & 0.000 & 0.002 & 0.000 & 0.000 & 0.000 \\
\cmidrule(l{1pt}r{0pt}){2-22}
& CP        & 0.008 & 0.012 & 0.018 & 0.018 & 0.006 & 0.009 & 0.015 & 0.021 & 0.005 & 0.004 & 0.004 & 0.003 & 0.003 & 0.001 & 0.001 & 0.001 & 0.002 & 0.001 & 0.002 & 0.003 \\
\cmidrule(l{1pt}r{0pt}){2-22}
& EP           & 0.004 & 0.005 & 0.006 & 0.005 & 0.005 & 0.006 & 0.007 & 0.008 & 0.004 & 0.005 & 0.005 & 0.005 & 0.004 & 0.004 & 0.004 & 0.004 & 0.004 & 0.004 & 0.004 & 0.004 \\
\cmidrule(l{1pt}r{0pt}){2-22}
& $B^2$ CP  & 0.006 & 0.007 & 0.008 & 0.007 & 0.005 & 0.005 & 0.007 & 0.008 & 0.004 & 0.003 & 0.003 & 0.003 & 0.003 & 0.003 & 0.002 & 0.002 & 0.002 & 0.002 & 0.002 & 0.003 \\
\cmidrule(l{1pt}r{0pt}){2-22}
& $B^2$ EP     & 0.008 & 0.007 & 0.007 & 0.006 & 0.006 & 0.007 & 0.010 & 0.010 & 0.005 & 0.006 & 0.005 & 0.005 & 0.004 & 0.004 & 0.004 & 0.004 & 0.003 & 0.004 & 0.004 & 0.004 \\
\bottomrule
\end{tabular}%
}%
}

\begin{minipage}{\textwidth}
\footnotesize
GSD: group sequential design; Conj.: conjugate prior; CP: commensurate prior; EP: elastic prior; $B^2$ CP: $B^2$-FIC Commensurate prior; $B^2$ EP: $B^2$-FIC Elastic prior. Values are cumulative type I error estimated from 10000 null simulation replicates. Phase II sample sizes are 20, 60, 120, and 200 per group.
\end{minipage}
\end{table}

\restoregeometry
\end{landscape}

\clearpage
\newgeometry{margin=0.1in, landscape}
\begin{landscape}

\begin{table}[H]
\centering
\tiny
\caption{\textbf{Final-analysis cumulative type I error across design and borrowing scenarios by Median PFS}}
\label{tab:type1_by_timepoint_fa_median_pfs}
\setlength{\tabcolsep}{1.2pt}
\renewcommand{\arraystretch}{1.6}

\resizebox{1.05\textwidth}{!}{%
{\setlength{\tabcolsep}{2.8pt}%
\begin{tabular}{@{}c c
  *{4}{c}@{\hspace{8pt}}
  *{4}{c}@{\hspace{8pt}}
  *{4}{c}@{\hspace{8pt}}
  *{4}{c}@{\hspace{8pt}}
  *{4}{c}@{}}
\toprule
\multirow{3}{*}[-0.3em]{\textbf{\begin{tabular}{@{}c@{}}Median PFS\\(Control)\end{tabular}}} &
\multirow{3}{*}[-0.3em]{\textbf{Method}} &
\multicolumn{20}{c}{\textbf{Discrepancy $\delta$}} \\
\cmidrule(lr){3-22}
& & \multicolumn{4}{c}{\textbf{$\mathbf{-0.4}$}} & \multicolumn{4}{c}{\textbf{$\mathbf{-0.2}$}} & \multicolumn{4}{c}{\textbf{$\mathbf{0}$}} & \multicolumn{4}{c}{\textbf{$\mathbf{0.2}$}} & \multicolumn{4}{c}{\textbf{$\mathbf{0.4}$}} \\
\cmidrule(l{0pt}r{5.5pt}){3-6}
\cmidrule(l{0pt}r{5.5pt}){7-10}
\cmidrule(l{0pt}r{5.5pt}){11-14}
\cmidrule(l{0pt}r{5.5pt}){15-18}
\cmidrule(l{0pt}r{0pt}){19-22}
& & \textbf{20} & \textbf{60} & \textbf{120} & \textbf{200} &
    \textbf{20} & \textbf{60} & \textbf{120} & \textbf{200} &
    \textbf{20} & \textbf{60} & \textbf{120} & \textbf{200} &
    \textbf{20} & \textbf{60} & \textbf{120} & \textbf{200} &
    \textbf{20} & \textbf{60} & \textbf{120} & \textbf{200} \\
\midrule

\multirow{6}{*}{4 months}
& GSD          & 0.026 & 0.026 & 0.026 & 0.026 & 0.026 & 0.026 & 0.026 & 0.026 & 0.026 & 0.026 & 0.026 & 0.026 & 0.026 & 0.026 & 0.026 & 0.026 & 0.026 & 0.026 & 0.026 & 0.026 \\
\cmidrule(l{1pt}r{0pt}){2-22}
& Conj.        & 0.040 & 0.141 & 0.411 & 0.767 & 0.025 & 0.050 & 0.106 & 0.220 & 0.015 & 0.015 & 0.014 & 0.013 & 0.010 & 0.004 & 0.001 & 0.000 & 0.007 & 0.001 & 0.000 & 0.000 \\
\cmidrule(l{1pt}r{0pt}){2-22}
& CP        & 0.027 & 0.041 & 0.042 & 0.039 & 0.024 & 0.035 & 0.054 & 0.073 & 0.016 & 0.014 & 0.013 & 0.013 & 0.011 & 0.007 & 0.005 & 0.006 & 0.010 & 0.009 & 0.011 & 0.013 \\
\cmidrule(l{1pt}r{0pt}){2-22}
& EP           & 0.018 & 0.016 & 0.016 & 0.016 & 0.017 & 0.017 & 0.017 & 0.020 & 0.016 & 0.015 & 0.017 & 0.017 & 0.015 & 0.015 & 0.015 & 0.015 & 0.015 & 0.015 & 0.015 & 0.015 \\
\cmidrule(l{1pt}r{0pt}){2-22}
& $B^2$ CP  & 0.021 & 0.026 & 0.025 & 0.026 & 0.018 & 0.022 & 0.025 & 0.027 & 0.016 & 0.015 & 0.014 & 0.014 & 0.013 & 0.011 & 0.010 & 0.009 & 0.012 & 0.011 & 0.011 & 0.011 \\
\cmidrule(l{1pt}r{0pt}){2-22}
& $B^2$ EP     & 0.024 & 0.022 & 0.018 & 0.016 & 0.021 & 0.025 & 0.024 & 0.020 & 0.016 & 0.017 & 0.017 & 0.017 & 0.013 & 0.015 & 0.015 & 0.015 & 0.014 & 0.015 & 0.015 & 0.015 \\
\midrule

\multirow{6}{*}{7 months}
& GSD          & 0.027 & 0.027 & 0.027 & 0.027 & 0.027 & 0.027 & 0.027 & 0.027 & 0.027 & 0.027 & 0.027 & 0.027 & 0.027 & 0.027 & 0.027 & 0.027 & 0.027 & 0.027 & 0.027 & 0.027 \\
\cmidrule(l{1pt}r{0pt}){2-22}
& Conj.        & 0.036 & 0.113 & 0.341 & 0.685 & 0.025 & 0.047 & 0.091 & 0.193 & 0.016 & 0.016 & 0.015 & 0.014 & 0.010 & 0.004 & 0.001 & 0.000 & 0.006 & 0.001 & 0.000 & 0.000 \\
\cmidrule(l{1pt}r{0pt}){2-22}
& CP        & 0.027 & 0.040 & 0.042 & 0.040 & 0.021 & 0.034 & 0.050 & 0.069 & 0.017 & 0.013 & 0.014 & 0.013 & 0.012 & 0.006 & 0.005 & 0.005 & 0.010 & 0.009 & 0.011 & 0.013 \\
\cmidrule(l{1pt}r{0pt}){2-22}
& EP           & 0.021 & 0.017 & 0.019 & 0.017 & 0.019 & 0.019 & 0.023 & 0.021 & 0.016 & 0.017 & 0.018 & 0.018 & 0.015 & 0.016 & 0.015 & 0.016 & 0.015 & 0.016 & 0.016 & 0.016 \\
\cmidrule(l{1pt}r{0pt}){2-22}
& $B^2$ CP  & 0.021 & 0.025 & 0.026 & 0.024 & 0.019 & 0.022 & 0.025 & 0.028 & 0.016 & 0.015 & 0.015 & 0.013 & 0.014 & 0.011 & 0.010 & 0.009 & 0.012 & 0.011 & 0.011 & 0.011 \\
\cmidrule(l{1pt}r{0pt}){2-22}
& $B^2$ EP     & 0.024 & 0.022 & 0.020 & 0.017 & 0.022 & 0.025 & 0.025 & 0.021 & 0.016 & 0.017 & 0.018 & 0.018 & 0.014 & 0.014 & 0.016 & 0.016 & 0.014 & 0.016 & 0.016 & 0.016 \\
\midrule

\multirow{6}{*}{12 months}
& GSD          & 0.027 & 0.027 & 0.027 & 0.027 & 0.027 & 0.027 & 0.027 & 0.027 & 0.027 & 0.027 & 0.027 & 0.027 & 0.027 & 0.027 & 0.027 & 0.027 & 0.027 & 0.027 & 0.027 & 0.027 \\
\cmidrule(l{1pt}r{0pt}){2-22}
& Conj.        & 0.035 & 0.098 & 0.285 & 0.601 & 0.024 & 0.043 & 0.083 & 0.167 & 0.016 & 0.016 & 0.015 & 0.015 & 0.011 & 0.004 & 0.001 & 0.000 & 0.007 & 0.001 & 0.000 & 0.000 \\
\cmidrule(l{1pt}r{0pt}){2-22}
& CP        & 0.028 & 0.039 & 0.044 & 0.043 & 0.023 & 0.034 & 0.048 & 0.063 & 0.017 & 0.015 & 0.015 & 0.013 & 0.012 & 0.007 & 0.004 & 0.005 & 0.009 & 0.009 & 0.011 & 0.013 \\
\cmidrule(l{1pt}r{0pt}){2-22}
& EP           & 0.017 & 0.018 & 0.019 & 0.017 & 0.018 & 0.019 & 0.022 & 0.021 & 0.016 & 0.017 & 0.017 & 0.019 & 0.016 & 0.016 & 0.016 & 0.016 & 0.016 & 0.016 & 0.016 & 0.016 \\
\cmidrule(l{1pt}r{0pt}){2-22}
& $B^2$ CP  & 0.022 & 0.025 & 0.026 & 0.026 & 0.019 & 0.022 & 0.025 & 0.029 & 0.016 & 0.015 & 0.015 & 0.014 & 0.013 & 0.011 & 0.010 & 0.009 & 0.013 & 0.011 & 0.011 & 0.012 \\
\cmidrule(l{1pt}r{0pt}){2-22}
& $B^2$ EP     & 0.026 & 0.022 & 0.021 & 0.018 & 0.022 & 0.024 & 0.025 & 0.024 & 0.016 & 0.018 & 0.018 & 0.019 & 0.015 & 0.015 & 0.016 & 0.016 & 0.014 & 0.016 & 0.016 & 0.016 \\
\bottomrule
\end{tabular}%
}%
}

\begin{minipage}{\textwidth}
\footnotesize
GSD: group sequential design; Conj.: conjugate prior; CP: commensurate prior; EP: elastic prior; $B^2$ CP: $B^2$-FIC Commensurate prior; $B^2$ EP: $B^2$-FIC Elastic prior. Values are cumulative type I error estimated from 10000 null simulation replicates. Phase II sample sizes are 20, 60, 120, and 200 per group.
\end{minipage}
\end{table}

\restoregeometry
\end{landscape}

\clearpage
\newgeometry{margin=0.1in, landscape}
\begin{landscape}

\begin{table}[H]
\centering
\tiny
\caption{\textbf{Cumulative type I error through interim 1 across design and borrowing scenarios by IA schedule}}
\label{tab:type1_by_timepoint_ia1_ia_schedule}
\setlength{\tabcolsep}{1.2pt}
\renewcommand{\arraystretch}{1.6}

\resizebox{1.05\textwidth}{!}{%
{\setlength{\tabcolsep}{2.8pt}%
\begin{tabular}{@{}c c
  *{4}{c}@{\hspace{8pt}}
  *{4}{c}@{\hspace{8pt}}
  *{4}{c}@{\hspace{8pt}}
  *{4}{c}@{\hspace{8pt}}
  *{4}{c}@{}}
\toprule
\multirow{3}{*}[-0.3em]{\textbf{IA schedule}} &
\multirow{3}{*}[-0.3em]{\textbf{Method}} &
\multicolumn{20}{c}{\textbf{Discrepancy $\delta$}} \\
\cmidrule(lr){3-22}
& & \multicolumn{4}{c}{\textbf{$\mathbf{-0.4}$}} & \multicolumn{4}{c}{\textbf{$\mathbf{-0.2}$}} & \multicolumn{4}{c}{\textbf{$\mathbf{0}$}} & \multicolumn{4}{c}{\textbf{$\mathbf{0.2}$}} & \multicolumn{4}{c}{\textbf{$\mathbf{0.4}$}} \\
\cmidrule(l{0pt}r{5.5pt}){3-6}
\cmidrule(l{0pt}r{5.5pt}){7-10}
\cmidrule(l{0pt}r{5.5pt}){11-14}
\cmidrule(l{0pt}r{5.5pt}){15-18}
\cmidrule(l{0pt}r{0pt}){19-22}
& & \textbf{20} & \textbf{60} & \textbf{120} & \textbf{200} &
    \textbf{20} & \textbf{60} & \textbf{120} & \textbf{200} &
    \textbf{20} & \textbf{60} & \textbf{120} & \textbf{200} &
    \textbf{20} & \textbf{60} & \textbf{120} & \textbf{200} &
    \textbf{20} & \textbf{60} & \textbf{120} & \textbf{200} \\
\midrule

\multirow{6}{*}{(0.4, 0.6, 1.0)}
& GSD          & 0.000 & 0.000 & 0.000 & 0.000 & 0.000 & 0.000 & 0.000 & 0.000 & 0.000 & 0.000 & 0.000 & 0.000 & 0.000 & 0.000 & 0.000 & 0.000 & 0.000 & 0.000 & 0.000 & 0.000 \\
\cmidrule(l{1pt}r{0pt}){2-22}
& Conj.        & 0.001 & 0.009 & 0.058 & 0.243 & 0.001 & 0.002 & 0.005 & 0.016 & 0.000 & 0.000 & 0.000 & 0.000 & 0.000 & 0.000 & 0.000 & 0.000 & 0.000 & 0.000 & 0.000 & 0.000 \\
\cmidrule(l{1pt}r{0pt}){2-22}
& CP        & 0.001 & 0.002 & 0.002 & 0.002 & 0.000 & 0.001 & 0.001 & 0.002 & 0.000 & 0.000 & 0.000 & 0.000 & 0.000 & 0.000 & 0.000 & 0.000 & 0.000 & 0.000 & 0.000 & 0.000 \\
\cmidrule(l{1pt}r{0pt}){2-22}
& EP           & 0.000 & 0.001 & 0.001 & 0.002 & 0.000 & 0.001 & 0.001 & 0.002 & 0.000 & 0.000 & 0.000 & 0.000 & 0.000 & 0.000 & 0.000 & 0.000 & 0.000 & 0.000 & 0.000 & 0.000 \\
\cmidrule(l{1pt}r{0pt}){2-22}
& $B^2$ CP  & 0.000 & 0.000 & 0.000 & 0.001 & 0.000 & 0.000 & 0.000 & 0.000 & 0.000 & 0.000 & 0.000 & 0.000 & 0.000 & 0.000 & 0.000 & 0.000 & 0.000 & 0.000 & 0.000 & 0.000 \\
\cmidrule(l{1pt}r{0pt}){2-22}
& $B^2$ EP     & 0.000 & 0.001 & 0.001 & 0.001 & 0.000 & 0.001 & 0.001 & 0.001 & 0.000 & 0.000 & 0.000 & 0.000 & 0.000 & 0.000 & 0.000 & 0.000 & 0.000 & 0.000 & 0.000 & 0.000 \\
\midrule

\multirow{6}{*}{(0.4, 0.7, 1.0)}
& GSD          & 0.000 & 0.000 & 0.000 & 0.000 & 0.000 & 0.000 & 0.000 & 0.000 & 0.000 & 0.000 & 0.000 & 0.000 & 0.000 & 0.000 & 0.000 & 0.000 & 0.000 & 0.000 & 0.000 & 0.000 \\
\cmidrule(l{1pt}r{0pt}){2-22}
& Conj.        & 0.001 & 0.007 & 0.057 & 0.237 & 0.000 & 0.002 & 0.004 & 0.014 & 0.000 & 0.000 & 0.000 & 0.000 & 0.000 & 0.000 & 0.000 & 0.000 & 0.000 & 0.000 & 0.000 & 0.000 \\
\cmidrule(l{1pt}r{0pt}){2-22}
& CP        & 0.000 & 0.000 & 0.002 & 0.003 & 0.000 & 0.000 & 0.001 & 0.001 & 0.000 & 0.000 & 0.000 & 0.000 & 0.000 & 0.000 & 0.000 & 0.000 & 0.000 & 0.000 & 0.000 & 0.000 \\
\cmidrule(l{1pt}r{0pt}){2-22}
& EP           & 0.000 & 0.000 & 0.001 & 0.000 & 0.000 & 0.000 & 0.001 & 0.001 & 0.000 & 0.000 & 0.000 & 0.000 & 0.000 & 0.000 & 0.000 & 0.000 & 0.000 & 0.000 & 0.000 & 0.000 \\
\cmidrule(l{1pt}r{0pt}){2-22}
& $B^2$ CP  & 0.000 & 0.000 & 0.000 & 0.000 & 0.000 & 0.000 & 0.000 & 0.000 & 0.000 & 0.000 & 0.000 & 0.000 & 0.000 & 0.000 & 0.000 & 0.000 & 0.000 & 0.000 & 0.000 & 0.000 \\
\cmidrule(l{1pt}r{0pt}){2-22}
& $B^2$ EP     & 0.000 & 0.001 & 0.001 & 0.000 & 0.000 & 0.000 & 0.001 & 0.001 & 0.000 & 0.000 & 0.000 & 0.000 & 0.000 & 0.000 & 0.000 & 0.000 & 0.000 & 0.000 & 0.000 & 0.000 \\
\midrule

\multirow{6}{*}{(0.4, 0.8, 1.0)}
& GSD          & 0.000 & 0.000 & 0.000 & 0.000 & 0.000 & 0.000 & 0.000 & 0.000 & 0.000 & 0.000 & 0.000 & 0.000 & 0.000 & 0.000 & 0.000 & 0.000 & 0.000 & 0.000 & 0.000 & 0.000 \\
\cmidrule(l{1pt}r{0pt}){2-22}
& Conj.        & 0.001 & 0.007 & 0.059 & 0.248 & 0.000 & 0.002 & 0.004 & 0.015 & 0.000 & 0.000 & 0.000 & 0.000 & 0.000 & 0.000 & 0.000 & 0.000 & 0.000 & 0.000 & 0.000 & 0.000 \\
\cmidrule(l{1pt}r{0pt}){2-22}
& CP        & 0.000 & 0.001 & 0.001 & 0.002 & 0.000 & 0.000 & 0.001 & 0.002 & 0.000 & 0.000 & 0.000 & 0.000 & 0.000 & 0.000 & 0.000 & 0.000 & 0.000 & 0.000 & 0.000 & 0.000 \\
\cmidrule(l{1pt}r{0pt}){2-22}
& EP           & 0.000 & 0.001 & 0.001 & 0.000 & 0.000 & 0.001 & 0.000 & 0.001 & 0.000 & 0.000 & 0.000 & 0.000 & 0.000 & 0.000 & 0.000 & 0.000 & 0.000 & 0.000 & 0.000 & 0.000 \\
\cmidrule(l{1pt}r{0pt}){2-22}
& $B^2$ CP  & 0.000 & 0.000 & 0.001 & 0.000 & 0.000 & 0.000 & 0.000 & 0.000 & 0.000 & 0.000 & 0.000 & 0.000 & 0.000 & 0.000 & 0.000 & 0.000 & 0.000 & 0.000 & 0.000 & 0.000 \\
\cmidrule(l{1pt}r{0pt}){2-22}
& $B^2$ EP     & 0.000 & 0.001 & 0.001 & 0.001 & 0.000 & 0.001 & 0.001 & 0.001 & 0.000 & 0.000 & 0.000 & 0.000 & 0.000 & 0.000 & 0.000 & 0.000 & 0.000 & 0.000 & 0.000 & 0.000 \\
\bottomrule
\end{tabular}%
}%
}

\begin{minipage}{\textwidth}
\footnotesize
GSD: group sequential design; Conj.: conjugate prior; CP: commensurate prior; EP: elastic prior; $B^2$ CP: $B^2$-FIC Commensurate prior; $B^2$ EP: $B^2$-FIC Elastic prior. Values are cumulative type I error estimated from 10000 null simulation replicates. Phase II sample sizes are 20, 60, 120, and 200 per group. Median PFS (control) is fixed at 7 months.
\end{minipage}
\end{table}

\restoregeometry
\end{landscape}

\clearpage
\newgeometry{margin=0.1in, landscape}
\begin{landscape}

\begin{table}[H]
\centering
\tiny
\caption{\textbf{Cumulative type I error through interim 2 across design and borrowing scenarios by IA schedule}}
\label{tab:type1_by_timepoint_ia2_ia_schedule}
\setlength{\tabcolsep}{1.2pt}
\renewcommand{\arraystretch}{1.6}

\resizebox{1.05\textwidth}{!}{%
{\setlength{\tabcolsep}{2.8pt}%
\begin{tabular}{@{}c c
  *{4}{c}@{\hspace{8pt}}
  *{4}{c}@{\hspace{8pt}}
  *{4}{c}@{\hspace{8pt}}
  *{4}{c}@{\hspace{8pt}}
  *{4}{c}@{}}
\toprule
\multirow{3}{*}[-0.3em]{\textbf{IA schedule}} &
\multirow{3}{*}[-0.3em]{\textbf{Method}} &
\multicolumn{20}{c}{\textbf{Discrepancy $\delta$}} \\
\cmidrule(lr){3-22}
& & \multicolumn{4}{c}{\textbf{$\mathbf{-0.4}$}} & \multicolumn{4}{c}{\textbf{$\mathbf{-0.2}$}} & \multicolumn{4}{c}{\textbf{$\mathbf{0}$}} & \multicolumn{4}{c}{\textbf{$\mathbf{0.2}$}} & \multicolumn{4}{c}{\textbf{$\mathbf{0.4}$}} \\
\cmidrule(l{0pt}r{5.5pt}){3-6}
\cmidrule(l{0pt}r{5.5pt}){7-10}
\cmidrule(l{0pt}r{5.5pt}){11-14}
\cmidrule(l{0pt}r{5.5pt}){15-18}
\cmidrule(l{0pt}r{0pt}){19-22}
& & \textbf{20} & \textbf{60} & \textbf{120} & \textbf{200} &
    \textbf{20} & \textbf{60} & \textbf{120} & \textbf{200} &
    \textbf{20} & \textbf{60} & \textbf{120} & \textbf{200} &
    \textbf{20} & \textbf{60} & \textbf{120} & \textbf{200} &
    \textbf{20} & \textbf{60} & \textbf{120} & \textbf{200} \\
\midrule

\multirow{6}{*}{(0.4, 0.6, 1.0)}
& GSD          & 0.003 & 0.003 & 0.003 & 0.003 & 0.003 & 0.003 & 0.003 & 0.003 & 0.003 & 0.003 & 0.003 & 0.003 & 0.003 & 0.003 & 0.003 & 0.003 & 0.003 & 0.003 & 0.003 & 0.003 \\
\cmidrule(l{1pt}r{0pt}){2-22}
& Conj.        & 0.006 & 0.037 & 0.159 & 0.472 & 0.004 & 0.010 & 0.025 & 0.063 & 0.002 & 0.002 & 0.002 & 0.002 & 0.001 & 0.001 & 0.000 & 0.000 & 0.001 & 0.000 & 0.000 & 0.000 \\
\cmidrule(l{1pt}r{0pt}){2-22}
& CP        & 0.006 & 0.009 & 0.011 & 0.012 & 0.003 & 0.006 & 0.011 & 0.015 & 0.002 & 0.002 & 0.002 & 0.002 & 0.002 & 0.001 & 0.000 & 0.001 & 0.001 & 0.001 & 0.002 & 0.002 \\
\cmidrule(l{1pt}r{0pt}){2-22}
& EP           & 0.002 & 0.005 & 0.003 & 0.004 & 0.002 & 0.004 & 0.004 & 0.007 & 0.002 & 0.002 & 0.003 & 0.003 & 0.002 & 0.002 & 0.002 & 0.002 & 0.002 & 0.002 & 0.002 & 0.002 \\
\cmidrule(l{1pt}r{0pt}){2-22}
& $B^2$ CP  & 0.003 & 0.004 & 0.005 & 0.005 & 0.002 & 0.003 & 0.004 & 0.005 & 0.002 & 0.002 & 0.002 & 0.002 & 0.001 & 0.001 & 0.001 & 0.001 & 0.002 & 0.002 & 0.001 & 0.001 \\
\cmidrule(l{1pt}r{0pt}){2-22}
& $B^2$ EP     & 0.005 & 0.006 & 0.004 & 0.003 & 0.003 & 0.004 & 0.005 & 0.005 & 0.002 & 0.002 & 0.003 & 0.003 & 0.002 & 0.002 & 0.002 & 0.002 & 0.002 & 0.002 & 0.002 & 0.002 \\
\midrule

\multirow{6}{*}{(0.4, 0.7, 1.0)}
& GSD          & 0.007 & 0.007 & 0.007 & 0.007 & 0.007 & 0.007 & 0.007 & 0.007 & 0.007 & 0.007 & 0.007 & 0.007 & 0.007 & 0.007 & 0.007 & 0.007 & 0.007 & 0.007 & 0.007 & 0.007 \\
\cmidrule(l{1pt}r{0pt}){2-22}
& Conj.        & 0.011 & 0.051 & 0.205 & 0.528 & 0.007 & 0.013 & 0.039 & 0.095 & 0.004 & 0.004 & 0.004 & 0.003 & 0.002 & 0.001 & 0.000 & 0.000 & 0.001 & 0.000 & 0.000 & 0.000 \\
\cmidrule(l{1pt}r{0pt}){2-22}
& CP        & 0.007 & 0.013 & 0.016 & 0.017 & 0.006 & 0.010 & 0.015 & 0.025 & 0.005 & 0.003 & 0.004 & 0.003 & 0.003 & 0.001 & 0.001 & 0.001 & 0.002 & 0.002 & 0.003 & 0.003 \\
\cmidrule(l{1pt}r{0pt}){2-22}
& EP           & 0.007 & 0.006 & 0.007 & 0.006 & 0.006 & 0.006 & 0.008 & 0.009 & 0.005 & 0.005 & 0.005 & 0.005 & 0.004 & 0.005 & 0.005 & 0.005 & 0.004 & 0.005 & 0.005 & 0.005 \\
\cmidrule(l{1pt}r{0pt}){2-22}
& $B^2$ CP  & 0.005 & 0.008 & 0.008 & 0.008 & 0.005 & 0.006 & 0.007 & 0.008 & 0.004 & 0.004 & 0.003 & 0.003 & 0.003 & 0.003 & 0.002 & 0.002 & 0.003 & 0.002 & 0.003 & 0.003 \\
\cmidrule(l{1pt}r{0pt}){2-22}
& $B^2$ EP     & 0.008 & 0.009 & 0.008 & 0.006 & 0.006 & 0.007 & 0.009 & 0.009 & 0.005 & 0.005 & 0.005 & 0.005 & 0.004 & 0.005 & 0.005 & 0.005 & 0.004 & 0.005 & 0.005 & 0.005 \\
\midrule

\multirow{6}{*}{(0.4, 0.8, 1.0)}
& GSD          & 0.014 & 0.014 & 0.014 & 0.014 & 0.014 & 0.014 & 0.014 & 0.014 & 0.014 & 0.014 & 0.014 & 0.014 & 0.014 & 0.014 & 0.014 & 0.014 & 0.014 & 0.014 & 0.014 & 0.014 \\
\cmidrule(l{1pt}r{0pt}){2-22}
& Conj.        & 0.018 & 0.068 & 0.248 & 0.585 & 0.012 & 0.024 & 0.055 & 0.124 & 0.007 & 0.007 & 0.006 & 0.006 & 0.004 & 0.001 & 0.001 & 0.000 & 0.003 & 0.000 & 0.000 & 0.000 \\
\cmidrule(l{1pt}r{0pt}){2-22}
& CP        & 0.014 & 0.020 & 0.023 & 0.021 & 0.010 & 0.016 & 0.025 & 0.035 & 0.007 & 0.006 & 0.006 & 0.005 & 0.005 & 0.002 & 0.002 & 0.002 & 0.004 & 0.004 & 0.004 & 0.006 \\
\cmidrule(l{1pt}r{0pt}){2-22}
& EP           & 0.010 & 0.010 & 0.009 & 0.009 & 0.010 & 0.010 & 0.012 & 0.011 & 0.008 & 0.009 & 0.009 & 0.009 & 0.007 & 0.008 & 0.008 & 0.008 & 0.008 & 0.008 & 0.008 & 0.008 \\
\cmidrule(l{1pt}r{0pt}){2-22}
& $B^2$ CP  & 0.010 & 0.012 & 0.012 & 0.013 & 0.009 & 0.010 & 0.012 & 0.014 & 0.008 & 0.007 & 0.006 & 0.005 & 0.006 & 0.004 & 0.004 & 0.004 & 0.005 & 0.004 & 0.005 & 0.005 \\
\cmidrule(l{1pt}r{0pt}){2-22}
& $B^2$ EP     & 0.010 & 0.011 & 0.010 & 0.009 & 0.010 & 0.011 & 0.013 & 0.013 & 0.008 & 0.009 & 0.009 & 0.009 & 0.007 & 0.007 & 0.008 & 0.008 & 0.008 & 0.008 & 0.008 & 0.008 \\
\bottomrule
\end{tabular}%
}%
}

\begin{minipage}{\textwidth}
\footnotesize
GSD: group sequential design; Conj.: conjugate prior; CP: commensurate prior; EP: elastic prior; $B^2$ CP: $B^2$-FIC Commensurate prior; $B^2$ EP: $B^2$-FIC Elastic prior. Values are cumulative type I error estimated from 10000 null simulation replicates. Phase II sample sizes are 20, 60, 120, and 200 per group. Median PFS (control) is fixed at 7 months.
\end{minipage}
\end{table}

\restoregeometry
\end{landscape}

\clearpage
\newgeometry{margin=0.1in, landscape}
\begin{landscape}

\begin{table}[H]
\centering
\tiny
\caption{\textbf{Final-analysis cumulative type I error across design and borrowing scenarios by IA schedule}}
\label{tab:type1_by_timepoint_fa_ia_schedule}
\setlength{\tabcolsep}{1.2pt}
\renewcommand{\arraystretch}{1.6}

\resizebox{1.05\textwidth}{!}{%
{\setlength{\tabcolsep}{2.8pt}%
\begin{tabular}{@{}c c
  *{4}{c}@{\hspace{8pt}}
  *{4}{c}@{\hspace{8pt}}
  *{4}{c}@{\hspace{8pt}}
  *{4}{c}@{\hspace{8pt}}
  *{4}{c}@{}}
\toprule
\multirow{3}{*}[-0.3em]{\textbf{IA schedule}} &
\multirow{3}{*}[-0.3em]{\textbf{Method}} &
\multicolumn{20}{c}{\textbf{Discrepancy $\delta$}} \\
\cmidrule(lr){3-22}
& & \multicolumn{4}{c}{\textbf{$\mathbf{-0.4}$}} & \multicolumn{4}{c}{\textbf{$\mathbf{-0.2}$}} & \multicolumn{4}{c}{\textbf{$\mathbf{0}$}} & \multicolumn{4}{c}{\textbf{$\mathbf{0.2}$}} & \multicolumn{4}{c}{\textbf{$\mathbf{0.4}$}} \\
\cmidrule(l{0pt}r{5.5pt}){3-6}
\cmidrule(l{0pt}r{5.5pt}){7-10}
\cmidrule(l{0pt}r{5.5pt}){11-14}
\cmidrule(l{0pt}r{5.5pt}){15-18}
\cmidrule(l{0pt}r{0pt}){19-22}
& & \textbf{20} & \textbf{60} & \textbf{120} & \textbf{200} &
    \textbf{20} & \textbf{60} & \textbf{120} & \textbf{200} &
    \textbf{20} & \textbf{60} & \textbf{120} & \textbf{200} &
    \textbf{20} & \textbf{60} & \textbf{120} & \textbf{200} &
    \textbf{20} & \textbf{60} & \textbf{120} & \textbf{200} \\
\midrule

\multirow{6}{*}{(0.4, 0.6, 1.0)}
& GSD          & 0.027 & 0.027 & 0.027 & 0.027 & 0.027 & 0.027 & 0.027 & 0.027 & 0.027 & 0.027 & 0.027 & 0.027 & 0.027 & 0.027 & 0.027 & 0.027 & 0.027 & 0.027 & 0.027 & 0.027 \\
\cmidrule(l{1pt}r{0pt}){2-22}
& Conj.        & 0.038 & 0.121 & 0.344 & 0.688 & 0.025 & 0.047 & 0.093 & 0.193 & 0.017 & 0.016 & 0.015 & 0.015 & 0.012 & 0.004 & 0.002 & 0.000 & 0.007 & 0.001 & 0.000 & 0.000 \\
\cmidrule(l{1pt}r{0pt}){2-22}
& CP        & 0.028 & 0.041 & 0.043 & 0.041 & 0.023 & 0.032 & 0.053 & 0.068 & 0.020 & 0.017 & 0.015 & 0.015 & 0.014 & 0.009 & 0.005 & 0.006 & 0.010 & 0.011 & 0.013 & 0.013 \\
\cmidrule(l{1pt}r{0pt}){2-22}
& EP           & 0.017 & 0.021 & 0.018 & 0.019 & 0.017 & 0.023 & 0.021 & 0.026 & 0.017 & 0.018 & 0.018 & 0.020 & 0.017 & 0.016 & 0.017 & 0.017 & 0.017 & 0.017 & 0.017 & 0.017 \\
\cmidrule(l{1pt}r{0pt}){2-22}
& $B^2$ CP  & 0.022 & 0.026 & 0.028 & 0.028 & 0.021 & 0.024 & 0.026 & 0.029 & 0.017 & 0.015 & 0.016 & 0.014 & 0.014 & 0.011 & 0.010 & 0.010 & 0.013 & 0.012 & 0.013 & 0.012 \\
\cmidrule(l{1pt}r{0pt}){2-22}
& $B^2$ EP     & 0.026 & 0.022 & 0.019 & 0.017 & 0.023 & 0.023 & 0.023 & 0.021 & 0.018 & 0.019 & 0.018 & 0.019 & 0.015 & 0.016 & 0.016 & 0.017 & 0.015 & 0.017 & 0.017 & 0.017 \\
\midrule

\multirow{6}{*}{(0.4, 0.7, 1.0)}
& GSD          & 0.027 & 0.027 & 0.027 & 0.027 & 0.027 & 0.027 & 0.027 & 0.027 & 0.027 & 0.027 & 0.027 & 0.027 & 0.027 & 0.027 & 0.027 & 0.027 & 0.027 & 0.027 & 0.027 & 0.027 \\
\cmidrule(l{1pt}r{0pt}){2-22}
& Conj.        & 0.036 & 0.113 & 0.341 & 0.685 & 0.025 & 0.047 & 0.091 & 0.193 & 0.016 & 0.016 & 0.015 & 0.014 & 0.010 & 0.004 & 0.001 & 0.000 & 0.006 & 0.001 & 0.000 & 0.000 \\
\cmidrule(l{1pt}r{0pt}){2-22}
& CP        & 0.027 & 0.040 & 0.042 & 0.040 & 0.021 & 0.034 & 0.050 & 0.069 & 0.017 & 0.013 & 0.014 & 0.013 & 0.012 & 0.006 & 0.005 & 0.005 & 0.010 & 0.009 & 0.011 & 0.013 \\
\cmidrule(l{1pt}r{0pt}){2-22}
& EP           & 0.021 & 0.017 & 0.019 & 0.017 & 0.019 & 0.019 & 0.023 & 0.021 & 0.016 & 0.017 & 0.018 & 0.018 & 0.015 & 0.016 & 0.015 & 0.016 & 0.015 & 0.016 & 0.016 & 0.016 \\
\cmidrule(l{1pt}r{0pt}){2-22}
& $B^2$ CP  & 0.021 & 0.025 & 0.026 & 0.024 & 0.019 & 0.022 & 0.025 & 0.028 & 0.016 & 0.015 & 0.015 & 0.013 & 0.014 & 0.011 & 0.010 & 0.009 & 0.012 & 0.011 & 0.011 & 0.011 \\
\cmidrule(l{1pt}r{0pt}){2-22}
& $B^2$ EP     & 0.024 & 0.022 & 0.020 & 0.017 & 0.022 & 0.025 & 0.025 & 0.021 & 0.016 & 0.017 & 0.018 & 0.018 & 0.014 & 0.014 & 0.016 & 0.016 & 0.014 & 0.016 & 0.016 & 0.016 \\
\midrule

\multirow{6}{*}{(0.4, 0.8, 1.0)}
& GSD          & 0.027 & 0.027 & 0.027 & 0.027 & 0.027 & 0.027 & 0.027 & 0.027 & 0.027 & 0.027 & 0.027 & 0.027 & 0.027 & 0.027 & 0.027 & 0.027 & 0.027 & 0.027 & 0.027 & 0.027 \\
\cmidrule(l{1pt}r{0pt}){2-22}
& Conj.        & 0.036 & 0.112 & 0.334 & 0.680 & 0.025 & 0.047 & 0.093 & 0.187 & 0.016 & 0.016 & 0.014 & 0.016 & 0.011 & 0.005 & 0.002 & 0.000 & 0.007 & 0.001 & 0.000 & 0.000 \\
\cmidrule(l{1pt}r{0pt}){2-22}
& CP        & 0.030 & 0.038 & 0.043 & 0.039 & 0.025 & 0.036 & 0.050 & 0.066 & 0.019 & 0.017 & 0.014 & 0.014 & 0.013 & 0.008 & 0.006 & 0.007 & 0.011 & 0.011 & 0.012 & 0.014 \\
\cmidrule(l{1pt}r{0pt}){2-22}
& EP           & 0.022 & 0.020 & 0.019 & 0.018 & 0.021 & 0.022 & 0.022 & 0.022 & 0.018 & 0.020 & 0.019 & 0.019 & 0.016 & 0.018 & 0.018 & 0.018 & 0.017 & 0.018 & 0.018 & 0.018 \\
\cmidrule(l{1pt}r{0pt}){2-22}
& $B^2$ CP  & 0.023 & 0.026 & 0.027 & 0.027 & 0.021 & 0.024 & 0.026 & 0.029 & 0.018 & 0.016 & 0.014 & 0.016 & 0.014 & 0.012 & 0.011 & 0.011 & 0.013 & 0.012 & 0.012 & 0.012 \\
\cmidrule(l{1pt}r{0pt}){2-22}
& $B^2$ EP     & 0.022 & 0.022 & 0.020 & 0.019 & 0.021 & 0.024 & 0.025 & 0.023 & 0.018 & 0.019 & 0.019 & 0.020 & 0.016 & 0.017 & 0.018 & 0.018 & 0.017 & 0.018 & 0.018 & 0.018 \\
\bottomrule
\end{tabular}%
}%
}

\begin{minipage}{\textwidth}
\footnotesize
GSD: group sequential design; Conj.: conjugate prior; CP: commensurate prior; EP: elastic prior; $B^2$ CP: $B^2$-FIC Commensurate prior; $B^2$ EP: $B^2$-FIC Elastic prior. Values are cumulative type I error estimated from 10000 null simulation replicates. Phase II sample sizes are 20, 60, 120, and 200 per group. Median PFS (control) is fixed at 7 months.
\end{minipage}
\end{table}

\restoregeometry
\end{landscape}

\section{Power at IA1, IA2 and FA across design and borrowing scenarios}
\label{sec:power-ia1-ia2-fa}
\clearpage
\newgeometry{margin=0.1in}
\begin{landscape}
\begin{table}[p]
\centering
\caption{Interim-1 power across sensitivity scenarios}
\label{tab:ia1_power_sensitivity}
\scriptsize
\setlength{\tabcolsep}{1.0pt}
\renewcommand{\arraystretch}{1.15}

\resizebox{1.05\textwidth}{!}{%
{\setlength{\tabcolsep}{2.8pt}%
\begin{tabular}{@{}c c
  *{4}{c}@{\hspace{8pt}}
  *{4}{c}@{\hspace{8pt}}
  *{4}{c}@{\hspace{8pt}}
  *{4}{c}@{\hspace{8pt}}
  *{4}{c}@{}}
\toprule
\multirow{3}{*}[-0.3em]{\textbf{\begin{tabular}{@{}c@{}}Median PFS\\(Control)\\HR=0.8\end{tabular}}} &
\multirow{3}{*}[-0.3em]{\textbf{Method}} &
\multicolumn{20}{c}{\textbf{Discrepancy $\delta$}} \\
\cmidrule(lr){3-22}
& & \multicolumn{4}{c}{\textbf{$\mathbf{-0.4}$}} & \multicolumn{4}{c}{\textbf{$\mathbf{-0.2}$}} & \multicolumn{4}{c}{\textbf{$\mathbf{0}$}} & \multicolumn{4}{c}{\textbf{$\mathbf{0.2}$}} & \multicolumn{4}{c}{\textbf{$\mathbf{0.4}$}} \\
\cmidrule(l{0pt}r{5.5pt}){3-6}
\cmidrule(l{0pt}r{5.5pt}){7-10}
\cmidrule(l{0pt}r{5.5pt}){11-14}
\cmidrule(l{0pt}r{5.5pt}){15-18}
\cmidrule(l{0pt}r{0pt}){19-22}
& & \textbf{20} & \textbf{60} & \textbf{120} & \textbf{200} &
    \textbf{20} & \textbf{60} & \textbf{120} & \textbf{200} &
    \textbf{20} & \textbf{60} & \textbf{120} & \textbf{200} &
    \textbf{20} & \textbf{60} & \textbf{120} & \textbf{200} &
    \textbf{20} & \textbf{60} & \textbf{120} & \textbf{200} \\
\midrule

\multirow{4}{*}{4 months}
& GSD          & 0.101 & 0.101 & 0.101 & 0.101 & 0.101 & 0.101 & 0.101 & 0.101 & 0.101 & 0.101 & 0.101 & 0.101 & 0.101 & 0.101 & 0.101 & 0.101 & 0.101 & 0.101 & 0.101 & 0.101 \\
\cmidrule(l{1pt}r{0pt}){2-22}
& EP           & 0.234 & 0.205 & 0.198 & 0.198 & 0.231 & 0.212 & 0.229 & 0.229 & 0.210 & 0.214 & 0.241 & 0.292 & 0.200 & 0.196 & 0.198 & 0.200 & 0.190 & 0.197 & 0.197 & 0.197 \\
\cmidrule(l{1pt}r{0pt}){2-22}
& $B^2$ EP     & 0.308 & 0.260 & 0.236 & 0.198 & 0.274 & 0.320 & 0.322 & 0.229 & 0.228 & 0.285 & 0.301 & 0.292 & 0.184 & 0.190 & 0.195 & 0.200 & 0.176 & 0.190 & 0.190 & 0.197 \\
\cmidrule(l{1pt}r{0pt}){2-22}
& $B^2$ CP  & 0.242 & 0.255 & 0.279 & 0.260 & 0.236 & 0.278 & 0.280 & 0.293 & 0.214 & 0.231 & 0.236 & 0.251 & 0.185 & 0.166 & 0.144 & 0.136 & 0.145 & 0.123 & 0.127 & 0.123 \\
\midrule

\multirow{4}{*}{7 months}
& GSD          & 0.101 & 0.101 & 0.101 & 0.101 & 0.101 & 0.101 & 0.101 & 0.101 & 0.101 & 0.101 & 0.101 & 0.101 & 0.101 & 0.101 & 0.101 & 0.101 & 0.101 & 0.101 & 0.101 & 0.101 \\
\cmidrule(l{1pt}r{0pt}){2-22}
& EP           & 0.273 & 0.241 & 0.248 & 0.220 & 0.265 & 0.264 & 0.303 & 0.252 & 0.229 & 0.244 & 0.273 & 0.300 & 0.203 & 0.216 & 0.219 & 0.223 & 0.201 & 0.217 & 0.212 & 0.218 \\
\cmidrule(l{1pt}r{0pt}){2-22}
& $B^2$ EP     & 0.303 & 0.296 & 0.248 & 0.220 & 0.284 & 0.308 & 0.322 & 0.252 & 0.237 & 0.263 & 0.294 & 0.300 & 0.197 & 0.209 & 0.219 & 0.223 & 0.192 & 0.207 & 0.211 & 0.218 \\
\cmidrule(l{1pt}r{0pt}){2-22}
& $B^2$ CP  & 0.247 & 0.255 & 0.273 & 0.278 & 0.220 & 0.261 & 0.292 & 0.299 & 0.207 & 0.235 & 0.220 & 0.238 & 0.182 & 0.165 & 0.148 & 0.144 & 0.160 & 0.151 & 0.133 & 0.123 \\
\midrule

\multirow{4}{*}{12 months}
& GSD          & 0.104 & 0.104 & 0.104 & 0.104 & 0.104 & 0.104 & 0.104 & 0.104 & 0.104 & 0.104 & 0.104 & 0.104 & 0.104 & 0.104 & 0.104 & 0.104 & 0.104 & 0.104 & 0.104 & 0.104 \\
\cmidrule(l{1pt}r{0pt}){2-22}
& EP           & 0.227 & 0.240 & 0.238 & 0.217 & 0.226 & 0.252 & 0.289 & 0.263 & 0.220 & 0.236 & 0.278 & 0.290 & 0.213 & 0.215 & 0.207 & 0.212 & 0.213 & 0.208 & 0.210 & 0.212 \\
\cmidrule(l{1pt}r{0pt}){2-22}
& $B^2$ EP     & 0.304 & 0.276 & 0.258 & 0.219 & 0.271 & 0.300 & 0.332 & 0.288 & 0.231 & 0.256 & 0.292 & 0.310 & 0.196 & 0.200 & 0.200 & 0.214 & 0.186 & 0.195 & 0.207 & 0.212 \\
\cmidrule(l{1pt}r{0pt}){2-22}
& $B^2$ CP  & 0.236 & 0.258 & 0.277 & 0.276 & 0.229 & 0.267 & 0.273 & 0.285 & 0.212 & 0.238 & 0.241 & 0.252 & 0.185 & 0.164 & 0.154 & 0.136 & 0.181 & 0.137 & 0.139 & 0.138 \\
\bottomrule
\end{tabular}%
}%
}

\medskip

\resizebox{1.05\textwidth}{!}{%
{\setlength{\tabcolsep}{2.8pt}%
\begin{tabular}{@{}c c
  *{4}{c}@{\hspace{8pt}}
  *{4}{c}@{\hspace{8pt}}
  *{4}{c}@{\hspace{8pt}}
  *{4}{c}@{\hspace{8pt}}
  *{4}{c}@{}}
\toprule
\multirow{3}{*}[-0.3em]{\textbf{\begin{tabular}{@{}c@{}}Median PFS\\(Control)\\HR=0.6\end{tabular}}} &
\multirow{3}{*}[-0.3em]{\textbf{Method}} &
\multicolumn{20}{c}{\textbf{Discrepancy $\delta$}} \\
\cmidrule(lr){3-22}
& & \multicolumn{4}{c}{\textbf{$\mathbf{-0.4}$}} & \multicolumn{4}{c}{\textbf{$\mathbf{-0.2}$}} & \multicolumn{4}{c}{\textbf{$\mathbf{0}$}} & \multicolumn{4}{c}{\textbf{$\mathbf{0.2}$}} & \multicolumn{4}{c}{\textbf{$\mathbf{0.4}$}} \\
\cmidrule(l{0pt}r{5.5pt}){3-6}
\cmidrule(l{0pt}r{5.5pt}){7-10}
\cmidrule(l{0pt}r{5.5pt}){11-14}
\cmidrule(l{0pt}r{5.5pt}){15-18}
\cmidrule(l{0pt}r{0pt}){19-22}
& & \textbf{20} & \textbf{60} & \textbf{120} & \textbf{200} &
    \textbf{20} & \textbf{60} & \textbf{120} & \textbf{200} &
    \textbf{20} & \textbf{60} & \textbf{120} & \textbf{200} &
    \textbf{20} & \textbf{60} & \textbf{120} & \textbf{200} &
    \textbf{20} & \textbf{60} & \textbf{120} & \textbf{200} \\
\midrule

\multirow{4}{*}{4 months}
& GSD          & 0.102 & 0.102 & 0.102 & 0.102 & 0.102 & 0.102 & 0.102 & 0.102 & 0.102 & 0.102 & 0.102 & 0.102 & 0.102 & 0.102 & 0.102 & 0.102 & 0.102 & 0.102 & 0.102 & 0.102 \\
\cmidrule(l{1pt}r{0pt}){2-22}
& EP           & 0.436 & 0.300 & 0.285 & 0.288 & 0.446 & 0.322 & 0.368 & 0.408 & 0.402 & 0.375 & 0.459 & 0.572 & 0.324 & 0.313 & 0.347 & 0.414 & 0.276 & 0.273 & 0.269 & 0.268 \\
\cmidrule(l{1pt}r{0pt}){2-22}
& $B^2$ EP     & 0.623 & 0.439 & 0.503 & 0.288 & 0.578 & 0.624 & 0.626 & 0.408 & 0.463 & 0.655 & 0.733 & 0.572 & 0.344 & 0.457 & 0.475 & 0.414 & 0.255 & 0.291 & 0.260 & 0.268 \\
\cmidrule(l{1pt}r{0pt}){2-22}
& $B^2$ CP  & 0.399 & 0.445 & 0.411 & 0.437 & 0.391 & 0.431 & 0.438 & 0.453 & 0.368 & 0.390 & 0.412 & 0.419 & 0.286 & 0.319 & 0.313 & 0.316 & 0.215 & 0.202 & 0.179 & 0.180 \\
\midrule

\multirow{4}{*}{7 months}
& GSD          & 0.082 & 0.082 & 0.082 & 0.082 & 0.082 & 0.082 & 0.082 & 0.082 & 0.082 & 0.082 & 0.082 & 0.082 & 0.082 & 0.082 & 0.082 & 0.082 & 0.082 & 0.082 & 0.082 & 0.082 \\
\cmidrule(l{1pt}r{0pt}){2-22}
& EP           & 0.490 & 0.385 & 0.440 & 0.296 & 0.457 & 0.462 & 0.569 & 0.396 & 0.395 & 0.461 & 0.629 & 0.557 & 0.316 & 0.357 & 0.422 & 0.384 & 0.263 & 0.270 & 0.263 & 0.265 \\
\cmidrule(l{1pt}r{0pt}){2-22}
& $B^2$ EP     & 0.564 & 0.536 & 0.459 & 0.296 & 0.508 & 0.643 & 0.610 & 0.396 & 0.416 & 0.596 & 0.696 & 0.557 & 0.321 & 0.409 & 0.459 & 0.384 & 0.246 & 0.268 & 0.265 & 0.265 \\
\cmidrule(l{1pt}r{0pt}){2-22}
& $B^2$ CP  & 0.392 & 0.427 & 0.436 & 0.437 & 0.364 & 0.412 & 0.442 & 0.456 & 0.331 & 0.374 & 0.398 & 0.421 & 0.272 & 0.275 & 0.306 & 0.295 & 0.230 & 0.187 & 0.181 & 0.163 \\
\midrule

\multirow{4}{*}{12 months}
& GSD          & 0.077 & 0.077 & 0.077 & 0.077 & 0.077 & 0.077 & 0.077 & 0.077 & 0.077 & 0.077 & 0.077 & 0.077 & 0.077 & 0.077 & 0.077 & 0.077 & 0.077 & 0.077 & 0.077 & 0.077 \\
\cmidrule(l{1pt}r{0pt}){2-22}
& EP           & 0.325 & 0.393 & 0.416 & 0.316 & 0.320 & 0.453 & 0.547 & 0.452 & 0.302 & 0.439 & 0.601 & 0.595 & 0.283 & 0.335 & 0.390 & 0.388 & 0.261 & 0.253 & 0.247 & 0.260 \\
\cmidrule(l{1pt}r{0pt}){2-22}
& $B^2$ EP     & 0.569 & 0.536 & 0.488 & 0.335 & 0.497 & 0.592 & 0.633 & 0.496 & 0.405 & 0.565 & 0.683 & 0.668 & 0.301 & 0.389 & 0.445 & 0.445 & 0.245 & 0.248 & 0.251 & 0.265 \\
\cmidrule(l{1pt}r{0pt}){2-22}
& $B^2$ CP  & 0.382 & 0.397 & 0.416 & 0.418 & 0.348 & 0.416 & 0.414 & 0.438 & 0.323 & 0.358 & 0.395 & 0.401 & 0.263 & 0.296 & 0.303 & 0.299 & 0.220 & 0.194 & 0.172 & 0.162 \\
\bottomrule
\end{tabular}%
}%
}

\begin{minipage}{\textwidth}
\footnotesize
GSD: group sequential design; EP: elastic prior; \(B^2\)-EP: calibrated elastic prior; \(B^2\)-CP: calibrated commensurate prior. Values are empirical power estimated from 1000 simulation replicates.
\end{minipage}

\end{table}
\end{landscape}
\restoregeometry
\clearpage

\clearpage
\newgeometry{margin=0.1in}
\begin{landscape}
\begin{table}[p]
\centering
\caption{Interim-2 power across sensitivity scenarios}
\label{tab:ia2_power_sensitivity}
\scriptsize
\setlength{\tabcolsep}{1.0pt}
\renewcommand{\arraystretch}{1.15}

\resizebox{1.05\textwidth}{!}{%
{\setlength{\tabcolsep}{2.8pt}%
\begin{tabular}{@{}c c
  *{4}{c}@{\hspace{8pt}}
  *{4}{c}@{\hspace{8pt}}
  *{4}{c}@{\hspace{8pt}}
  *{4}{c}@{\hspace{8pt}}
  *{4}{c}@{}}
\toprule
\multirow{3}{*}[-0.3em]{\textbf{\begin{tabular}{@{}c@{}}Median PFS\\(Control)\\HR=0.8\end{tabular}}} &
\multirow{3}{*}[-0.3em]{\textbf{Method}} &
\multicolumn{20}{c}{\textbf{Discrepancy $\delta$}} \\
\cmidrule(lr){3-22}
& & \multicolumn{4}{c}{\textbf{$\mathbf{-0.4}$}} & \multicolumn{4}{c}{\textbf{$\mathbf{-0.2}$}} & \multicolumn{4}{c}{\textbf{$\mathbf{0}$}} & \multicolumn{4}{c}{\textbf{$\mathbf{0.2}$}} & \multicolumn{4}{c}{\textbf{$\mathbf{0.4}$}} \\
\cmidrule(l{0pt}r{5.5pt}){3-6}
\cmidrule(l{0pt}r{5.5pt}){7-10}
\cmidrule(l{0pt}r{5.5pt}){11-14}
\cmidrule(l{0pt}r{5.5pt}){15-18}
\cmidrule(l{0pt}r{0pt}){19-22}
& & \textbf{20} & \textbf{60} & \textbf{120} & \textbf{200} &
    \textbf{20} & \textbf{60} & \textbf{120} & \textbf{200} &
    \textbf{20} & \textbf{60} & \textbf{120} & \textbf{200} &
    \textbf{20} & \textbf{60} & \textbf{120} & \textbf{200} &
    \textbf{20} & \textbf{60} & \textbf{120} & \textbf{200} \\
\midrule

\multirow{4}{*}{4 months}
& GSD          & 0.609 & 0.609 & 0.609 & 0.609 & 0.609 & 0.609 & 0.609 & 0.609 & 0.609 & 0.609 & 0.609 & 0.609 & 0.609 & 0.609 & 0.609 & 0.609 & 0.609 & 0.609 & 0.609 & 0.609 \\
\cmidrule(l{1pt}r{0pt}){2-22}
& EP           & 0.841 & 0.827 & 0.826 & 0.826 & 0.847 & 0.829 & 0.830 & 0.830 & 0.839 & 0.830 & 0.838 & 0.861 & 0.827 & 0.827 & 0.833 & 0.827 & 0.823 & 0.825 & 0.826 & 0.826 \\
\cmidrule(l{1pt}r{0pt}){2-22}
& $B^2$ EP     & 0.871 & 0.831 & 0.841 & 0.826 & 0.868 & 0.853 & 0.862 & 0.830 & 0.841 & 0.858 & 0.873 & 0.861 & 0.819 & 0.823 & 0.819 & 0.827 & 0.802 & 0.820 & 0.819 & 0.826 \\
\cmidrule(l{1pt}r{0pt}){2-22}
& $B^2$ CP  & 0.866 & 0.872 & 0.869 & 0.871 & 0.863 & 0.877 & 0.885 & 0.877 & 0.839 & 0.852 & 0.861 & 0.876 & 0.818 & 0.808 & 0.784 & 0.783 & 0.804 & 0.775 & 0.759 & 0.760 \\
\midrule

\multirow{4}{*}{7 months}
& GSD          & 0.633 & 0.633 & 0.633 & 0.633 & 0.633 & 0.633 & 0.633 & 0.633 & 0.633 & 0.633 & 0.633 & 0.633 & 0.633 & 0.633 & 0.633 & 0.633 & 0.633 & 0.633 & 0.633 & 0.633 \\
\cmidrule(l{1pt}r{0pt}){2-22}
& EP           & 0.845 & 0.828 & 0.828 & 0.823 & 0.848 & 0.838 & 0.838 & 0.828 & 0.841 & 0.835 & 0.866 & 0.843 & 0.819 & 0.822 & 0.819 & 0.827 & 0.812 & 0.822 & 0.818 & 0.823 \\
\cmidrule(l{1pt}r{0pt}){2-22}
& $B^2$ EP     & 0.861 & 0.838 & 0.826 & 0.823 & 0.855 & 0.860 & 0.840 & 0.828 & 0.843 & 0.847 & 0.874 & 0.843 & 0.815 & 0.813 & 0.817 & 0.827 & 0.804 & 0.813 & 0.820 & 0.823 \\
\cmidrule(l{1pt}r{0pt}){2-22}
& $B^2$ CP  & 0.856 & 0.864 & 0.862 & 0.854 & 0.850 & 0.865 & 0.873 & 0.867 & 0.836 & 0.845 & 0.860 & 0.867 & 0.825 & 0.806 & 0.811 & 0.797 & 0.807 & 0.787 & 0.770 & 0.772 \\
\midrule

\multirow{4}{*}{12 months}
& GSD          & 0.631 & 0.631 & 0.631 & 0.631 & 0.631 & 0.631 & 0.631 & 0.631 & 0.631 & 0.631 & 0.631 & 0.631 & 0.631 & 0.631 & 0.631 & 0.631 & 0.631 & 0.631 & 0.631 & 0.631 \\
\cmidrule(l{1pt}r{0pt}){2-22}
& EP           & 0.819 & 0.823 & 0.821 & 0.814 & 0.820 & 0.828 & 0.827 & 0.819 & 0.819 & 0.825 & 0.851 & 0.844 & 0.816 & 0.813 & 0.817 & 0.819 & 0.815 & 0.814 & 0.813 & 0.814 \\
\cmidrule(l{1pt}r{0pt}){2-22}
& $B^2$ EP     & 0.860 & 0.831 & 0.825 & 0.814 & 0.853 & 0.851 & 0.838 & 0.823 & 0.829 & 0.840 & 0.862 & 0.856 & 0.810 & 0.801 & 0.813 & 0.818 & 0.790 & 0.809 & 0.811 & 0.814 \\
\cmidrule(l{1pt}r{0pt}){2-22}
& $B^2$ CP  & 0.845 & 0.860 & 0.856 & 0.854 & 0.842 & 0.861 & 0.868 & 0.877 & 0.827 & 0.837 & 0.857 & 0.867 & 0.819 & 0.802 & 0.796 & 0.779 & 0.799 & 0.778 & 0.771 & 0.765 \\
\bottomrule
\end{tabular}%
}%
}

\medskip

\resizebox{1.05\textwidth}{!}{%
{\setlength{\tabcolsep}{2.8pt}%
\begin{tabular}{@{}c c
  *{4}{c}@{\hspace{8pt}}
  *{4}{c}@{\hspace{8pt}}
  *{4}{c}@{\hspace{8pt}}
  *{4}{c}@{\hspace{8pt}}
  *{4}{c}@{}}
\toprule
\multirow{3}{*}[-0.3em]{\textbf{\begin{tabular}{@{}c@{}}Median PFS\\(Control)\\HR=0.6\end{tabular}}} &
\multirow{3}{*}[-0.3em]{\textbf{Method}} &
\multicolumn{20}{c}{\textbf{Discrepancy $\delta$}} \\
\cmidrule(lr){3-22}
& & \multicolumn{4}{c}{\textbf{$\mathbf{-0.4}$}} & \multicolumn{4}{c}{\textbf{$\mathbf{-0.2}$}} & \multicolumn{4}{c}{\textbf{$\mathbf{0}$}} & \multicolumn{4}{c}{\textbf{$\mathbf{0.2}$}} & \multicolumn{4}{c}{\textbf{$\mathbf{0.4}$}} \\
\cmidrule(l{0pt}r{5.5pt}){3-6}
\cmidrule(l{0pt}r{5.5pt}){7-10}
\cmidrule(l{0pt}r{5.5pt}){11-14}
\cmidrule(l{0pt}r{5.5pt}){15-18}
\cmidrule(l{0pt}r{0pt}){19-22}
& & \textbf{20} & \textbf{60} & \textbf{120} & \textbf{200} &
    \textbf{20} & \textbf{60} & \textbf{120} & \textbf{200} &
    \textbf{20} & \textbf{60} & \textbf{120} & \textbf{200} &
    \textbf{20} & \textbf{60} & \textbf{120} & \textbf{200} &
    \textbf{20} & \textbf{60} & \textbf{120} & \textbf{200} \\
\midrule

\multirow{4}{*}{4 months}
& GSD          & 0.618 & 0.618 & 0.618 & 0.618 & 0.618 & 0.618 & 0.618 & 0.618 & 0.618 & 0.618 & 0.618 & 0.618 & 0.618 & 0.618 & 0.618 & 0.618 & 0.618 & 0.618 & 0.618 & 0.618 \\
\cmidrule(l{1pt}r{0pt}){2-22}
& EP           & 0.898 & 0.876 & 0.874 & 0.875 & 0.905 & 0.881 & 0.880 & 0.882 & 0.911 & 0.890 & 0.896 & 0.901 & 0.900 & 0.889 & 0.914 & 0.938 & 0.880 & 0.886 & 0.885 & 0.885 \\
\cmidrule(l{1pt}r{0pt}){2-22}
& $B^2$ EP     & 0.938 & 0.887 & 0.902 & 0.875 & 0.948 & 0.907 & 0.919 & 0.882 & 0.932 & 0.939 & 0.952 & 0.901 & 0.907 & 0.937 & 0.959 & 0.938 & 0.861 & 0.888 & 0.884 & 0.885 \\
\cmidrule(l{1pt}r{0pt}){2-22}
& $B^2$ CP  & 0.927 & 0.922 & 0.918 & 0.921 & 0.922 & 0.937 & 0.935 & 0.939 & 0.913 & 0.940 & 0.948 & 0.950 & 0.889 & 0.911 & 0.922 & 0.931 & 0.856 & 0.834 & 0.837 & 0.836 \\
\midrule

\multirow{4}{*}{7 months}
& GSD          & 0.600 & 0.600 & 0.600 & 0.600 & 0.600 & 0.600 & 0.600 & 0.600 & 0.600 & 0.600 & 0.600 & 0.600 & 0.600 & 0.600 & 0.600 & 0.600 & 0.600 & 0.600 & 0.600 & 0.600 \\
\cmidrule(l{1pt}r{0pt}){2-22}
& EP           & 0.910 & 0.884 & 0.892 & 0.876 & 0.915 & 0.895 & 0.907 & 0.880 & 0.923 & 0.919 & 0.935 & 0.899 & 0.902 & 0.918 & 0.953 & 0.944 & 0.884 & 0.888 & 0.891 & 0.891 \\
\cmidrule(l{1pt}r{0pt}){2-22}
& $B^2$ EP     & 0.927 & 0.904 & 0.889 & 0.876 & 0.942 & 0.925 & 0.907 & 0.880 & 0.930 & 0.955 & 0.941 & 0.899 & 0.905 & 0.938 & 0.960 & 0.944 & 0.872 & 0.883 & 0.892 & 0.891 \\
\cmidrule(l{1pt}r{0pt}){2-22}
& $B^2$ CP  & 0.928 & 0.919 & 0.922 & 0.925 & 0.923 & 0.937 & 0.929 & 0.936 & 0.909 & 0.940 & 0.946 & 0.944 & 0.889 & 0.904 & 0.925 & 0.931 & 0.863 & 0.848 & 0.834 & 0.833 \\
\midrule

\multirow{4}{*}{12 months}
& GSD          & 0.612 & 0.612 & 0.612 & 0.612 & 0.612 & 0.612 & 0.612 & 0.612 & 0.612 & 0.612 & 0.612 & 0.612 & 0.612 & 0.612 & 0.612 & 0.612 & 0.612 & 0.612 & 0.612 & 0.612 \\
\cmidrule(l{1pt}r{0pt}){2-22}
& EP           & 0.883 & 0.889 & 0.888 & 0.882 & 0.894 & 0.903 & 0.900 & 0.886 & 0.899 & 0.925 & 0.928 & 0.912 & 0.890 & 0.914 & 0.946 & 0.950 & 0.884 & 0.884 & 0.893 & 0.894 \\
\cmidrule(l{1pt}r{0pt}){2-22}
& $B^2$ EP     & 0.947 & 0.913 & 0.892 & 0.884 & 0.955 & 0.935 & 0.909 & 0.888 & 0.940 & 0.956 & 0.946 & 0.924 & 0.908 & 0.932 & 0.963 & 0.963 & 0.865 & 0.881 & 0.890 & 0.901 \\
\cmidrule(l{1pt}r{0pt}){2-22}
& $B^2$ CP  & 0.914 & 0.928 & 0.919 & 0.918 & 0.918 & 0.936 & 0.938 & 0.934 & 0.912 & 0.933 & 0.945 & 0.946 & 0.889 & 0.898 & 0.921 & 0.932 & 0.862 & 0.842 & 0.829 & 0.841 \\
\bottomrule
\end{tabular}%
}%
}

\begin{minipage}{\textwidth}
\footnotesize
GSD: group sequential design; EP: elastic prior; \(B^2\)-EP: calibrated elastic prior; \(B^2\)-CP: calibrated commensurate prior. Values are empirical power estimated from 1000 simulation replicates.
\end{minipage}

\end{table}
\end{landscape}
\restoregeometry
\clearpage

\clearpage
\newgeometry{margin=0.1in}
\begin{landscape}
\begin{table}[p]
\centering
\caption{Final-analysis power across sensitivity scenarios}
\label{tab:final_power_sensitivity}
\scriptsize
\setlength{\tabcolsep}{1.0pt}
\renewcommand{\arraystretch}{1.15}

\resizebox{1.05\textwidth}{!}{%
{\setlength{\tabcolsep}{2.8pt}%
\begin{tabular}{@{}c c
  *{4}{c}@{\hspace{8pt}}
  *{4}{c}@{\hspace{8pt}}
  *{4}{c}@{\hspace{8pt}}
  *{4}{c}@{\hspace{8pt}}
  *{4}{c}@{}}
\toprule
\multirow{3}{*}[-0.3em]{\textbf{\begin{tabular}{@{}c@{}}Median PFS\\(Control)\\HR=0.8\end{tabular}}} &
\multirow{3}{*}[-0.3em]{\textbf{Method}} &
\multicolumn{20}{c}{\textbf{Discrepancy $\delta$}} \\
\cmidrule(lr){3-22}
& & \multicolumn{4}{c}{\textbf{$\mathbf{-0.4}$}} & \multicolumn{4}{c}{\textbf{$\mathbf{-0.2}$}} & \multicolumn{4}{c}{\textbf{$\mathbf{0}$}} & \multicolumn{4}{c}{\textbf{$\mathbf{0.2}$}} & \multicolumn{4}{c}{\textbf{$\mathbf{0.4}$}} \\
\cmidrule(l{0pt}r{5.5pt}){3-6}
\cmidrule(l{0pt}r{5.5pt}){7-10}
\cmidrule(l{0pt}r{5.5pt}){11-14}
\cmidrule(l{0pt}r{5.5pt}){15-18}
\cmidrule(l{0pt}r{0pt}){19-22}
& & \textbf{20} & \textbf{60} & \textbf{120} & \textbf{200} &
    \textbf{20} & \textbf{60} & \textbf{120} & \textbf{200} &
    \textbf{20} & \textbf{60} & \textbf{120} & \textbf{200} &
    \textbf{20} & \textbf{60} & \textbf{120} & \textbf{200} &
    \textbf{20} & \textbf{60} & \textbf{120} & \textbf{200} \\
\midrule

\multirow{4}{*}{4 months}
& GSD          & 0.902 & 0.902 & 0.902 & 0.902 & 0.902 & 0.902 & 0.902 & 0.902 & 0.902 & 0.902 & 0.902 & 0.902 & 0.902 & 0.902 & 0.902 & 0.902 & 0.902 & 0.902 & 0.902 & 0.902 \\
\cmidrule(l{1pt}r{0pt}){2-22}
& EP           & 0.972 & 0.970 & 0.970 & 0.970 & 0.975 & 0.970 & 0.970 & 0.971 & 0.972 & 0.971 & 0.973 & 0.973 & 0.970 & 0.971 & 0.972 & 0.972 & 0.969 & 0.970 & 0.970 & 0.970 \\
\cmidrule(l{1pt}r{0pt}){2-22}
& $B^2$ EP     & 0.977 & 0.970 & 0.972 & 0.970 & 0.979 & 0.973 & 0.974 & 0.971 & 0.976 & 0.980 & 0.977 & 0.973 & 0.966 & 0.971 & 0.971 & 0.972 & 0.964 & 0.969 & 0.969 & 0.970 \\
\cmidrule(l{1pt}r{0pt}){2-22}
& $B^2$ CP  & 0.978 & 0.975 & 0.976 & 0.975 & 0.976 & 0.979 & 0.979 & 0.980 & 0.971 & 0.977 & 0.981 & 0.982 & 0.967 & 0.967 & 0.962 & 0.961 & 0.964 & 0.959 & 0.955 & 0.957 \\
\midrule

\multirow{4}{*}{7 months}
& GSD          & 0.902 & 0.902 & 0.902 & 0.902 & 0.902 & 0.902 & 0.902 & 0.902 & 0.902 & 0.902 & 0.902 & 0.902 & 0.902 & 0.902 & 0.902 & 0.902 & 0.902 & 0.902 & 0.902 & 0.902 \\
\cmidrule(l{1pt}r{0pt}){2-22}
& EP           & 0.974 & 0.969 & 0.972 & 0.969 & 0.975 & 0.969 & 0.973 & 0.970 & 0.973 & 0.972 & 0.975 & 0.974 & 0.967 & 0.969 & 0.966 & 0.969 & 0.968 & 0.969 & 0.968 & 0.969 \\
\cmidrule(l{1pt}r{0pt}){2-22}
& $B^2$ EP     & 0.975 & 0.971 & 0.971 & 0.969 & 0.977 & 0.972 & 0.974 & 0.970 & 0.975 & 0.978 & 0.977 & 0.974 & 0.963 & 0.968 & 0.966 & 0.969 & 0.966 & 0.966 & 0.969 & 0.969 \\
\cmidrule(l{1pt}r{0pt}){2-22}
& $B^2$ CP  & 0.977 & 0.977 & 0.979 & 0.976 & 0.977 & 0.977 & 0.980 & 0.980 & 0.971 & 0.980 & 0.979 & 0.980 & 0.965 & 0.969 & 0.964 & 0.961 & 0.961 & 0.958 & 0.957 & 0.953 \\
\midrule

\multirow{4}{*}{12 months}
& GSD          & 0.910 & 0.910 & 0.910 & 0.910 & 0.910 & 0.910 & 0.910 & 0.910 & 0.910 & 0.910 & 0.910 & 0.910 & 0.910 & 0.910 & 0.910 & 0.910 & 0.910 & 0.910 & 0.910 & 0.910 \\
\cmidrule(l{1pt}r{0pt}){2-22}
& EP           & 0.971 & 0.971 & 0.972 & 0.971 & 0.972 & 0.972 & 0.973 & 0.971 & 0.970 & 0.973 & 0.977 & 0.978 & 0.971 & 0.970 & 0.972 & 0.971 & 0.971 & 0.970 & 0.970 & 0.971 \\
\cmidrule(l{1pt}r{0pt}){2-22}
& $B^2$ EP     & 0.978 & 0.972 & 0.972 & 0.971 & 0.979 & 0.976 & 0.973 & 0.971 & 0.977 & 0.977 & 0.978 & 0.979 & 0.968 & 0.971 & 0.972 & 0.971 & 0.966 & 0.969 & 0.969 & 0.971 \\
\cmidrule(l{1pt}r{0pt}){2-22}
& $B^2$ CP  & 0.975 & 0.974 & 0.976 & 0.980 & 0.979 & 0.978 & 0.981 & 0.978 & 0.976 & 0.974 & 0.981 & 0.980 & 0.972 & 0.969 & 0.969 & 0.964 & 0.968 & 0.964 & 0.963 & 0.962 \\
\bottomrule
\end{tabular}%
}%
}

\medskip

\resizebox{1.05\textwidth}{!}{%
{\setlength{\tabcolsep}{2.8pt}%
\begin{tabular}{@{}c c
  *{4}{c}@{\hspace{8pt}}
  *{4}{c}@{\hspace{8pt}}
  *{4}{c}@{\hspace{8pt}}
  *{4}{c}@{\hspace{8pt}}
  *{4}{c}@{}}
\toprule
\multirow{3}{*}[-0.3em]{\textbf{\begin{tabular}{@{}c@{}}Median PFS\\(Control)\\HR=0.6\end{tabular}}} &
\multirow{3}{*}[-0.3em]{\textbf{Method}} &
\multicolumn{20}{c}{\textbf{Discrepancy $\delta$}} \\
\cmidrule(lr){3-22}
& & \multicolumn{4}{c}{\textbf{$\mathbf{-0.4}$}} & \multicolumn{4}{c}{\textbf{$\mathbf{-0.2}$}} & \multicolumn{4}{c}{\textbf{$\mathbf{0}$}} & \multicolumn{4}{c}{\textbf{$\mathbf{0.2}$}} & \multicolumn{4}{c}{\textbf{$\mathbf{0.4}$}} \\
\cmidrule(l{0pt}r{5.5pt}){3-6}
\cmidrule(l{0pt}r{5.5pt}){7-10}
\cmidrule(l{0pt}r{5.5pt}){11-14}
\cmidrule(l{0pt}r{5.5pt}){15-18}
\cmidrule(l{0pt}r{0pt}){19-22}
& & \textbf{20} & \textbf{60} & \textbf{120} & \textbf{200} &
    \textbf{20} & \textbf{60} & \textbf{120} & \textbf{200} &
    \textbf{20} & \textbf{60} & \textbf{120} & \textbf{200} &
    \textbf{20} & \textbf{60} & \textbf{120} & \textbf{200} &
    \textbf{20} & \textbf{60} & \textbf{120} & \textbf{200} \\
\midrule

\multirow{4}{*}{4 months}
& GSD          & 0.895 & 0.895 & 0.895 & 0.895 & 0.895 & 0.895 & 0.895 & 0.895 & 0.895 & 0.895 & 0.895 & 0.895 & 0.895 & 0.895 & 0.895 & 0.895 & 0.895 & 0.895 & 0.895 & 0.895 \\
\cmidrule(l{1pt}r{0pt}){2-22}
& EP           & 0.989 & 0.987 & 0.987 & 0.987 & 0.990 & 0.987 & 0.988 & 0.987 & 0.990 & 0.987 & 0.987 & 0.988 & 0.988 & 0.989 & 0.990 & 0.992 & 0.987 & 0.992 & 0.991 & 0.992 \\
\cmidrule(l{1pt}r{0pt}){2-22}
& $B^2$ EP     & 0.993 & 0.987 & 0.989 & 0.987 & 0.992 & 0.987 & 0.991 & 0.987 & 0.990 & 0.989 & 0.995 & 0.988 & 0.990 & 0.995 & 0.998 & 0.992 & 0.989 & 0.995 & 0.992 & 0.992 \\
\cmidrule(l{1pt}r{0pt}){2-22}
& $B^2$ CP  & 0.995 & 0.993 & 0.994 & 0.994 & 0.993 & 0.995 & 0.997 & 0.996 & 0.993 & 0.994 & 0.997 & 0.996 & 0.992 & 0.995 & 0.996 & 0.997 & 0.989 & 0.988 & 0.987 & 0.986 \\
\midrule

\multirow{4}{*}{7 months}
& GSD          & 0.888 & 0.888 & 0.888 & 0.888 & 0.888 & 0.888 & 0.888 & 0.888 & 0.888 & 0.888 & 0.888 & 0.888 & 0.888 & 0.888 & 0.888 & 0.888 & 0.888 & 0.888 & 0.888 & 0.888 \\
\cmidrule(l{1pt}r{0pt}){2-22}
& EP           & 0.993 & 0.990 & 0.992 & 0.990 & 0.996 & 0.992 & 0.993 & 0.990 & 0.991 & 0.992 & 0.995 & 0.991 & 0.992 & 0.995 & 0.999 & 0.993 & 0.992 & 0.996 & 0.989 & 0.994 \\
\cmidrule(l{1pt}r{0pt}){2-22}
& $B^2$ EP     & 0.993 & 0.992 & 0.992 & 0.990 & 0.996 & 0.992 & 0.993 & 0.990 & 0.991 & 0.995 & 0.995 & 0.991 & 0.991 & 0.997 & 0.999 & 0.993 & 0.992 & 0.994 & 0.988 & 0.994 \\
\cmidrule(l{1pt}r{0pt}){2-22}
& $B^2$ CP  & 0.996 & 0.996 & 0.997 & 0.995 & 0.996 & 0.998 & 0.997 & 0.997 & 0.994 & 0.998 & 0.999 & 0.999 & 0.992 & 0.996 & 0.997 & 0.999 & 0.990 & 0.989 & 0.985 & 0.983 \\
\midrule

\multirow{4}{*}{12 months}
& GSD          & 0.892 & 0.892 & 0.892 & 0.892 & 0.892 & 0.892 & 0.892 & 0.892 & 0.892 & 0.892 & 0.892 & 0.892 & 0.892 & 0.892 & 0.892 & 0.892 & 0.892 & 0.892 & 0.892 & 0.892 \\
\cmidrule(l{1pt}r{0pt}){2-22}
& EP           & 0.989 & 0.989 & 0.989 & 0.989 & 0.990 & 0.989 & 0.990 & 0.989 & 0.989 & 0.991 & 0.993 & 0.989 & 0.990 & 0.995 & 0.998 & 0.996 & 0.990 & 0.994 & 0.990 & 0.993 \\
\cmidrule(l{1pt}r{0pt}){2-22}
& $B^2$ EP     & 0.995 & 0.989 & 0.989 & 0.989 & 0.994 & 0.989 & 0.990 & 0.989 & 0.994 & 0.995 & 0.995 & 0.990 & 0.992 & 0.997 & 0.998 & 0.997 & 0.987 & 0.995 & 0.989 & 0.992 \\
\cmidrule(l{1pt}r{0pt}){2-22}
& $B^2$ CP  & 0.995 & 0.995 & 0.997 & 0.995 & 0.995 & 0.997 & 0.997 & 0.996 & 0.993 & 0.997 & 0.998 & 0.998 & 0.993 & 0.996 & 0.995 & 0.997 & 0.988 & 0.988 & 0.989 & 0.984 \\
\bottomrule
\end{tabular}%
}%
}

\begin{minipage}{\textwidth}
\footnotesize
GSD: group sequential design; EP: elastic prior; \(B^2\)-EP: calibrated elastic prior; \(B^2\)-CP: calibrated commensurate prior. Values are empirical power estimated from 1000 simulation replicates.
\end{minipage}

\end{table}
\end{landscape}
\restoregeometry
\clearpage

\section{ESS at IA1, IA2 and FA across design and borrowing scenarios}
\label{sec:ess-ia1-ia2-fa}
\clearpage
\newgeometry{margin=0.1in}
\begin{landscape}
\begin{table}[p]
\centering
\caption{Interim-1 treatment-arm event-scale ESS across sensitivity scenarios}
\label{tab:ia1_treatment_ess_sensitivity}
\scriptsize
\setlength{\tabcolsep}{1.0pt}
\renewcommand{\arraystretch}{1.15}

\resizebox{1.05\textwidth}{!}{%
{\setlength{\tabcolsep}{2.8pt}%
\begin{tabular}{@{}c c
  *{4}{c}@{\hspace{8pt}}
  *{4}{c}@{\hspace{8pt}}
  *{4}{c}@{\hspace{8pt}}
  *{4}{c}@{\hspace{8pt}}
  *{4}{c}@{}}
\toprule
\multirow{3}{*}[-0.3em]{\textbf{\begin{tabular}{@{}c@{}}Median PFS\\(Control)\\HR=0.8\end{tabular}}} &
\multirow{3}{*}[-0.3em]{\textbf{Method}} &
\multicolumn{20}{c}{\textbf{Discrepancy $\delta$}} \\
\cmidrule(lr){3-22}
& & \multicolumn{4}{c}{\textbf{$\mathbf{-0.4}$}} & \multicolumn{4}{c}{\textbf{$\mathbf{-0.2}$}} & \multicolumn{4}{c}{\textbf{$\mathbf{0}$}} & \multicolumn{4}{c}{\textbf{$\mathbf{0.2}$}} & \multicolumn{4}{c}{\textbf{$\mathbf{0.4}$}} \\
\cmidrule(l{0pt}r{5.5pt}){3-6}
\cmidrule(l{0pt}r{5.5pt}){7-10}
\cmidrule(l{0pt}r{5.5pt}){11-14}
\cmidrule(l{0pt}r{5.5pt}){15-18}
\cmidrule(l{0pt}r{0pt}){19-22}
& & \textbf{20} & \textbf{60} & \textbf{120} & \textbf{200} &
    \textbf{20} & \textbf{60} & \textbf{120} & \textbf{200} &
    \textbf{20} & \textbf{60} & \textbf{120} & \textbf{200} &
    \textbf{20} & \textbf{60} & \textbf{120} & \textbf{200} &
    \textbf{20} & \textbf{60} & \textbf{120} & \textbf{200} \\
\midrule

\multirow{4}{*}{4 months}
& GSD          & 0.0 & 0.0 & 0.0 & 0.0 & 0.0 & 0.0 & 0.0 & 0.0 & 0.0 & 0.0 & 0.0 & 0.0 & 0.0 & 0.0 & 0.0 & 0.0 & 0.0 & 0.0 & 0.0 & 0.0 \\
\cmidrule(l{1pt}r{0pt}){2-22}
& EP           & 7.4 & 0.7 & 0.5 & $-$0.6 & 12.5 & 5.1 & 9.1 & 6.3 & 14.9 & 13.1 & 35.6 & 71.8 & 11.1 & 5.3 & 8.3 & 4.8 & 4.9 & 0.3 & $-$0.2 & 0.0 \\
\cmidrule(l{1pt}r{0pt}){2-22}
& $B^2$ EP     & 4.7 & $-$3.7 & 3.8 & $-$0.6 & 15.2 & 22.9 & 26.7 & 6.3 & 19.8 & 52.5 & 84.3 & 71.8 & 14.0 & 20.9 & 21.8 & 4.8 & 4.0 & $-$2.9 & 1.0 & 0.0 \\
\cmidrule(l{1pt}r{0pt}){2-22}
& $B^2$ CP  & 1.0 & $-$13.7 & $-$21.0 & $-$24.0 & 9.9 & 9.5 & 2.9 & $-$3.8 & 11.4 & 28.8 & 46.8 & 62.2 & 7.7 & 9.0 & 1.1 & $-$8.9 & $-$1.0 & $-$14.5 & $-$22.9 & $-$25.0 \\
\midrule

\multirow{4}{*}{7 months}
& GSD          & 0.0 & 0.0 & 0.0 & 0.0 & 0.0 & 0.0 & 0.0 & 0.0 & 0.0 & 0.0 & 0.0 & 0.0 & 0.0 & 0.0 & 0.0 & 0.0 & 0.0 & 0.0 & 0.0 & 0.0 \\
\cmidrule(l{1pt}r{0pt}){2-22}
& EP           & 8.4 & 3.9 & 2.7 & $-$0.2 & 14.0 & 15.4 & 24.2 & 8.9 & 16.2 & 25.8 & 64.7 & 69.7 & 12.3 & 14.5 & 19.8 & 5.0 & 5.0 & 1.9 & 0.8 & $-$0.1 \\
\cmidrule(l{1pt}r{0pt}){2-22}
& $B^2$ EP     & 5.9 & 4.9 & 0.6 & $-$0.2 & 14.3 & 27.4 & 26.0 & 8.9 & 17.2 & 45.0 & 77.7 & 69.7 & 12.1 & 22.3 & 20.7 & 5.0 & 3.3 & 1.4 & $-$0.6 & $-$0.1 \\
\cmidrule(l{1pt}r{0pt}){2-22}
& $B^2$ CP  & 2.2 & $-$9.5 & $-$19.3 & $-$23.3 & 7.9 & 10.6 & 4.8 & 0.5 & 9.9 & 24.3 & 40.7 & 57.1 & 7.4 & 9.6 & 3.5 & $-$6.1 & 0.3 & $-$12.1 & $-$21.7 & $-$24.1 \\
\midrule

\multirow{4}{*}{12 months}
& GSD          & 0.0 & 0.0 & 0.0 & 0.0 & 0.0 & 0.0 & 0.0 & 0.0 & 0.0 & 0.0 & 0.0 & 0.0 & 0.0 & 0.0 & 0.0 & 0.0 & 0.0 & 0.0 & 0.0 & 0.0 \\
\cmidrule(l{1pt}r{0pt}){2-22}
& EP           & 4.3 & 5.2 & 3.4 & $-$0.7 & 6.4 & 15.0 & 22.0 & 14.0 & 7.2 & 23.2 & 51.6 & 63.2 & 5.3 & 15.0 & 18.6 & 7.7 & 2.7 & 2.6 & 0.7 & $-$0.3 \\
\cmidrule(l{1pt}r{0pt}){2-22}
& $B^2$ EP     & 4.5 & 7.4 & 3.2 & $-$1.4 & 11.3 & 24.4 & 27.3 & 15.8 & 13.6 & 37.0 & 66.8 & 83.4 & 9.6 & 21.5 & 22.8 & 9.7 & 1.8 & 2.3 & $-$0.3 & $-$1.0 \\
\cmidrule(l{1pt}r{0pt}){2-22}
& $B^2$ CP  & 2.9 & $-$6.3 & $-$16.4 & $-$21.2 & 7.4 & 10.9 & 6.9 & 3.6 & 8.8 & 21.9 & 36.7 & 49.7 & 6.5 & 9.9 & 5.8 & $-$4.4 & 0.5 & $-$10.5 & $-$21.1 & $-$24.4 \\
\bottomrule
\end{tabular}%
}%
}

\vspace{2pt}

\resizebox{1.05\textwidth}{!}{%
{\setlength{\tabcolsep}{2.8pt}%
\begin{tabular}{@{}c c
  *{4}{c}@{\hspace{8pt}}
  *{4}{c}@{\hspace{8pt}}
  *{4}{c}@{\hspace{8pt}}
  *{4}{c}@{\hspace{8pt}}
  *{4}{c}@{}}
\toprule
\multirow{3}{*}[-0.3em]{\textbf{\begin{tabular}{@{}c@{}}Median PFS\\(Control)\\HR=0.6\end{tabular}}} &
\multirow{3}{*}[-0.3em]{\textbf{Method}} &
\multicolumn{20}{c}{\textbf{Discrepancy $\delta$}} \\
\cmidrule(lr){3-22}
& & \multicolumn{4}{c}{\textbf{$\mathbf{-0.4}$}} & \multicolumn{4}{c}{\textbf{$\mathbf{-0.2}$}} & \multicolumn{4}{c}{\textbf{$\mathbf{0}$}} & \multicolumn{4}{c}{\textbf{$\mathbf{0.2}$}} & \multicolumn{4}{c}{\textbf{$\mathbf{0.4}$}} \\
\cmidrule(l{0pt}r{5.5pt}){3-6}
\cmidrule(l{0pt}r{5.5pt}){7-10}
\cmidrule(l{0pt}r{5.5pt}){11-14}
\cmidrule(l{0pt}r{5.5pt}){15-18}
\cmidrule(l{0pt}r{0pt}){19-22}
& & \textbf{20} & \textbf{60} & \textbf{120} & \textbf{200} &
    \textbf{20} & \textbf{60} & \textbf{120} & \textbf{200} &
    \textbf{20} & \textbf{60} & \textbf{120} & \textbf{200} &
    \textbf{20} & \textbf{60} & \textbf{120} & \textbf{200} &
    \textbf{20} & \textbf{60} & \textbf{120} & \textbf{200} \\
\midrule

\multirow{4}{*}{4 months}
& GSD          & 0.0 & 0.0 & 0.0 & 0.0 & 0.0 & 0.0 & 0.0 & 0.0 & 0.0 & 0.0 & 0.0 & 0.0 & 0.0 & 0.0 & 0.0 & 0.0 & 0.0 & 0.0 & 0.0 & 0.0 \\
\cmidrule(l{1pt}r{0pt}){2-22}
& EP           & 6.8 & 0.7 & $-$1.2 & $-$2.1 & 11.4 & 6.4 & 9.8 & 14.7 & 13.1 & 12.2 & 36.3 & 82.9 & 9.8 & 5.8 & 8.1 & 8.8 & 4.2 & $-$0.5 & $-$1.5 & $-$2.5 \\
\cmidrule(l{1pt}r{0pt}){2-22}
& $B^2$ EP     & 6.8 & $-$1.3 & $-$0.2 & $-$2.1 & 15.4 & 27.4 & 26.5 & 14.7 & 18.5 & 49.7 & 83.6 & 82.9 & 12.7 & 21.4 & 22.6 & 8.8 & 2.6 & $-$5.3 & $-$4.4 & $-$2.5 \\
\cmidrule(l{1pt}r{0pt}){2-22}
& $B^2$ CP  & 3.2 & $-$1.2 & $-$5.0 & $-$7.0 & 7.6 & 12.2 & 13.4 & 14.6 & 9.5 & 20.4 & 28.7 & 34.0 & 6.4 & 10.4 & 10.5 & 9.3 & 0.5 & $-$5.3 & $-$10.1 & $-$12.3 \\
\midrule

\multirow{4}{*}{7 months}
& GSD          & 0.0 & 0.0 & 0.0 & 0.0 & 0.0 & 0.0 & 0.0 & 0.0 & 0.0 & 0.0 & 0.0 & 0.0 & 0.0 & 0.0 & 0.0 & 0.0 & 0.0 & 0.0 & 0.0 & 0.0 \\
\cmidrule(l{1pt}r{0pt}){2-22}
& EP           & 7.4 & 4.2 & 1.6 & $-$1.7 & 12.1 & 15.4 & 24.0 & 13.1 & 13.8 & 23.1 & 58.7 & 64.1 & 10.6 & 14.0 & 18.7 & 8.7 & 4.1 & 0.7 & $-$2.6 & $-$2.7 \\
\cmidrule(l{1pt}r{0pt}){2-22}
& $B^2$ EP     & 7.0 & 5.1 & $-$0.9 & $-$1.7 & 13.0 & 26.1 & 27.2 & 13.1 & 15.2 & 40.5 & 72.3 & 64.1 & 11.1 & 22.4 & 21.3 & 8.7 & 2.7 & $-$1.2 & $-$5.0 & $-$2.7 \\
\cmidrule(l{1pt}r{0pt}){2-22}
& $B^2$ CP  & 3.1 & 0.2 & $-$3.4 & $-$5.7 & 6.6 & 11.4 & 13.1 & 14.5 & 8.0 & 17.9 & 26.4 & 32.9 & 5.9 & 10.2 & 11.0 & 9.9 & 1.0 & $-$3.9 & $-$8.7 & $-$11.6 \\
\midrule

\multirow{4}{*}{12 months}
& GSD          & 0.0 & 0.0 & 0.0 & 0.0 & 0.0 & 0.0 & 0.0 & 0.0 & 0.0 & 0.0 & 0.0 & 0.0 & 0.0 & 0.0 & 0.0 & 0.0 & 0.0 & 0.0 & 0.0 & 0.0 \\
\cmidrule(l{1pt}r{0pt}){2-22}
& EP           & 4.0 & 5.3 & 3.6 & $-$0.6 & 5.2 & 14.9 & 23.1 & 18.5 & 5.8 & 20.1 & 45.1 & 61.4 & 4.8 & 13.2 & 19.5 & 12.8 & 2.6 & 2.0 & $-$1.3 & $-$3.2 \\
\cmidrule(l{1pt}r{0pt}){2-22}
& $B^2$ EP     & 6.1 & 7.3 & 2.8 & $-$2.2 & 10.8 & 22.8 & 28.9 & 22.1 & 12.4 & 32.5 & 59.8 & 81.5 & 9.0 & 20.1 & 24.5 & 15.9 & 2.7 & 1.3 & $-$3.7 & $-$5.0 \\
\cmidrule(l{1pt}r{0pt}){2-22}
& $B^2$ CP  & 3.4 & 1.5 & $-$1.6 & $-$4.1 & 5.8 & 10.8 & 13.0 & 14.3 & 6.9 & 15.8 & 23.6 & 30.0 & 5.2 & 9.5 & 11.3 & 10.4 & 1.3 & $-$2.7 & $-$7.1 & $-$10.5 \\
\bottomrule
\end{tabular}%
}%
}

\vspace{2pt}
\noindent\makebox[\textwidth][c]{%
\begin{minipage}{1.05\textwidth}
\footnotesize
GSD: group sequential design; EP: elastic prior; \(B^2\)-EP: calibrated elastic prior; \(B^2\)-CP: calibrated commensurate prior. Values are mean interim-1 treatment-arm event-scale ESS across 1000 simulation replicates.
\end{minipage}%
}

\end{table}
\end{landscape}
\restoregeometry
\clearpage

\clearpage
\newgeometry{margin=0.1in}
\begin{landscape}
\begin{table}[p]
\centering
\caption{Interim-2 treatment-arm event-scale ESS across sensitivity scenarios}
\label{tab:ia2_treatment_ess_sensitivity}
\scriptsize
\setlength{\tabcolsep}{1.0pt}
\renewcommand{\arraystretch}{1.15}

\resizebox{1.05\textwidth}{!}{%
{\setlength{\tabcolsep}{2.8pt}%
\begin{tabular}{@{}c c
  *{4}{c}@{\hspace{8pt}}
  *{4}{c}@{\hspace{8pt}}
  *{4}{c}@{\hspace{8pt}}
  *{4}{c}@{\hspace{8pt}}
  *{4}{c}@{}}
\toprule
\multirow{3}{*}[-0.3em]{\textbf{\begin{tabular}{@{}c@{}}Median PFS\\(Control)\\HR=0.8\end{tabular}}} &
\multirow{3}{*}[-0.3em]{\textbf{Method}} &
\multicolumn{20}{c}{\textbf{Discrepancy $\delta$}} \\
\cmidrule(lr){3-22}
& & \multicolumn{4}{c}{\textbf{$\mathbf{-0.4}$}} & \multicolumn{4}{c}{\textbf{$\mathbf{-0.2}$}} & \multicolumn{4}{c}{\textbf{$\mathbf{0}$}} & \multicolumn{4}{c}{\textbf{$\mathbf{0.2}$}} & \multicolumn{4}{c}{\textbf{$\mathbf{0.4}$}} \\
\cmidrule(l{0pt}r{5.5pt}){3-6}
\cmidrule(l{0pt}r{5.5pt}){7-10}
\cmidrule(l{0pt}r{5.5pt}){11-14}
\cmidrule(l{0pt}r{5.5pt}){15-18}
\cmidrule(l{0pt}r{0pt}){19-22}
& & \textbf{20} & \textbf{60} & \textbf{120} & \textbf{200} &
    \textbf{20} & \textbf{60} & \textbf{120} & \textbf{200} &
    \textbf{20} & \textbf{60} & \textbf{120} & \textbf{200} &
    \textbf{20} & \textbf{60} & \textbf{120} & \textbf{200} &
    \textbf{20} & \textbf{60} & \textbf{120} & \textbf{200} \\
\midrule

\multirow{4}{*}{4 months}
& GSD          & 0.0 & 0.0 & 0.0 & 0.0 & 0.0 & 0.0 & 0.0 & 0.0 & 0.0 & 0.0 & 0.0 & 0.0 & 0.0 & 0.0 & 0.0 & 0.0 & 0.0 & 0.0 & 0.0 & 0.0 \\
\cmidrule(l{1pt}r{0pt}){2-22}
& EP           & 7.7 & 0.5 & 0.1 & 0.0 & 13.5 & 6.7 & 8.7 & 8.4 & 16.5 & 14.4 & 41.1 & 78.4 & 12.3 & 6.3 & 7.7 & 4.0 & 5.0 & 0.5 & $-$0.2 & 0.0 \\
\cmidrule(l{1pt}r{0pt}){2-22}
& $B^2$ EP     & 6.4 & $-$0.5 & 4.0 & 0.0 & 16.7 & 26.4 & 27.4 & 8.4 & 21.6 & 57.5 & 93.1 & 78.4 & 15.7 & 24.2 & 24.2 & 4.0 & 5.0 & 1.1 & 1.3 & 0.0 \\
\cmidrule(l{1pt}r{0pt}){2-22}
& $B^2$ CP  & 1.9 & $-$14.6 & $-$24.3 & $-$25.2 & 8.3 & 9.3 & $-$1.3 & $-$11.2 & 12.3 & 31.7 & 53.2 & 71.5 & 8.9 & 8.0 & $-$4.3 & $-$15.2 & $-$1.2 & $-$17.6 & $-$25.7 & $-$24.3 \\
\midrule

\multirow{4}{*}{7 months}
& GSD          & 0.0 & 0.0 & 0.0 & 0.0 & 0.0 & 0.0 & 0.0 & 0.0 & 0.0 & 0.0 & 0.0 & 0.0 & 0.0 & 0.0 & 0.0 & 0.0 & 0.0 & 0.0 & 0.0 & 0.0 \\
\cmidrule(l{1pt}r{0pt}){2-22}
& EP           & 9.2 & 3.0 & 3.3 & 0.0 & 15.7 & 17.2 & 26.0 & 10.6 & 18.1 & 29.6 & 74.5 & 71.5 & 13.7 & 15.1 & 21.0 & 5.7 & 5.9 & 1.5 & 1.3 & 0.0 \\
\cmidrule(l{1pt}r{0pt}){2-22}
& $B^2$ EP     & 7.6 & 5.9 & 2.7 & 0.0 & 15.8 & 30.3 & 28.3 & 10.6 & 18.8 & 51.0 & 90.5 & 71.5 & 13.9 & 26.2 & 23.5 & 5.7 & 4.9 & 3.4 & 0.9 & 0.0 \\
\cmidrule(l{1pt}r{0pt}){2-22}
& $B^2$ CP  & 2.7 & $-$11.0 & $-$22.1 & $-$24.3 & 8.4 & 8.5 & 1.8 & $-$7.8 & 11.6 & 29.4 & 48.4 & 67.8 & 8.2 & 9.1 & $-$1.0 & $-$13.9 & 0.2 & $-$14.4 & $-$24.0 & $-$25.8 \\
\midrule

\multirow{4}{*}{12 months}
& GSD          & 0.0 & 0.0 & 0.0 & 0.0 & 0.0 & 0.0 & 0.0 & 0.0 & 0.0 & 0.0 & 0.0 & 0.0 & 0.0 & 0.0 & 0.0 & 0.0 & 0.0 & 0.0 & 0.0 & 0.0 \\
\cmidrule(l{1pt}r{0pt}){2-22}
& EP           & 4.2 & 4.3 & 2.8 & 0.4 & 7.1 & 17.6 & 24.2 & 14.4 & 8.5 & 28.5 & 61.7 & 78.9 & 6.1 & 16.1 & 21.4 & 10.6 & 3.1 & 2.4 & 1.3 & 0.1 \\
\cmidrule(l{1pt}r{0pt}){2-22}
& $B^2$ EP     & 5.5 & 7.8 & 3.8 & 0.3 & 13.5 & 29.0 & 31.0 & 17.3 & 16.3 & 45.4 & 81.2 & 103.8 & 11.5 & 24.7 & 27.4 & 12.1 & 3.6 & 4.0 & 1.5 & 0.0 \\
\cmidrule(l{1pt}r{0pt}){2-22}
& $B^2$ CP  & 3.7 & $-$9.5 & $-$19.6 & $-$24.8 & 8.3 & 10.0 & 3.8 & $-$2.7 & 10.6 & 25.8 & 43.5 & 61.9 & 7.3 & 10.0 & 1.8 & $-$9.6 & 0.8 & $-$13.6 & $-$23.3 & $-$24.2 \\
\bottomrule
\end{tabular}%
}%
}

\vspace{2pt}

\resizebox{1.05\textwidth}{!}{%
{\setlength{\tabcolsep}{2.8pt}%
\begin{tabular}{@{}c c
  *{4}{c}@{\hspace{8pt}}
  *{4}{c}@{\hspace{8pt}}
  *{4}{c}@{\hspace{8pt}}
  *{4}{c}@{\hspace{8pt}}
  *{4}{c}@{}}
\toprule
\multirow{3}{*}[-0.3em]{\textbf{\begin{tabular}{@{}c@{}}Median PFS\\(Control)\\HR=0.6\end{tabular}}} &
\multirow{3}{*}[-0.3em]{\textbf{Method}} &
\multicolumn{20}{c}{\textbf{Discrepancy $\delta$}} \\
\cmidrule(lr){3-22}
& & \multicolumn{4}{c}{\textbf{$\mathbf{-0.4}$}} & \multicolumn{4}{c}{\textbf{$\mathbf{-0.2}$}} & \multicolumn{4}{c}{\textbf{$\mathbf{0}$}} & \multicolumn{4}{c}{\textbf{$\mathbf{0.2}$}} & \multicolumn{4}{c}{\textbf{$\mathbf{0.4}$}} \\
\cmidrule(l{0pt}r{5.5pt}){3-6}
\cmidrule(l{0pt}r{5.5pt}){7-10}
\cmidrule(l{0pt}r{5.5pt}){11-14}
\cmidrule(l{0pt}r{5.5pt}){15-18}
\cmidrule(l{0pt}r{0pt}){19-22}
& & \textbf{20} & \textbf{60} & \textbf{120} & \textbf{200} &
    \textbf{20} & \textbf{60} & \textbf{120} & \textbf{200} &
    \textbf{20} & \textbf{60} & \textbf{120} & \textbf{200} &
    \textbf{20} & \textbf{60} & \textbf{120} & \textbf{200} &
    \textbf{20} & \textbf{60} & \textbf{120} & \textbf{200} \\
\midrule

\multirow{4}{*}{4 months}
& GSD          & 0.0 & 0.0 & 0.0 & 0.0 & 0.0 & 0.0 & 0.0 & 0.0 & 0.0 & 0.0 & 0.0 & 0.0 & 0.0 & 0.0 & 0.0 & 0.0 & 0.0 & 0.0 & 0.0 & 0.0 \\
\cmidrule(l{1pt}r{0pt}){2-22}
& EP           & 7.7 & 0.7 & $-$0.3 & $-$0.5 & 12.6 & 7.6 & 10.6 & 13.3 & 14.8 & 14.3 & 44.1 & 102.8 & 11.1 & 5.5 & 8.2 & 9.5 & 5.0 & 0.0 & $-$0.5 & $-$0.9 \\
\cmidrule(l{1pt}r{0pt}){2-22}
& $B^2$ EP     & 8.2 & 0.7 & 2.8 & $-$0.5 & 17.0 & 28.8 & 30.2 & 13.3 & 20.3 & 57.4 & 97.8 & 102.8 & 14.4 & 24.6 & 23.8 & 9.5 & 4.6 & $-$1.2 & $-$0.1 & $-$0.9 \\
\cmidrule(l{1pt}r{0pt}){2-22}
& $B^2$ CP  & 2.6 & $-$4.5 & $-$9.8 & $-$12.9 & 8.5 & 13.1 & 12.6 & 12.4 & 11.0 & 24.5 & 37.9 & 47.2 & 7.2 & 11.4 & 9.9 & 7.5 & 0.4 & $-$7.8 & $-$13.5 & $-$16.2 \\
\midrule

\multirow{4}{*}{7 months}
& GSD          & 0.0 & 0.0 & 0.0 & 0.0 & 0.0 & 0.0 & 0.0 & 0.0 & 0.0 & 0.0 & 0.0 & 0.0 & 0.0 & 0.0 & 0.0 & 0.0 & 0.0 & 0.0 & 0.0 & 0.0 \\
\cmidrule(l{1pt}r{0pt}){2-22}
& EP           & 8.8 & 4.4 & 3.7 & $-$0.6 & 14.2 & 18.0 & 27.0 & 14.0 & 16.5 & 27.6 & 73.1 & 80.3 & 12.3 & 14.2 & 20.9 & 11.6 & 5.5 & 1.8 & 0.3 & $-$0.8 \\
\cmidrule(l{1pt}r{0pt}){2-22}
& $B^2$ EP     & 8.0 & 6.7 & 2.2 & $-$0.6 & 14.9 & 29.4 & 30.2 & 14.0 & 17.3 & 48.6 & 89.0 & 80.3 & 12.5 & 24.9 & 24.1 & 11.6 & 4.5 & 1.7 & $-$1.2 & $-$0.8 \\
\cmidrule(l{1pt}r{0pt}){2-22}
& $B^2$ CP  & 3.1 & $-$2.8 & $-$8.3 & $-$11.3 & 7.6 & 12.9 & 13.1 & 14.1 & 9.4 & 22.6 & 34.9 & 43.8 & 6.6 & 10.9 & 9.9 & 7.8 & 0.4 & $-$6.7 & $-$12.7 & $-$15.5 \\
\midrule

\multirow{4}{*}{12 months}
& GSD          & 0.0 & 0.0 & 0.0 & 0.0 & 0.0 & 0.0 & 0.0 & 0.0 & 0.0 & 0.0 & 0.0 & 0.0 & 0.0 & 0.0 & 0.0 & 0.0 & 0.0 & 0.0 & 0.0 & 0.0 \\
\cmidrule(l{1pt}r{0pt}){2-22}
& EP           & 4.2 & 5.5 & 3.7 & $-$0.1 & 6.7 & 17.0 & 25.4 & 20.2 & 7.4 & 26.0 & 56.7 & 80.5 & 6.0 & 14.9 & 20.9 & 13.9 & 2.9 & 2.6 & 0.9 & $-$0.6 \\
\cmidrule(l{1pt}r{0pt}){2-22}
& $B^2$ EP     & 7.1 & 7.9 & 3.0 & $-$1.0 & 12.9 & 27.0 & 31.5 & 24.6 & 15.0 & 41.6 & 77.0 & 103.8 & 10.9 & 23.7 & 26.5 & 16.8 & 3.3 & 3.0 & $-$0.1 & $-$1.3 \\
\cmidrule(l{1pt}r{0pt}){2-22}
& $B^2$ CP  & 3.0 & $-$1.3 & $-$6.8 & $-$10.4 & 6.9 & 12.0 & 13.4 & 13.8 & 8.4 & 20.0 & 31.6 & 41.4 & 6.1 & 10.5 & 10.9 & 8.4 & 1.0 & $-$5.6 & $-$11.6 & $-$14.9 \\
\bottomrule
\end{tabular}%
}%
}

\vspace{2pt}
\noindent\makebox[\textwidth][c]{%
\begin{minipage}{1.05\textwidth}
\footnotesize
GSD: group sequential design; EP: elastic prior; \(B^2\)-EP: calibrated elastic prior; \(B^2\)-CP: calibrated commensurate prior. Values are mean interim-1 treatment-arm event-scale ESS across 1000 simulation replicates.
\end{minipage}%
}

\end{table}
\end{landscape}
\restoregeometry
\clearpage

\clearpage
\newgeometry{margin=0.1in}
\begin{landscape}
\begin{table}[p]
\centering
\caption{Final-analysis treatment-arm event-scale ESS across sensitivity scenarios}
\label{tab:final_treatment_ess_sensitivity}
\scriptsize
\setlength{\tabcolsep}{1.0pt}
\renewcommand{\arraystretch}{1.15}

\resizebox{1.05\textwidth}{!}{%
{\setlength{\tabcolsep}{2.8pt}%
\begin{tabular}{@{}c c
  *{4}{c}@{\hspace{8pt}}
  *{4}{c}@{\hspace{8pt}}
  *{4}{c}@{\hspace{8pt}}
  *{4}{c}@{\hspace{8pt}}
  *{4}{c}@{}}
\toprule
\multirow{3}{*}[-0.3em]{\textbf{\begin{tabular}{@{}c@{}}Median PFS\\(Control)\\HR=0.8\end{tabular}}} &
\multirow{3}{*}[-0.3em]{\textbf{Method}} &
\multicolumn{20}{c}{\textbf{Discrepancy $\delta$}} \\
\cmidrule(lr){3-22}
& & \multicolumn{4}{c}{\textbf{$\mathbf{-0.4}$}} & \multicolumn{4}{c}{\textbf{$\mathbf{-0.2}$}} & \multicolumn{4}{c}{\textbf{$\mathbf{0}$}} & \multicolumn{4}{c}{\textbf{$\mathbf{0.2}$}} & \multicolumn{4}{c}{\textbf{$\mathbf{0.4}$}} \\
\cmidrule(l{0pt}r{5.5pt}){3-6}
\cmidrule(l{0pt}r{5.5pt}){7-10}
\cmidrule(l{0pt}r{5.5pt}){11-14}
\cmidrule(l{0pt}r{5.5pt}){15-18}
\cmidrule(l{0pt}r{0pt}){19-22}
& & \textbf{20} & \textbf{60} & \textbf{120} & \textbf{200} &
    \textbf{20} & \textbf{60} & \textbf{120} & \textbf{200} &
    \textbf{20} & \textbf{60} & \textbf{120} & \textbf{200} &
    \textbf{20} & \textbf{60} & \textbf{120} & \textbf{200} &
    \textbf{20} & \textbf{60} & \textbf{120} & \textbf{200} \\
\midrule

\multirow{4}{*}{4 months}
& GSD          & 0.0 & 0.0 & 0.0 & 0.0 & 0.0 & 0.0 & 0.0 & 0.0 & 0.0 & 0.0 & 0.0 & 0.0 & 0.0 & 0.0 & 0.0 & 0.0 & 0.0 & 0.0 & 0.0 & 0.0 \\
\cmidrule(l{1pt}r{0pt}){2-22}
& EP           & 7.9 & 0.6 & 0.2 & 0.0 & 14.1 & 7.4 & 7.8 & 6.6 & 17.7 & 15.4 & 43.8 & 85.2 & 13.1 & 6.1 & 7.1 & 5.8 & 4.9 & 0.5 & 0.0 & 0.0 \\
\cmidrule(l{1pt}r{0pt}){2-22}
& $B^2$ EP     & 8.1 & 1.5 & 3.5 & 0.0 & 18.4 & 26.9 & 27.8 & 6.6 & 22.9 & 61.5 & 97.7 & 85.2 & 16.7 & 25.9 & 23.5 & 5.8 & 5.6 & 1.6 & 1.5 & 0.0 \\
\cmidrule(l{1pt}r{0pt}){2-22}
& $B^2$ CP  & 2.8 & $-$16.3 & $-$25.7 & $-$25.3 & 10.4 & 10.3 & $-$2.2 & $-$16.1 & 12.2 & 33.4 & 55.4 & 77.8 & 9.9 & 9.4 & $-$3.6 & $-$21.0 & $-$0.4 & $-$16.5 & $-$22.9 & $-$23.7 \\
\midrule

\multirow{4}{*}{7 months}
& GSD          & 0.0 & 0.0 & 0.0 & 0.0 & 0.0 & 0.0 & 0.0 & 0.0 & 0.0 & 0.0 & 0.0 & 0.0 & 0.0 & 0.0 & 0.0 & 0.0 & 0.0 & 0.0 & 0.0 & 0.0 \\
\cmidrule(l{1pt}r{0pt}){2-22}
& EP           & 10.4 & 2.2 & 2.6 & 0.1 & 17.5 & 17.5 & 24.2 & 7.0 & 20.8 & 33.8 & 83.4 & 83.1 & 15.5 & 15.0 & 21.5 & 5.0 & 6.6 & 1.5 & 1.1 & 0.0 \\
\cmidrule(l{1pt}r{0pt}){2-22}
& $B^2$ EP     & 9.0 & 6.4 & 2.1 & 0.1 & 17.4 & 33.1 & 28.1 & 7.0 & 21.5 & 58.5 & 100.7 & 83.1 & 15.7 & 28.3 & 24.1 & 5.0 & 5.9 & 3.7 & 0.8 & 0.0 \\
\cmidrule(l{1pt}r{0pt}){2-22}
& $B^2$ CP  & $-$0.3 & $-$15.9 & $-$22.2 & $-$23.0 & 12.5 & 6.5 & 0.2 & $-$14.4 & 13.2 & 31.9 & 51.1 & 75.6 & 10.0 & 8.9 & $-$5.5 & $-$18.2 & 0.1 & $-$17.3 & $-$23.2 & $-$22.9 \\
\midrule

\multirow{4}{*}{12 months}
& GSD          & 0.0 & 0.0 & 0.0 & 0.0 & 0.0 & 0.0 & 0.0 & 0.0 & 0.0 & 0.0 & 0.0 & 0.0 & 0.0 & 0.0 & 0.0 & 0.0 & 0.0 & 0.0 & 0.0 & 0.0 \\
\cmidrule(l{1pt}r{0pt}){2-22}
& EP           & 4.3 & 3.1 & 2.0 & 0.3 & 8.2 & 18.8 & 23.2 & 12.3 & 10.3 & 35.0 & 77.0 & 90.9 & 7.7 & 17.2 & 20.4 & 9.1 & 2.9 & 1.8 & 0.9 & 0.0 \\
\cmidrule(l{1pt}r{0pt}){2-22}
& $B^2$ EP     & 6.8 & 6.9 & 2.8 & 0.3 & 16.0 & 32.6 & 31.1 & 15.1 & 19.4 & 54.8 & 100.0 & 120.7 & 13.9 & 28.1 & 26.4 & 11.1 & 4.7 & 4.1 & 1.3 & 0.0 \\
\cmidrule(l{1pt}r{0pt}){2-22}
& $B^2$ CP  & 4.0 & $-$11.8 & $-$24.0 & $-$24.6 & 10.3 & 9.5 & $-$1.0 & $-$10.6 & 13.0 & 27.5 & 50.0 & 70.7 & 10.2 & 8.2 & $-$3.0 & $-$17.1 & $-$0.1 & $-$16.4 & $-$23.9 & $-$23.6 \\
\bottomrule
\end{tabular}%
}%
}

\vspace{2pt}

\resizebox{1.05\textwidth}{!}{%
{\setlength{\tabcolsep}{2.8pt}%
\begin{tabular}{@{}c c
  *{4}{c}@{\hspace{8pt}}
  *{4}{c}@{\hspace{8pt}}
  *{4}{c}@{\hspace{8pt}}
  *{4}{c}@{\hspace{8pt}}
  *{4}{c}@{}}
\toprule
\multirow{3}{*}[-0.3em]{\textbf{\begin{tabular}{@{}c@{}}Median PFS\\(Control)\\HR=0.6\end{tabular}}} &
\multirow{3}{*}[-0.3em]{\textbf{Method}} &
\multicolumn{20}{c}{\textbf{Discrepancy $\delta$}} \\
\cmidrule(lr){3-22}
& & \multicolumn{4}{c}{\textbf{$\mathbf{-0.4}$}} & \multicolumn{4}{c}{\textbf{$\mathbf{-0.2}$}} & \multicolumn{4}{c}{\textbf{$\mathbf{0}$}} & \multicolumn{4}{c}{\textbf{$\mathbf{0.2}$}} & \multicolumn{4}{c}{\textbf{$\mathbf{0.4}$}} \\
\cmidrule(l{0pt}r{5.5pt}){3-6}
\cmidrule(l{0pt}r{5.5pt}){7-10}
\cmidrule(l{0pt}r{5.5pt}){11-14}
\cmidrule(l{0pt}r{5.5pt}){15-18}
\cmidrule(l{0pt}r{0pt}){19-22}
& & \textbf{20} & \textbf{60} & \textbf{120} & \textbf{200} &
    \textbf{20} & \textbf{60} & \textbf{120} & \textbf{200} &
    \textbf{20} & \textbf{60} & \textbf{120} & \textbf{200} &
    \textbf{20} & \textbf{60} & \textbf{120} & \textbf{200} &
    \textbf{20} & \textbf{60} & \textbf{120} & \textbf{200} \\
\midrule

\multirow{4}{*}{4 months}
& GSD          & 0.0 & 0.0 & 0.0 & 0.0 & 0.0 & 0.0 & 0.0 & 0.0 & 0.0 & 0.0 & 0.0 & 0.0 & 0.0 & 0.0 & 0.0 & 0.0 & 0.0 & 0.0 & 0.0 & 0.0 \\
\cmidrule(l{1pt}r{0pt}){2-22}
& EP           & 8.1 & 1.1 & 0.0 & $-$0.4 & 13.5 & 6.6 & 12.5 & 12.3 & 16.3 & 16.4 & 47.6 & 102.1 & 12.8 & 7.0 & 9.3 & 8.9 & 5.4 & 0.4 & 0.0 & $-$0.4 \\
\cmidrule(l{1pt}r{0pt}){2-22}
& $B^2$ EP     & 9.3 & 1.8 & 4.1 & $-$0.4 & 18.3 & 31.4 & 32.4 & 12.3 & 22.8 & 65.3 & 106.3 & 102.1 & 16.2 & 26.7 & 27.2 & 8.9 & 5.4 & 0.6 & 1.8 & $-$0.4 \\
\cmidrule(l{1pt}r{0pt}){2-22}
& $B^2$ CP  & 2.7 & $-$6.9 & $-$13.4 & $-$16.5 & 9.0 & 13.2 & 11.3 & 9.5 & 12.0 & 28.4 & 44.6 & 57.2 & 8.3 & 11.5 & 8.3 & 4.3 & 0.1 & $-$9.6 & $-$15.9 & $-$18.3 \\
\midrule

\multirow{4}{*}{7 months}
& GSD          & 0.0 & 0.0 & 0.0 & 0.0 & 0.0 & 0.0 & 0.0 & 0.0 & 0.0 & 0.0 & 0.0 & 0.0 & 0.0 & 0.0 & 0.0 & 0.0 & 0.0 & 0.0 & 0.0 & 0.0 \\
\cmidrule(l{1pt}r{0pt}){2-22}
& EP           & 10.1 & 3.8 & 3.2 & $-$0.4 & 16.7 & 18.6 & 29.7 & 11.1 & 19.0 & 32.6 & 87.5 & 94.1 & 14.6 & 16.2 & 23.6 & 8.5 & 6.2 & 2.4 & 1.4 & $-$0.2 \\
\cmidrule(l{1pt}r{0pt}){2-22}
& $B^2$ EP     & 9.6 & 6.8 & 2.1 & $-$0.4 & 17.4 & 33.3 & 32.7 & 11.1 & 20.2 & 57.4 & 108.1 & 94.1 & 14.9 & 27.5 & 26.5 & 8.5 & 5.4 & 3.5 & 0.5 & $-$0.2 \\
\cmidrule(l{1pt}r{0pt}){2-22}
& $B^2$ CP  & 2.7 & $-$5.8 & $-$12.7 & $-$15.5 & 8.6 & 13.3 & 11.6 & 10.0 & 10.9 & 26.3 & 42.3 & 53.9 & 7.5 & 10.8 & 8.9 & 4.9 & 0.1 & $-$9.3 & $-$15.6 & $-$17.9 \\
\midrule

\multirow{4}{*}{12 months}
& GSD          & 0.0 & 0.0 & 0.0 & 0.0 & 0.0 & 0.0 & 0.0 & 0.0 & 0.0 & 0.0 & 0.0 & 0.0 & 0.0 & 0.0 & 0.0 & 0.0 & 0.0 & 0.0 & 0.0 & 0.0 \\
\cmidrule(l{1pt}r{0pt}){2-22}
& EP           & 4.5 & 4.7 & 2.9 & 0.1 & 7.4 & 19.6 & 27.6 & 17.1 & 9.2 & 33.1 & 77.9 & 107.5 & 7.5 & 17.4 & 22.5 & 11.6 & 3.4 & 2.8 & 1.3 & 0.0 \\
\cmidrule(l{1pt}r{0pt}){2-22}
& $B^2$ EP     & 8.3 & 7.6 & 3.1 & $-$0.3 & 16.1 & 31.7 & 34.1 & 20.9 & 18.7 & 52.8 & 103.1 & 140.3 & 13.7 & 27.2 & 28.1 & 14.6 & 4.5 & 3.8 & 0.9 & $-$0.2 \\
\cmidrule(l{1pt}r{0pt}){2-22}
& $B^2$ CP  & 3.1 & $-$4.9 & $-$12.1 & $-$15.0 & 7.7 & 12.9 & 11.8 & 10.7 & 10.5 & 24.4 & 39.5 & 51.9 & 7.3 & 11.2 & 9.5 & 5.4 & 0.3 & $-$8.6 & $-$15.1 & $-$17.4 \\
\bottomrule
\end{tabular}%
}%
}

\vspace{2pt}
\noindent\makebox[\textwidth][c]{%
\begin{minipage}{1.05\textwidth}
\footnotesize
GSD: group sequential design; EP: elastic prior; \(B^2\)-EP: calibrated elastic prior; \(B^2\)-CP: calibrated commensurate prior. Values are mean interim-1 treatment-arm event-scale ESS across 1000 simulation replicates.
\end{minipage}%
}

\end{table}
\end{landscape}
\restoregeometry
\clearpage

\section{ IA2 scheduling decisions for control median PFS of 4 months}
\begin{table}[H]
\centering
\caption{IA2 scheduling decision table}
\label{tab:decision_rules_compare}
\footnotesize
\setlength{\tabcolsep}{3pt}
\renewcommand{\arraystretch}{1.15}

\begin{tabularx}{\textwidth}{
>{\centering\arraybackslash}p{2.6cm}
*{5}{>{\centering\arraybackslash}X}
}
\toprule
\multirow{2}{*}{\makecell{\textbf{Phase II}\\ \textbf{HR interval}}} &
\multicolumn{5}{c}{\textbf{Phase III IA1 HR interval}} \\
\cmidrule(lr){2-6}
&
\makecell{$\leq 0.50$} &
\makecell{$>0.50$\\ to $\leq 0.60$} &
\makecell{$>0.60$\\ to $\leq 0.70$} &
\makecell{$>0.70$\\ to $\leq 0.80$} &
\makecell{$>0.80$\\ to $\leq 0.90$} \\
\midrule

\rule{0pt}{2.3ex}$\leq 0.50$
& IA1 & $0.50$ & $0.70$ & $0.80$ & $1.0$ \\

\rule{0pt}{2.3ex}$>0.50$ to $\leq 0.60$
& $0.60$ & $0.60$ & $0.70$ & $0.80$ & $1.0$ \\

\rule{0pt}{2.3ex}$>0.60$ to $\leq 0.70$
& $0.60$ & $0.60$ & $0.70$ & $0.80$ & $1.0$ \\

\rule{0pt}{2.3ex}$>0.70$ to $\leq 0.80$
& $0.70$ & $0.70$ & $0.80$ & $0.80$ & $1.0$ \\

\rule{0pt}{2.3ex}$>0.80$ to $\leq 0.90$
& $1.0$ & $1.0$ & $1.0$ & $1.0$ & $1.0$ \\

\bottomrule
\end{tabularx}

\vspace{0.5ex}
\parbox{\textwidth}{%
\footnotesize
Results are shown for \(m_C^{(III)}=4\) months. 0.50--0.80
denote IA2 information fractions, and 1.0 denotes final analysis.
}
\end{table}

\begin{table}[H]
\centering
\caption{Recommended IA2 event counts}
\label{tab:decision_rules_compare_events}
\footnotesize
\setlength{\tabcolsep}{3pt}
\renewcommand{\arraystretch}{1.15}

\begin{tabularx}{\textwidth}{
>{\centering\arraybackslash}p{2.6cm}
*{5}{>{\centering\arraybackslash}X}
}
\toprule
\multirow{2}{*}{\makecell{\textbf{Phase II}\\ \textbf{HR interval}}} &
\multicolumn{5}{c}{\textbf{Phase III IA1 HR interval}} \\
\cmidrule(lr){2-6}
&
\makecell{$\leq 0.50$} &
\makecell{$>0.50$\\ to $\leq 0.60$} &
\makecell{$>0.60$\\ to $\leq 0.70$} &
\makecell{$>0.70$\\ to $\leq 0.80$} &
\makecell{$>0.80$\\ to $\leq 0.90$} \\
\midrule

\rule{0pt}{2.3ex}$\leq 0.50$
& 36 & 82 & 236 & 688 & 3853 \\

\rule{0pt}{2.3ex}$>0.50$ to $\leq 0.60$
& 54 & 98 & 236 & 688 & 3853 \\

\rule{0pt}{2.3ex}$>0.60$ to $\leq 0.70$
& 54 & 98 & 236 & 688 & 3853 \\

\rule{0pt}{2.3ex}$>0.70$ to $\leq 0.80$
& 63 & 115 & 269 & 688 & 3853 \\

\rule{0pt}{2.3ex}$>0.80$ to $\leq 0.90$
& 89 & 163 & 336 & 859 & 3853 \\

\bottomrule
\end{tabularx}

\vspace{0.5ex}
\parbox{\textwidth}{%
\footnotesize
Results are shown for \(m_C^{(III)}=4\) months. Entries are the IA2 event
targets corresponding to Table~\ref{tab:decision_rules_compare}.
}
\end{table}

\section{ IA2 scheduling decisions for control median PFS of 12 months}
\begin{table}[H]
\centering
\caption{IA2 scheduling decision table}
\label{tab:decision_rules_compare}
\footnotesize
\setlength{\tabcolsep}{3pt}
\renewcommand{\arraystretch}{1.15}

\begin{tabularx}{\textwidth}{
>{\centering\arraybackslash}p{2.6cm}
*{5}{>{\centering\arraybackslash}X}
}
\toprule
\multirow{2}{*}{\makecell{\textbf{Phase II}\\ \textbf{HR interval}}} &
\multicolumn{5}{c}{\textbf{Phase III IA1 HR interval}} \\
\cmidrule(lr){2-6}
&
\makecell{$\leq 0.50$} &
\makecell{$>0.50$\\ to $\leq 0.60$} &
\makecell{$>0.60$\\ to $\leq 0.70$} &
\makecell{$>0.70$\\ to $\leq 0.80$} &
\makecell{$>0.80$\\ to $\leq 0.90$} \\
\midrule

\rule{0pt}{2.3ex}$\leq 0.50$
& $0.50$ & $0.50$ & $0.70$ & $0.80$ & $1.0$ \\

\rule{0pt}{2.3ex}$>0.50$ to $\leq 0.60$
& $0.60$ & $0.60$ & $0.70$ & $0.80$ & $1.0$ \\

\rule{0pt}{2.3ex}$>0.60$ to $\leq 0.70$
& $0.70$ & $0.70$ & $0.70$ & $0.80$ & $1.0$ \\

\rule{0pt}{2.3ex}$>0.70$ to $\leq 0.80$
& $0.80$ & $0.80$ & $0.80$ & $0.80$ & $1.0$ \\

\rule{0pt}{2.3ex}$>0.80$ to $\leq 0.90$
& $1.0$ & $1.0$ & $1.0$ & $1.0$ & $1.0$ \\

\bottomrule
\end{tabularx}

\vspace{0.5ex}
\parbox{\textwidth}{%
\footnotesize
Results are shown for \(m_C^{(III)}=12\) months. 0.50--0.80
denote IA2 information fractions, and 1.0 denotes final analysis.
}
\end{table}

\begin{table}[H]
\centering
\caption{Recommended IA2 event counts}
\label{tab:decision_rules_compare_events}
\footnotesize
\setlength{\tabcolsep}{3pt}
\renewcommand{\arraystretch}{1.15}

\begin{tabularx}{\textwidth}{
>{\centering\arraybackslash}p{2.6cm}
*{5}{>{\centering\arraybackslash}X}
}
\toprule
\multirow{2}{*}{\makecell{\textbf{Phase II}\\ \textbf{HR interval}}} &
\multicolumn{5}{c}{\textbf{Phase III IA1 HR interval}} \\
\cmidrule(lr){2-6}
&
\makecell{$\leq 0.50$} &
\makecell{$>0.50$\\ to $\leq 0.60$} &
\makecell{$>0.60$\\ to $\leq 0.70$} &
\makecell{$>0.70$\\ to $\leq 0.80$} &
\makecell{$>0.80$\\ to $\leq 0.90$} \\
\midrule

\rule{0pt}{2.3ex}$\leq 0.50$
& 45 & 82 & 236 & 688 & 3853 \\

\rule{0pt}{2.3ex}$>0.50$ to $\leq 0.60$
& 54 & 98 & 236 & 688 & 3853 \\

\rule{0pt}{2.3ex}$>0.60$ to $\leq 0.70$
& 63 & 115 & 236 & 688 & 3853 \\

\rule{0pt}{2.3ex}$>0.70$ to $\leq 0.80$
& 72 & 131 & 269 & 688 & 3853 \\

\rule{0pt}{2.3ex}$>0.80$ to $\leq 0.90$
& 89 & 163 & 336 & 859 & 3853 \\

\bottomrule
\end{tabularx}

\vspace{0.5ex}
\parbox{\textwidth}{%
\footnotesize
Results are shown for \(m_C^{(III)}=12\) months. Entries are the IA2 event
targets corresponding to Table~\ref{tab:decision_rules_compare}.
}
\end{table}


\end{document}